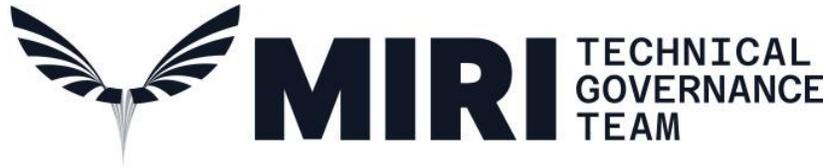

November 2024

# Mechanisms to Verify International Agreements About AI Development

Aaron Scher, Lisa Thiergart

# Abstract

International agreements about AI development may be required to reduce catastrophic risks from advanced AI systems. However, agreements about such a high-stakes technology must be backed by verification mechanisms—processes or tools that give one party greater confidence that another is following the agreed-upon rules, typically by detecting violations. This report gives an overview of potential verification approaches for three example policy goals, aiming to demonstrate how countries could practically verify claims about each other's AI development and deployment. The focus is on international agreements and state-involved AI development, but these approaches could also be applied to domestic regulation of companies. While many of the ideal solutions for verification are not yet technologically feasible, we emphasize that increased access (e.g., physical inspections of data centers) can often substitute for these technical approaches. Therefore, we remain hopeful that significant political will could enable ambitious international coordination, with strong verification mechanisms, to reduce catastrophic AI risks.

Correspondence to: aaron.scher@intelligence.org, lisa@intelligence.org

# Executive Summary

## About this Report

This report provides an overview of mechanisms and approaches that could be used to verify compliance with future international agreements about AI development. Drawing on a review of the literature and discussions with relevant experts, this report includes rough feasibility estimates and R&D timelines for various mechanisms, with in-depth discussion of some mechanisms that are underexplored or particularly promising: AI-enabled approaches, model behavior specifications, interconnect bandwidth limits, and more. Given the nascent state of AI verification research and the rapidly evolving landscape of AI development, this analysis serves as a foundation for further work.

Most readers will benefit from reading the executive summary. Those working on international or domestic verification around AI will benefit from reading the entire report and appendix. The perspectives shared in this report are those of the authors and are not intended to represent the views of their organization.

## Background and Motivation

A growing number of experts believe the development of advanced AI systems could cause catastrophic harm, such as human extinction (Bengio et al., 2024; Carlsmith, 2024; Grace et al., 2024; Hendrycks et al., 2023). If such risks become more apparent, governments may wish to create international agreements to govern AI development or deployment. However, some actors may perceive a strategic benefit in violating such an agreement, so there is a substantial risk of defection. Therefore, these agreements will benefit from governments being able to verify the compliance of foreign (and domestic) actors rather than relying on goodwill. This report provides an overview of approaches and mechanisms that could be used for such verification.

This report makes a few scope and framing choices: it focuses on verification and accompanying self-reporting measures, rather than other aspects of international agreements (such as benefit sharing, enforcement, scientific collaboration, etc.), and it focuses on international rather than domestic regulation. Conveniently, **many of the mechanisms discussed are also applicable to verification of domestic regulations**. Additionally, this report assumes substantial political will (i.e., committed support from key decision-makers) for international cooperation to prevent major risks from general-purpose, advanced AI systems. We focus on a situation where governments are playing a major role in AI development, and **attempts to subvert verification regimes are coming from well-resourced state actors**. We note that international cooperation in AI development may not be necessary if, for instance, the United States' lead in AI is upheld via strong export controls for AI chips and very strong cybersecurity around AI model weights (Nevo et al., 2024). In that case, the U.S. federal government could control frontier AI development via domestic regulation, without the fear of frontier AI development happening elsewhere—at least in the short term. This report investigates the mechanisms available for verification, if international cooperation is needed. Verification mechanisms are processes or tools that give one party greater confidence that the other is following the agreed-upon rules, typically by detecting violations.

Given the potential for existential risks and the complexity of fully addressing them, this report examines selected policy goals as a means to illustrate the broader challenge of verification. These goals, while not necessarily optimal, are used to highlight the kinds of technical and institutional measures that could enhance compliance with a variety of policy goals. This report examines **three policy goals: [locating AI compute](#) (the physical computers used for AI activities), [verifying that some known compute is not being used for a large training run](#), and [verifying the authenticity of model evaluations](#).** These goals are primarily chosen because they help highlight the space of verification methods available, rather than because they are the most desirable goals. For instance, ideal international coordination might restrict dangerous AI development and deployment, perhaps based on international approval of a safety case (Clymer et al., 2024), but this is a difficult goal to discuss with the field's current, underdeveloped understanding of AI safety, so we use the proxy of pre-training compute thresholds (we briefly discuss verifying that a safety case is followed in [the appendix](#)). For each goal, we describe multiple verification approaches, often relying on particular technical mechanisms, and provide analysis of the technical and operational [feasibility](#) of these approaches.

## Verifying the location of AI compute

Tracking AI-relevant compute is crucial for AI governance, as chips are the fundamental substrate of AI development. We discuss two potential approaches: **1) Low-tech physical inspections, 2) High-tech on-chip mechanisms for remote location attestation**. Physical inspections could be implemented immediately but require physical access to data centers, i.e., international inspectors visiting data centers, counting existing chips, and setting up security cameras, with the primary goal of ensuring declared chips do not leave their data centers. On-chip mechanisms would involve newly manufactured AI chips, or a secondary processor added alongside existing chips, remotely attesting to their location (e.g., via measuring the response time to a series of servers located throughout the world), but they likely **require improved chip security to prevent tampering** (Aarne et al., 2024; Brass & Aarne, 2024; Kulp et al., 2024; Petrie et al., 2024). In particular, the main security issue is protecting a chip's private key, which it uses for this location attestation: if the key were extracted, the chip's location could be spoofed. On-chip location attestation has the benefits of not requiring physical access to chips and potentially revealing only a chip's general location.

In both approaches, locating AI compute is primarily being done by **tracking AI chips and their supply chain rather than hoping to detect secret data centers**, a choice we discuss in the appendix. Tracking chips appears more promising because advances in distributed training will allow many data centers to effectively act as one, detecting covert data centers may be difficult, AI chips are closer to the target of regulation (AI activities), and the chip supply chain is narrow and amenable to monitoring. In practice, monitoring should focus both on tracking AI chips and data centers, but we expect focusing on AI chips to be more reliable.

Both approaches rely on **securing the AI chip supply chain**. The first requires preventing chip smuggling, and the second requires that manufactured chips have the correct on-chip mechanisms and are sufficiently tamper-proofed. The high-level approach to verifying the location of AI compute is to locate AI chips at an initial point in time (e.g., self-report and physical inspections or by monitoring fabrication plants) and then ensure they remain monitored (e.g., security cameras or on-chip mechanisms with strong security). Compute monitoring may be politically difficult, given the access requirements. This challenging task is made easier by centralizing AI compute—e.g., via export controls—because it reduces the number of governments whose participation is crucial to the success of such a verification regime.

## Verifying that known compute is not being used for a large training run

This section addresses the policy goal of verifying that data centers are not conducting large AI training runs, i.e., total training compute does not exceed some agreed-upon threshold. This policy goal is included because the size of a pre-training run is sometimes a proxy for risk from AI systems (Heim & Koessler, 2024), it is a major focus in AI governance research and policy (EU AI Act, 2024;

Executive Order 14110, 2023), and it allows for a clear decomposition of the space of verification mechanisms. While this report focuses on pre-training, it may be necessary in the future for international coordination to focus on inference or post-training. The verification approach is as follows: **locate AI compute via the first policy goal, exclude compute facilities incapable of supporting large training runs, and verify that capable facilities are not engaging in such activities**.

Exclusion of data centers will likely involve identifying certain chips as "AI chips" and excluding other chips—as is done in current U.S. export controls (Dohmen & Feldgoise, 2023). Data centers could also potentially be excluded based on having only a small number of AI chips. This approach would involve self-reporting and physical inspections to confirm these accounts. However, **distributed training likely poses significant issues for compute exclusion because it could enable multiple small-to-medium amounts of compute (e.g., 5,000 chips) to collectively carry out a large training run**. If there are no restrictions on frontier AI development, large training runs are likely to occur in a few large data centers, but if there are international agreements to prohibit large training runs, subversion attempts will likely be covert, for instance, doing highly distributed training. Given this consideration and the threat model of well-resourced state actors, **it may be necessary to monitor even relatively small data centers**.

Multiple verification approaches can build confidence that some data center that could be used for a large training run is not being used for that purpose. Data centers can engage in "compute accounting" by **keeping a registry of their chip activities, allowing verification that these registries are correct, and then demonstrating there are not enough chip-hours left for a violation**. This verification could include **partial re-running of declared workloads** by a verifier in a mutually trusted data center (i.e., one which all parties are confident is secure enough to not leak sensitive information about the workloads being re-run) (Baker et al., Forthcoming; Shavit, 2023). It could also include **classifying workloads based on high-level measures of AI chips**. Classifying chip activities on the basis of non-invasive measures such as power consumption is likely difficult, given our threat model, but it could be feasible, especially if temporary code access is granted (this would provide in-distribution, correctly labeled data). One classification method that appears especially promising is **interconnect bandwidth limits**—substantially reducing the amount of data that can flow in and out of a set of chips in order to prevent these chips from efficiently participating in a distributed training run while still allowing efficient text inference. For example, 128 chips with high bandwidth interconnect to each other but very low external bandwidth, enough for inference tokens but not training gradients. This is likely implementable with minor modifications to existing technology, physical access and monitoring of compute, but without any code access; this approach is discussed in the appendix.

Workload classification could also involve access to chip activities in a zero-knowledge manner (e.g., sensitive information is not leaked), such as by using **Flexible Hardware-Enabled Guarantee, FlexHEG, mechanisms** (Petrie et al., 2024). The general FlexHEG approach is to have a secure processor that interacts with an AI chip and carries out governance operations, such as saving snapshots of the chip's memory or controlling which other chips the chip can interact with, which is secure against tampering (e.g., with an enclosure which destroys the chip if it detects tampering). One particular implementation of FlexHEG mechanisms could likely be designed and tested in less than two years and then retrofitted to existing chips, making it especially promising. FlexHEG mechanisms could aid with workload classification in numerous ways, e.g., by comparing a chip's memory contents to a declared workload and showing the chip is doing what it should be.

After verifying its chips' activities using a combination of the above methods, a data center would make the argument that there are not enough unverified chip hours remaining to carry out a violation; this is a non-trivial argument for most classification methods, and forthcoming work from Baker and colleagues makes progress on it. Overall, verification of this policy goal appears difficult: partial re-running of workloads requires mutually trusted computing infrastructure, a tall order; many other approaches will be difficult to make adversarially robust, especially given likely advances in algorithmic efficiency and distributed training.

## Verifying the authenticity of model evaluations

Countries may want to evaluate each other's AI models for various reasons, such as ensuring AI capabilities progress is moving slowly or asserting some quality about a model. The main difficulties in doing this are **ensuring the correct model is evaluated**, **ensuring the evaluation process is secure for both parties**, **and ensuring evaluations are effective**. One technical mechanism that can assist with security is **Trusted Execution Environments (TEEs)**—enabled via "Confidential Computing" on NVIDIA GPUs—which could allow only mutually approved code to run, but without giving either party access to sensitive data. Model authenticity could be established by **ensuring the model code and weights run in a TEE are the same during evaluation as deployment or training**. While minimal versions of these techniques will likely be available soon, existing AI chips may not be sufficiently secure against well-resourced nation-state actors. Unfortunately, the science of evaluations is still in its early stages, so effectively verifying many important properties of AI systems may remain elusive even if these technical challenges in security and model authenticity are solved.

## Verifying various policy goals

A few verification mechanisms will be useful for a wide range of policy goals. **Whistleblower programs** and **interviews** are likely to be effective, they have strong precedent, and they **can verify a wide range of claims**, e.g., claims about AI models, development practices, and AI chip production. AI-enabled verification approaches (such as a mutually trusted AI system conducting model evaluations) could allow zero-knowledge verification of complex properties, which would be very useful, but these require favorable developments in AI technology.

Monitoring AI inference (i.e., all copies of a deployed model) is critical to many policy goals, such as AI deployment following a safety case. This is a difficult task due to the requirement that all inference be known and monitored. One promising approach is **using strong security to prevent model weights from leaving a data center, and then monitoring the data center closely**.

Model behavior specifications could also contribute to verification. If AI systems are doing a substantial amount of intellectual labor, it may be desirable to **design AI systems that follow a general set of agreed-upon principles, such as refusing to violate international agreements**. The technical feasibility of such an approach likely varies significantly with AI capability level: we expect it is probably feasible for current AI systems given a few years of R&D (and given substantial effort), but it may be extremely difficult for future AI systems, akin to the general AI alignment problem (Ngo et al., 2024).

## Takeaways

Verifying compliance with many international agreements on AI is likely feasible, even if it is needed very soon, but it will **require substantial political will** and some participation from monitored countries. Core difficulties with verification include advancements in algorithmic progress, advancements in distributed training, the difficulty of classifying AI chip activities in an adversarial setting, and the novel threat landscape with highly capable AI systems. Furthermore, **the vast majority of verification approaches in this report are not ready to be implemented**, requiring years of R&D first. Some mechanisms whose development must be started early include FlexHEGs, mutually trusted data centers and computing infrastructure, centralizing and tracking existing AI chips, centralizing and tracking the chip supply chain, developing inference-only chips or other chips that can only be used for selective AI activities, and securing AI model weights against theft. Many verification mechanisms are **applicable to both international and domestic regulation**, so early work could be crucial even if only domestic regulation is needed. The present analysis points to a few high-priority areas of future work other than designing and prototyping various verification mechanisms discussed: fostering the international political will necessary for effective verification regimes, analyzing the verification space with a specific focus on domestic regulation and less capable adversaries, and designing verification regimes aimed at key policy goals such as frontier model training following a safety case (discussed briefly in the appendix).

Effective verification mechanisms could be a catalyst for international agreements, as has been the case previously (Toivanen, 2017). Strong verification is crucial, regardless of whether the international situation is one of "trust but verify" or "distrust and verify".

# Table of Contents

# About This Report

The perspectives shared in this report are those of the authors and are not intended to represent the views of their organization.

Our main contributions are as follows:

- We lay out the range of verification mechanisms that have been discussed previously, organized around practically verifying particular policy goals. Our work is informed by reviewing previous literature and discussion with relevant experts. Given the nascent state of AI verification research and the rapidly evolving landscape of AI development, this analysis serves as a foundation for further work.
- We discuss some verification mechanisms that have not received significant attention previously: AI-enabled approaches, model behavior specifications, signatures of high-level chip measures, inference classifiers akin to Trust and Safety filters, and more.
- We provide detailed analysis of the challenges and implementation details of interconnect bandwidth limits, an approach discussed previously (Kulp et al., 2024), which we think is relatively promising.
- Our description of the verification landscape indicates numerous areas for future research, some of which require substantial serial time and thus work must begin early.

Most readers will benefit from reading the executive summary. Those working on international or domestic verification around AI will benefit from reading the entire report and appendix.

# Background and Motivation

As artificial intelligence (AI) systems become more capable and large-scale risks become more salient, the international community may come together to regulate AI development through treaties and other agreements. Central to the effectiveness and feasibility of such agreements is whether compliance can be verified by other countries rather than merely being trust-backed. This report provides an overview of verification mechanisms—processes or tools that give one party greater confidence that the other is following the agreed-upon rules, typically by detecting violations—for international agreements about AI development.

This report makes a handful of framing choices:

This report focuses on **verifying international agreements rather than domestic regulations**. This choice is based on the requirement that effective coordination to reduce catastrophic AI risk be global and broadly encompassing. This report also imagines the **major AI development projects of the future to be state-involved efforts**, but many of the mechanisms discussed can also be applied to domestic regulation. While the ideas are more generally applicable, readers could view this document as answering the question, "How would the governments of the United States and China verify that the other is following a bilateral agreement on AI development?"

Therefore, the primary **threat to violating international agreements comes from sophisticated and well-resourced nation-state actors**. When considering if a verification mechanism, such as firmware licensing for AI chips, will stand up to circumvention efforts, readers should imagine said efforts are carried out by the most advanced state actors, such as OC5 as described by Nevo et al. (2024), with physical access to the AI chips. Because the relevant threat actors are well-resourced, verification and enforcement mechanisms should aim to increase the cost of treaty violations by multiple orders of magnitude (e.g., increase the time needed for a prohibited AI training run by >100x). Nevertheless, verification measures that are less effective than this could be useful, especially when stacked together. The benefits of AI development and of violating these international agreements may appear very large, so our threat model includes nation-states spending billions of dollars on AI projects and potentially billions of dollars to subvert verification mechanisms (i.e., covertly violate agreements and bypass the verification mechanisms aimed at detecting violations). It's key to avoid conflating "current AI companies are doing method $X$ rather than $Y$ because it saves some money" and "nation-states could only do method $X$ and not $Y$ in the future".

This report **focuses on the technical and institutional mechanisms needed to verify compliance with policy goals** rather than other aspects of international agreements. International agreements for AI could have other components, such as benefit sharing, scientific collaborations, withdrawal conditions, and enforcement mechanisms used in response to violations. International agreements could also have collective decision-making procedures (Hausenloy et al., 2023), which some verification mechanisms could facilitate, but which are out of scope for this report. Many of the verification approaches in this report include the monitored party making some claim about their AI activities being acceptable and granting enough access for a verifying party to confirm that the claim is correct. For instance, verifying the location of AI chips may involve physical inspections to count chips, and verifying that a particular workload was run on some AI chips may involve re-running that workload to confirm it happened as stated. Some approaches require the participation of the monitored party but do not assume the monitored party is providing truthful information. Instead, verification mechanisms should make it difficult to lie without getting caught. Some verification mechanisms would be effective at preventing overt violations of an international agreement, but this report generally focuses on a situation where violations are covert, and the goal is to detect them, while enforcement is out of scope for this report.

This report focuses on regulations pertaining to **risks from frontier AI systems**—highly capable, general-purpose AI systems which could have dangerous capabilities, i.e., which pose a threat to global security (Anderljung et al., 2023). The term "frontier" sometimes refers to the most capable models at a given time and sometimes refers to models that could pose large-scale risks due to their high capabilities (i.e., which are above some capability threshold, regardless of other models' capabilities). This work pertains to both definitions, though verification will likely be much easier if only the most capable AI systems are of concern. While we focus on general-purpose AI systems, international coordination on AI could have many other targets, for instance, regulating the use of Lethal Autonomous Weapons Systems. For discussion of verification related to narrow AI systems, see forthcoming work from Harack and colleagues.

This report **assumes a future world where there is substantial political will** among global powers for coordination on AI (i.e., similar to the U.S. response to 9/11, but globally). Political will is not sufficient for coordination, however, because of the risk that a country will break its agreements and engage in unsafe development. There are numerous reasons a country could decide to violate an international agreement about AI development, such as the benefits of increased AI capabilities being concentrated to the developers while catastrophic risks are diffused across humanity. If these risks are perceived to be fairly low, the cost-benefit trade-off could incentivize each actor, individually, to race ahead; however, coordinating to avoid this might have been preferable. Verification mechanisms enable meaningful coordination even when there is substantial distrust between countries.

International AI governance is sometimes framed as a race between countries (Aschenbrenner, 2024). This report instead targets a world where there is **international cooperation to reduce AI risks**. Depending on various factors, such as the cybersecurity of U.S. AI companies, the effectiveness of U.S. chip export controls, and when particular AI capabilities are developed, such cooperation may or may not be necessary to different degrees. This report assumes some cooperation is desired and aims to lay out the verification approaches that could enable such cooperation. To motivate this and the above framing choices, we sketch a few hypothetical futures in the appendix.

In order to organize the space of verification mechanisms, this report investigates three policy goals that international agreements may seek to accomplish. These policy goals are primarily chosen for their relevance to previous work and the feasibility of applying verification to them rather than because they are the most important policy goals. The policy goal of locating AI-relevant compute is chosen because it may be a necessary requirement for many other goals (such as monitoring AI development and deployment). The policy goal of verifying that known compute is not being used for a large training run is chosen because it may be a rough proxy for dangerous AI development while being more tractable to discuss, i.e., it is unclear how to classify future AI development as dangerous or safe, but, currently, large training runs appear to be a proxy for danger (Heim & Koessler, 2024). The policy goal of verifying the authenticity of model evaluations is chosen because of its salience in the International Dialogues on AI Safety ("IDAIS-Venice," 2024) and because external evaluations are included in some AI regulations (EU AI Act, 2024; Executive Order 14110, 2023). A fourth section includes a brief discussion of various other policy goals.

The policy goals discussed here help decompose the space of verification mechanisms, but we do not claim these should be the primary policy goals of international coordination. More ambitious and risk-reducing policy goals might include:

- All frontier AI development follows a safety case (Clymer et al., 2024)—a goal we discuss briefly in the appendix

- Given a data center, know it is not being used for any dangerous AI *development or deployment* (i.e., covering pre-training, post-training, and inference)

- Frontier AI development does not happen until humanity's understanding of advanced AI system safety is much more mature (i.e., global pause)

Many of the technical building blocks in this report are relevant to achieving those policy goals, but we leave specific discussions about them for future work.

We assume general familiarity with common AI and AI governance concepts; "compute" is used to refer to both the hardware used in AI development and as a measurement of operations done by these chips (Gupta & Ranjan, 2024).

# Verifying the Location of AI Compute

## Why This Policy Goal is Included

AI chips (compute) are the main physical instantiation of AI development and deployment. They are highly specialized, and their supply chain is concentrated, so they are a strong node for AI governance (Sastry et al., 2024). Knowing where AI compute is located is often a necessary step to verifying other properties of that compute, e.g., that it's not being used for something dangerous. Compute is also necessary for AI development and excludable (access to compute can be restricted), although the types of compute that could be used for AI development could change in the future. Governments can be confident there is no covert, dangerous AI development occurring if nearly all AI chips are accounted for (this section) and are verifiably doing something else (the next section). In the future, strong on-chip mechanisms may allow for flexible governance of chips without locating them, obviating the need for this policy goal.

### Key Takeaways

- There are two approaches to tracking AI chips: physical inspections and on-chip mechanisms.

- In the absence of secure on-chip mechanisms (at least 2 years away), political will is needed to facilitate the physical inspection and monitoring regime that would successfully track the location of AI chips.

- Tracking AI chips from their manufacturing and throughout their lifecycle appears more effective than monitoring for undeclared data centers.

## Verification Approach

"AI compute" refers to the computer chips used in frontier AI workloads, e.g., NVIDIA H100 GPUs and Google TPUs. For the sake of simplicity, we defer to the criteria for defining advanced (AI) chips as used by the current U.S. export controls (Dohmen & Feldgoise, 2023), based on total performance, performance density, or marketing for data center use. Currently, the specific hardware used for frontier AI workloads is highly specific, being designed and fabricated by a handful of companies (Sastry et al., 2024). The viability of locating AI compute is sensitive to questions of what counts as an AI chip and how many chips are concerning—considerations we discuss in the appendix.

Data centers are the buildings, or parts of buildings, which primarily house computer systems such as AI chips. Locating AI compute will, therefore. make use of data centers, e.g., involving the inspection of them or their power draw. However, it is sometimes more useful to focus specifically on AI chips rather than on the data centers that house them. Using this level of abstraction is useful for a few reasons: data centers are much broader than AI—there are estimated to be tens of thousands of data centers around the world (Pilz & Heim, 2023)—but only a small fraction of these are likely to house AI chips; advances in distributed training will likely allow geographically separate data centers to effectively act as one; and the AI chip supply chain is both specialized and concentrated, thus amenable to regulation. We discuss these considerations and more in the appendix. As a result, the mechanisms in this section apply mainly to tracking AI chips themselves rather than data centers. Both approaches will be useful in practice, but we caution against focusing too strongly on the data center level of abstraction.

Tracking AI-relevant compute is a difficult task, and there are two primary approaches. The first approach is a low-tech solution that is implementable in the near future. It consists of physical inspections of declared data centers and a registry that tracks the ownership and location of AI compute. The second approach uses on-chip mechanisms for AI governance. In particular, it involves AI chips being outfitted with tamper-proof security mechanisms and remote location attestation. It is desirable to pursue both approaches because they have different strengths and weaknesses. The first has the major downside of being politically difficult due to the access requirements for international inspectors or continuous monitoring. However, it is technically feasible today, whereas the second approach will likely not be feasible for at least two years.

In the first, low-tech, approach to tracking AI chips, the following components are necessary: countries self-report the location of their data centers that house AI chips, identify and count AI chips in those data centers, and keep an updated registry of this information; reported centers are physically inspected to ensure the presence of a random sample of the expected chips. If there is a worry that routine physical inspections would not be frequent enough to detect chips being moved and used, it may be necessary for reported data centers to have security to prevent chip theft and undergo continuous monitoring—security cameras watching chips and perimeter controls—in order to prevent chip smuggling. There is precedent for this level of shared monitoring in the case of nuclear weapons treaties (Baker, 2023). In order to simplify this data center monitoring process, it may be useful to move AI chips to a small number of data centers (i.e., individuals and corporations could still own AI chips, but these would be housed in specific facilities). Additionally, national intelligence operations could potentially be used to detect undisclosed data centers, including via human intelligence, power grid analysis, and cyber operations. Initially, implementing this approach may be quite difficult, as there are millions of AI-relevant chips already in existence, and no central tracking of them has taken place to our knowledge (besides public estimates, which are often very uncertain). However, substantial political will, international and domestic laws, and the small number of companies that produce these chips could aid in locating a significant fraction of existing chips and bringing them into a monitored data center regime.

Physically tracking **new** cutting-edge AI chips is comparatively easier, as these chips are fabricated in fewer than 20 facilities (though they undergo further development before being shipped to data centers) (Sastry et al., 2024; *TSMC Fabs - Taiwan Semiconductor Manufacturing Company Limited*, n.d.). International agreements can leverage this small number of fabrication facilities to track chips from their production to the data centers they are housed in. The narrow supply chain also enables tracking of critical components, such as EUV lithography machines, likely making it feasible to prevent covert chip production for years. The efficacy of tracking the compute supply chain would depend on the scope of monitoring and whether alternative chip production pathways become viable—something which may be particularly worrying if highly capable AI systems have been developed and are used for this purpose.

The second approach to tracking AI compute requires that chips have certain on-chip mechanisms that enable remote location attestation (Aarne et al., 2024; Kulp et al., 2024). This could include chips using a time-based ping to servers throughout the world in order to locate themselves (Brass & Aarne, 2024). Implementing such a mechanism on AI chips themselves is technically straightforward but would likely have security vulnerabilities, and it requires coordinating chip design across many companies. The main security issue is ensuring that a chip's private key cannot be extracted, as extraction would allow other chips to pretend to be the AI chip of note, resulting in its location being spoofed. Some AI chips, including NVIDIA H100s, have such a private key (Nertney, 2023), but it is unclear how well-secured this is. While it is difficult to be confident, existing chips are likely insecure: they do not appear to have undergone extremely rigorous testing, and the threat model assumed in this report is that of the most capable nation-state actors. If new chips with better security are needed, this would add substantial development time, likely at least two years, including extensive testing and mutual trust that the implementation has not been purposefully backdoored.

One approach to on-chip location attestation, which could be faster to implement, is using a Flexible Hardware-Enabled Guarantee, FlexHEG, mechanism that is implemented alongside the AI chip (Petrie et al., 2024). FlexHEGs are primarily discussed in a future section, and their application to location verification is narrow. One specific implementation of a FlexHEG mechanism involves a secure processor that is mutually trusted to carry out location attestation and a tamper-proof enclosure that encompasses the main AI accelerator and the secure processor: efforts to tamper with the location attestation mechanism or extract its private key would result in, e.g., the AI chip self-destructing. This particular mechanism could be applied to chips at any point in their lifecycle (e.g., at data centers or after chip fabrication but before shipping to data centers), but it **requires temporary physical access to chips**. The implementation time for this type of chip-adjacent location attestation is faster than for creating new AI accelerators—perhaps one to two years for development and a year for mass production and retrofitting existing chips. Implementation is faster because the secure processor could be much less performant than cutting-edge AI chips (i.e., many more could be produced quickly, and the chip can be simpler and more secure), and this mechanism could be applied to chips after fabrication (i.e., it is not delayed by production-time bottlenecks). On the other hand, approaches that require fabricating new AI accelerators will necessarily have the downside of taking over a year to saturate the compute stock (i.e., replace existing chips as a large fraction of compute) (Heim, 2024a). Approaches that involve fabricating new AI chips can utilize bottlenecks in the chip supply chain to ensure all new chips have these mechanisms applied, whereas retrofitting depends on being able to locate existing chips, e.g., via a combination of self-reporting, intelligence operations, and whistleblowers, which is likely less reliable. As with standard on-chip approaches, one difficulty with FlexHEG mechanisms is trusting that other parties have not backdoored the implementation, in this case of tamper-proof enclosures or the secure processor, in a way that allows private key extraction.

While on-chip mechanisms (including FlexHEGs) can be used for location verification, this is not their primary use. Rather, as discussed previously (Aarne et al., 2024; Kulp et al., 2024; Petrie et al., 2024), they could be designed to enable other forms of governance, such as having an operating license issued by an international regulator; we discuss these in the next section. In such a case, locating these chips may not be necessary because they can be governed remotely.

## Analysis

Whether location verification takes place via low-tech or high-tech methods, securing the AI chip supply chain will be crucial. For low-tech approaches, it will be necessary to track chips from their manufacturing (i.e., fabrication, if this remains a tight bottleneck) throughout their life cycle, ensuring chips are not smuggled. For high-tech approaches, it will be necessary to ensure the correct on-chip mechanisms are applied and that chips are sufficiently tamper-proofed. By default, there is a concern that state actors may try to subvert governance mechanisms during chip production, e.g., by implementing a backdoor. Securing the chip supply chain to prevent these subversions will likely involve the registration of key production facilities, the use of various methods to track chip production, physical inspections and continuous monitoring of fabrication facilities, and multilateral export controls on chip production equipment to make this monitoring easier by reducing the number of necessary countries. It could also require mutual evaluation of chip designs and testing of fabricated chips to ensure they have whatever on-chip mechanisms are mandated by the agreement.

Meeting this policy goal likely requires that the international community treat AI chips with a seriousness similar to nuclear materials—these are resources that pose an incredible danger in the wrong hands, and tracking them is of utmost importance. When lost or stolen, AI chips will likely be difficult to find due to them being hard to distinguish from other computers (from a distance) and lacking notable physical attributes. Attempting to detect secret data centers may help but is unlikely to be the load-bearing part of a verification regime: it will likely be too easy to hide AI compute among other compute or construct secret data centers, though whistleblowers could enable detection in such a scenario. If the amount of compute that must be monitored is very high, e.g., data centers with >50,000 AI chips, detecting data centers may be effective, but this approach is less robust than focusing on chips; we discuss this more in [the appendix](#).

Generally speaking, a multi-pronged approach will be most effective. Locating compute should involve using national intelligence operations to detect data centers that might house AI compute, physically inspecting AI data centers, implementing continuous monitoring of AI chips (e.g., security cameras), and investing in higher-tech on-chip mechanisms that could allow robust location verification with less access (or the implementation of governance mechanisms on a chip without knowing that chip's location). It should also involve centralizing and monitoring the chip supply chain in order to reduce the likelihood of secret AI development projects occurring.

While decentralized access to AI compute is often considered desirable, an international verification regime would likely benefit from compute being physically centralized, both nationally and internationally. The case for international compute centralization is fairly straightforward: if fewer countries have the computing capacity to do activities that are prohibited by an international agreement (such as training frontier AI models), fewer countries are necessary for successful coordination to regulate such AI activities. Domestic compute centralization likely makes verification easier by decreasing the number of physical facilities that need to reach the requisite level of monitoring and security. In both cases, the usefulness of centralizing compute is primarily about where AI chips are located, rather than who owns them or uses them. Generally speaking, compute can be made democratically accessible while still being physically centralized, as is the case for cloud compute providers (Heim et al., 2024). Such centralization would make verification easier in both the international and domestic case.

## Building Blocks Preview

13/37 rows displayed. See appendix for [complete table](#) and explanation of [feasibility estimates](#).

| Building Block | Mechanism | Details |
|---|---|---|
| Chip ownership and location registry | Chip ownership and location registry | AI chip owners / data center operators must register the chips they own, and where those chips are located, with a central regulator. Chip sales require updating the registry with the new owner and location. |

| | Feasibility | | Previous work (non-comprehensive) |
|---|---|---|---|
| | High ▾   <1 year | | Jones (2024); Shavit (2023); Baker (2023); Fist & Grunewald (2023) |

| Flexible Hardware-Enabled Guarantee (FlexHEG) mechanisms | Tamper-evident chips | Includes: [video surveillance](), tamper-evident seals or packaging on chips, and potentially remote attestation. Helps detect violators but does not on its own prevent tampering—needs to be paired with enforcement. |
|---|---|---|

| | **Feasibility** | **Previous work** (non-comprehensive) |
|---|---|---|
| | High ▾ 1–5 years | Aarne et al. (2024); TamperSec (n.d.) |

**Notes**
Some approaches here require replacing AI chips, but some may be applied to existing chips (with temporary physical access) or at the data center level.

| Chip location tracking | Ping-based location tracking | Set up a series of servers that ping chips and triangulate chip location based on response time to different servers. This is relatively low precision. |
|---|---|---|

| | **Feasibility** | **Previous work** (non-comprehensive) |
|---|---|---|
| | High ▾ 2–4 years | Aarne et al. (2024); Brass & Aarne (2024) |

**Notes**
May not require replacing hardware. The main security requirement is chips having a private key, however, it is unclear if current chips are secure against physical attacks to extract such a key, as would be needed here. May require chips to be connected to the internet.

| National intelligence operations | Satellites (visual) | Satellite imagery may be used to detect the construction of data centers, fabs, and power infrastructure. |
|---|---|---|

| | **Feasibility** | **Previous work** (non-comprehensive) |
|---|---|---|
| | High ▾ <1 year | Wasil, Reed, et al. (2024); Pilz & Heim (2024) |

**Notes**
Satellite information is unlikely to differentiate AI vs. non-AI data centers. It also may struggle to differentiate data centers from other industrial buildings (cooling may be a key differentiator).

------------------------------------------------------------

| | Tracking of relevant personnel | There are hundreds to thousands of experts who have the knowledge and skills to contribute to frontier AI development, chip design, or chip fabrication. Building state-of-the-art domestic chip production would be very difficult without these people. Various measures could be used to confirm that these people are not contributing to covert national projects in violation of an international agreement. |

**Feasibility**

High ▾ <1 year

---

| Security | AI data centers are secure against chip theft | Data center operators and their countries collaborate to boost data center security. Ideally, this would prevent the theft of model weights and algorithmic secrets, but preventing the theft of physical chips is the focus of this section and is probably much easier (current security may be sufficient if done with [international collaboration](#)). |

**Feasibility**

High ▾ <1 year

---

| Physical inspection of data centers | Verify chip identity and count | Use unique chip identifiers to ensure the right chips are present; count to make sure all are accounted for. |

**Feasibility**

High ▾ <1 year

| Continuous monitoring of AI data centers | Perimeter and portal continuous monitoring | Materials coming in and out of data centers (especially AI chips) flow through portals, which are jointly monitored to avoid chip theft. Existing physical security measures such as fencing and cameras are likely to be sufficient if they can be jointly monitored. |
|---|---|---|

**Feasibility** | **Previous work** (non-comprehensive)

High ▾ <2 year | *START: Annex to Protocol on Inspection and Continuous Monitoring Activities* (n.d.)

**Notes**
It may also be desirable for network traffic in and out of data centers to be jointly monitored (this can reduce the risk of model weights being improperly moved), but that is out of scope for this section.

| | Security cameras inside data centers | This is standard (Google Cloud Tech, 2020) but would involve giving international inspectors access or (more likely) having international inspectors install their own cameras. The primary goal of security cameras in data centers for verification is to ensure chips are not being removed, added, or modified to bypass governance mechanisms, so cameras should focus on chips. |
|---|---|---|

**Feasibility**

High ▾ <1 year

**Notes**
Cameras might confirm that unauthorized chips aren't being substituted into a data center, that interconnect limits aren't being changed, or that chips are not being tampered with.

| Register chip production | Supply chain registry | Key parts of the chip supply chain must register their production facilities with a central verifier. This should likely be based on a thorough assessment of the current bottlenecks and crucial actors for chip production. |
|---|---|---|

**Feasibility**

High ▾ <1 year

| Tracking chip production | Chain-of-custody | Chain-of-custody (CISA, n.d.) is implemented for the AI chip supply chain: firms document the transfer and storage of components along the chip supply chain. |
|---|---|---|

**Feasibility**

Medium ▾  1-3 years

| Physical inspection of fabs | Physical inspection and continuous monitoring of fabs | Check that chip production facilities are implementing the agreed-upon on-chip measures, have the manufacturing capacity claimed, and are not doing unauthorized production or distribution. Includes human inspectors, cameras, interviews with employees, etc. |
|---|---|---|

**Feasibility**

High ▾  <1 year

**Notes**
Includes inventory audits.

| Multilateral export controls | Multilateral export controls | Members of the international agreement have a shared list of export controlled countries for AI chips and major components, in effect requiring a license from an international authority. |
|---|---|---|

**Feasibility**

High ▾  <2 year

**Notes**
These rules could be "disallow" or "allow" based, depending on risk. Countries might agree to reduce the proliferation of AI chips via shared export controls. It appears difficult to make such agreements robust to a country later deciding to break such an agreement.

# Verifying That Known Compute is Not Being Used for a Large Training Run

## Why This Policy Goal is Included

This policy goal is included because it allows for a clear decomposition of the space of verification mechanisms, the size of a pre-training run is sometimes a proxy for risk from AI systems, and it is a major focus in AI governance research and policy. While this report focuses on pre-training, it may be necessary in the future for international coordination to focus on inference or post-training.

Ideally, international coordination should aim to restrict dangerous AI development and deployment (i.e., dangerous pre-training, post-training, and inference). However, humanity's limited understanding of the dangers from advanced AI systems makes discussing this goal somewhat intractable, so we presently discuss the policy goal of restricting large (pre)training runs. This is identical to restricting training runs that exceed some compute threshold—compute thresholds are used in various AI regulations (EU AI Act, 2024; Executive Order 14110, 2023). **The size of a training run is currently a proxy for danger (Heim & Koessler, 2024), but this is unlikely to remain a sufficient proxy in the future**. On the one hand, large training runs may eventually be safe when accompanied by a safety case or other strong argument for their safety. On the other hand, substantial danger will likely originate from post-training or inference on already-trained models, and algorithmic progress will likely allow dangerous training runs to occur with increasingly smaller amounts of compute.

During an intelligence explosion (rapid, AI-driven, AI progress), international relations are likely to be strained, and restricting the size of AI training runs may slow a dangerous international AI race, permitting more focus on safety. Compared to more detailed policy objectives—such as all frontier AI development following a safety case (Clymer et al., 2024), which we discuss briefly in the appendix—restricting large training runs could be more politically and technologically tractable, even in a low-trust setting. Restricting large training runs is an imprecise approach to AI risk reduction, but it may provide marginal safety improvements in high-stakes scenarios. Crucially, for the sake of this report, the verification mechanisms discussed in this section will apply to numerous other policy goals.

## Key Takeaways

The principal ways to show that some compute is not being used for a large training run are to show that it is either incapable, or that it is not (based on classifying a large enough share of the chip-hours as not part of a large training run).

- Data center operators could engage in "compute accounting" (Baker et al., Forthcoming), demonstrating that a sufficiently large portion of the available chip-hours are not being used in a large training run; this will include some form of registry where data center operators keep track of the workloads running.

- An important mechanism discussed previously is partial workload re-running (Choi et al., 2023; Jia et al., 2021; Shavit, 2023; Baker et al., Forthcoming), where a mutually trusted data center is used to confirm that a declared workload actually occurred.

- One promising approach to verifying that some compute is not being used for a large training run is to construct small pods of chips that have restrictions on their external interconnect bandwidth, i.e., restricting the rate of information flow in and out of the pod, such that sharing gradients between pods is too slow for (data) parallel training. This restriction has been discussed for newly manufactured chips (Kulp et al., 2024), but it could also be applied with networking equipment and verified with security cameras and physical inspections while requiring no code access (discussed in the appendix).

- Implementations of the FlexHEG design stack could monitor chip activity to ensure chips are not participating in a large training run, and these approaches are especially desirable because they could implement various governance operations which do not need to be determined in advance, but technical development is needed ahead of time to make these approaches available.

## Verification Approach

In this section, the goal is to verify that some AI compute (e.g., that was located with techniques discussed in the previous section) is not being used for some prohibited purpose. Ideally, international regulations would specifically ban dangerous AI development and deployment. However, this objective is currently intractable due to humanity's poor understanding of dangers from AI systems. Instead, we focus on the proxy objective of ensuring AI chips are not being used for large AI training runs.

We use "data center" to refer to a physical space that houses compute and "data center operator" to refer to some actor who has responsibility for it, though institutional details are not the focus of this report. The approach to this policy objective includes two key components. First, exclude as much compute as possible on the basis of it not being capable of supporting a large AI training run. This is principally to reduce the verification cost and compliance burden for actors who are very unlikely to contribute to risks, such as an individual with a personal computer. Second, verify that some compute, which could be used for large AI training, is not, in fact, being used for this purpose.

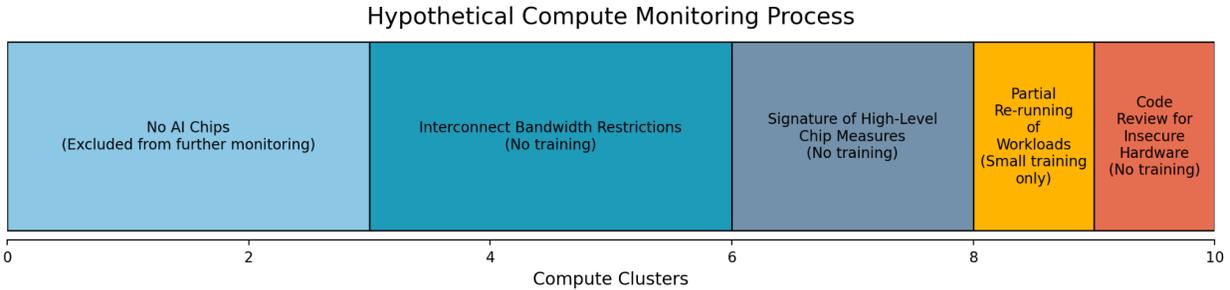

Figure 1: A hypothetical breakdown of how different amounts of compute could be monitored using mechanisms discussed in this section. Parentheses include what the mechanism could tell us about some particular compute.

## Excluding Compute From Further Monitoring

Compute may be excluded from further monitoring due to its incapability to be used for large training runs, based on the number of chips and type of chips in a data center. The quantity and types of chips in a data center can be confirmed with physical inspections and continuous monitoring. These exclusions face one core issue: progress in the field of distributed training may make even small AI compute clusters useful for large training runs when combined together, complicating compute-based verification. Hopefully, in the future, AI chips could be designed to specifically allow for certain operations while being ineffective at others, for example, chips that can execute AI inference but are highly inefficient for AI training; current inference-specialized AI chips are not all sufficiently limited. While the definition of "AI compute" may need to change and should therefore be flexible, using the operationalization from current U.S. export controls (Dohmen & Feldgoise, 2023) means most GPUs that are not intended for data center use and most CPUs (Feldgoise et al., 2024) can be excluded.

We now turn to compute that cannot be excluded on the basis of incapability. There are a few approaches to showing that a chip that could be useful for large training runs is not being used for this purpose. Trivially, chips could be shut down in emergency or transitory situations, drawing no power and thus not being useful for a violation. This is not desirable, however: AI systems could be creating large amounts of economic value or contributing to critical services, so completely shutting off AI compute may be very costly. Instead, we would like approaches that allow AI compute to continue doing activities that are permitted by an international agreement but not other activities—in this case, large AI training runs are not permitted. These approaches include partial re-running of chip activities to confirm that some declared use of compute happened as stated, workload classification based on high-level chip measures, classification using FlexHEGs, and other methods.

## Partial Re-Running of Chip Activities

Data center operators could keep registries of the activities of their chips, and these registries could be used to verify that chips were not used for a violation. In particular, the partial re-running of chip activities protocol (Choi et al., 2023; Jia et al., 2021; Shavit, 2023) and generalizations of it (Baker et al., Forthcoming) may be able to accomplish this goal. This approach involves a data center operator reporting enough information about a particular workload that this workload can be replicated in a mutually trusted data center, confirming that the workload re-run matches the declared workload and, therefore, that the declared workload was actually how some compute was used. Re-running is not the only implementation method available to confirm a declared training run is correct, but we refer to these methods as re-running for simplicity. Ideally, this protocol is zero-knowledge in that the trusted data center could report only whether the declared workload is correct. Additionally, other properties of the workload could be confirmed in the trusted data center, e.g., the architecture of the model used, whether particular reported evaluation results are correct, and effectively all other information about the workload; these could still be reported in a zero-knowledge manner, only revealing the concise answer to governance-related questions without revealing the underlying data.

Despite this approach being zero-knowledge in principle, there could be practical security challenges that make this difficult. The "trusted" data center needs to be mutually believed to not leak sensitive information, which could require new AI hardware or very strict security measures. This approach of keeping and verifying a registry of chip activities has a major downside in that it requires code access, even if this is intended to be secure. Code in the context of AI development is particularly sensitive, encapsulating model weights and architectures. In some of the potential futures where international verification is needed, this information is some of the most sensitive information a government has (i.e., akin to the location of nuclear weapons).

## Workload Classification With High-Level Chip Measures

Therefore, ideally, one could verify that AI chips are not taking part in a large training run based on just high-level information about these chips' activities without code access. This could be implemented alongside a registry (i.e., acting as a secondary verification of that registry) or on its own. The primary approach here is to measure a chip's activity with external sensors—e.g., a chip's power draw, within-server bandwidth, between-server bandwidth, core utilization, etc.—and then to classify whether the chip is participating in a large training run, on this basis.

One promising approach to this classification is to impose communication limits, which restrict chips to high communication only with a small number (e.g., 128) of other chips but low communication with chips outside this "pod". This is a generalization of the "fixed-sets" approach discussed by Kulp et al. (2024). Crucially, this approach could be implemented by using specialized networking equipment, physical inspections, and security cameras, but not requiring access to code or the manufacturing of new AI chips. A data center operator would claim that a pod of chips is not participating in a large training run because it does not have sufficient bandwidth with the rest of the world, and the pod could still be used for inference or various other uses. If international verification is needed soon, this may be one of the most effective mechanisms to verify that some compute is not being used in a large training run, given its feasibility with existing technology, limited access requirements, and relative robustness. Interconnect bandwidth limits have a number of downsides. For example, they may not be robust to shifts in AI development practices, and the relevant limits could be strict enough that they impose costs on permitted AI activity (e.g., only permitting text inference but not inference with bandwidth-heavy modalities such as images or videos). We discuss details of this approach in the appendix.

Various other methods could, speculatively, distinguish training and inference with just high-level information about chip activity (e.g., power draw, bandwidth utilization, numerical precision of computations). Many of these methods appear easily spoofable given the present threat model, e.g., by making training 20% slower in order to look like inference, within the budget of relevant actors. However, they could be made robust if they are paired with temporary code access. Specifically, an inspector could take measurements of some chips' power draw while reviewing code to ensure the chips are conducting inference, and this creates a baseline, or signature, of the chip's power draw while conducting an approved activity (in this case, inference on a specific model). Later, the inspector would only monitor the power draw to ensure it is sufficiently close to the baseline.

## Workload Classification With FlexHEG Mechanisms

Workload classification could also take place via AI chips either reporting on their activities or having their activities restricted. Flexible Hardware-Enabled Guarantee, FlexHEG, refers to a design stack for AI chips that can securely implement a variety of governance functions, such as monitoring or restricting chip activities (Petrie et al., 2024). Work on FlexHEG mechanisms is in its early stages, so specific implementation details have not yet been determined. Similar ideas have also been referred to as "on-chip governance mechanisms" (Aarne et al., 2024) or "hardware-enabled governance mechanisms, HEMs" (Kulp et al., 2024); these concepts are somewhat overlapping, so we refer to "FlexHEG" for simplicity and to gesture at the implementations most relevant to verification of chip activities.

One specific implementation worth discussing is "chip-adjacent" FlexHEGs. A chip-adjacent FlexHEG implementation would involve designing and building a secure process that sits next to an AI chip, both in a tamper-proof enclosure—e.g., a high-tech box encompassing a GPU or server, with an additional computer inside, and the box destroys the GPU if tampered with. Both the second processor and the enclosure must be mutually trusted, e.g., all relevant parties believe the mechanisms have not had security vulnerabilities inserted in them. The role of the secure processor is to implement various governance functions, which could include monitoring what operations are happening on a chip, monitoring the total number of operations happening on a chip, restricting chip activities, logging the contents of the chip's memory to compare with declared registries, controlling what other chips this chip interacts with via encrypting communication, and more. The secure processor needs to interact with the AI chip to various degrees in order to perform the governance functions desired; successful FlexHEG implementations would be sufficiently flexible that the governance functions carried out could be modified over time by an international authority. The role of the tamper-proof enclosure is to make interference with these governance functions extremely difficult (e.g., interference would result in a chip self-destructing).

This chip-adjacent implementation is particularly promising in the near term because designing a mutually trusted secure processor that carries out governance operations is likely much easier than redesigning all high-performance AI chips themselves to the same degree of security and mutual trust. In the longer run (e.g., 3-5 years), designing AI chips themselves to have the requisite security—FlexHEGs implemented directly on-chip—is likely desirable as it could be more secure; as mentioned, FlexHEG is a general design stack that is implementation agnostic.

Previous work has discussed "on-chip" or "hardware-enabled" governance mechanisms, which sometimes serve a similar purpose to FlexHEGs (Aarne et al., 2024; Kulp et al., 2024), and for simplicity, we refer to these under the umbrella of FlexHEGs because they usually have similar security requirements. A key feature of FlexHEGs is that, if they are well designed, the range of governance functions enabled can be quite broad and remotely updatable by an international authority. Here, we discuss how FlexHEGs could be applied to workload classification.

Existing proposals (Aarne et al., 2024; Kulp et al., 2024; Petrie et al., 2024) are mainly technical approaches for *how* to implement chip monitoring or restricting. However, for their application to workload classification, it is also crucial to determine on what basis chip activities would be restricted or flagged by monitoring, especially given our assumption of well-resourced adversaries who may attempt to disguise workloads. FlexHEG mechanisms could implement classification of workloads based on high-level chip information (as in the external sensor case), specific mathematical operations used (e.g., an "allow" or "disallow" list of chip operations), outsourcing this classification to inspectors, checking measured chip activity (e.g., chip memory snapshots) against declared chip activity registries, or something else. These different approaches to classification likely differ substantially in their ease of implementation and robustness, and more work is needed to prioritize among them.

FlexHEG approaches are malleable and could implement different restrictions over time, especially via mechanisms such as offline licensing (Aarne et al., 2024; Kulp et al., 2024) or workload approval. Offline licensing could operate similarly to general software licensing, where a regulator issues an AI chip a license for some period of time or amount of compute and renews the license if some governance-related criteria are met (e.g., the snapshots from the chip's memory match those expected from the declared registry of chip activity). Workload approval would be much more fine-grained, where each workload submitted to chips requires sign-off from a regulator. Fundamentally, these approaches rely on strong security around a general processor carrying out governance operations—whether it be the AI accelerator itself or a chip-adjacent auxiliary processor. The R&D time for these mechanisms is multiple years, perhaps only a couple in the case of chip-adjacent FlexHEGs, as they do not require designing and fabricating new cutting-edge AI chips.

FlexHEG mechanisms are especially promising because, if they are implemented well, they would not require that compute be located in order to be governed. Locating compute is necessary if AI chips are insecure and could likely be modified to bypass governance mechanisms (thus, physical inspections and security cameras can confirm this modification is not happening), but tamper-proofing, as part of the FlexHEG stack, would include chips being hardened against such attacks. Therefore, chips could be trusted to carry out verification operations without their location being known, and if the FlexHEG implementation is sufficiently flexible, these governance operations could change over time.

## Other Workload Classification

Trivially, [inspectors](#) could be granted access to large amounts of code, which would enable them to classify workloads as permitted or not permitted. That is, one way to verify that some AI chips are not being used for a large training run is to read the code being run on them. While this carries privacy and security concerns, it does not require significant R&D the way FlexHEG approaches do: this presents a case where *access* can trade off against the *technological maturity* of verification mechanisms.

AI chips are sometimes used for non-AI activities. Fortunately, many of the methods that could be used to differentiate between AI workloads can also differentiate AI and non-AI workloads, such as [re-running chip activities](#), [inspectors reviewing code](#), or [classifying chip activity with high-level chip measures](#). For verifying particular non-AI activities, such as cryptocurrency mining, [Proof-of-Work](#) approaches could be used. In the future, it may be possible to develop Proof-of-Work-style approaches for other workloads (ideally including AI workloads, as [partial re-running](#) attempts to do), which could be used to verify that compute was used for some declared purpose.

## Compute Accounting

At its core, distinguishing large training runs from other AI workloads at the chip level is a difficult problem because both consist of basic arithmetic operations. The general approach to verifying that some compute (i.e., some number of chips for some time), which could be used for a violation, is not being used for a violation is to classify a *significant portion of chip time* as non-violations. This could include a combination of the above methods, e.g., partial re-running of certain workloads to confirm they were correctly reported and classifying workloads on the basis of either high-level chip measures or aspects of the code they are running. Some of these approaches include an affirmative claim that chips were doing some specific other thing (e.g., for partial re-running of declared workloads), and some are negative claims (e.g., interconnect limits mean some chips could not be participating in a large training run).

The key argument a data center operator would make in compute accounting is that there are not enough unverified chip operations (or hours, depending on the implementation) remaining to carry out a violation. This is often a non-trivial argument—especially for affirmative claims—and forthcoming work from Baker and colleagues makes progress on it. For example, chips often do redundant operations during training ("recomputation") in order to save on memory, leading the number of operations actually performed by a chip to exceed the quantity required by the workload, a gap that could potentially be exploited to do illicit computation. Fortunately, negative claims may not have this issue: if a pod of chips has interconnect limits placed on them, it is infeasible for them to be acting as part of a larger, parallel training run, regardless of the efficiency or completeness of declared workloads. Figure 1 gives an example of how an AI developer might demonstrate that they are not violating an agreement by piecing together different methods for verifying that compute is not being used for a large training workload.

## Analysis

Verifying that chips are not being used for large training runs presents significant challenges. For instance, an adversary could split a large training workload across multiple smaller, seemingly innocuous data centers, or state-level adversaries controlling data centers might obscure the distinctions between training and inference, complicating detection. While granting inspectors greater access could mitigate these risks, such access introduces privacy and security concerns. Moreover, the verification mechanisms in this section are primarily being ideated and designed in today's threat landscape, so they are unlikely to be robust to subversion attempts from highly capable (i.e., superintelligent) AI systems.

An ideal verification regime might exclude data centers incapable of supporting large-scale training, but this is difficult due to algorithmic progress, which makes smaller amounts of compute increasingly capable, and distributed training, which allows multiple small amounts of compute to be combined for large workloads. As a result, thresholds for exclusion must be sensitive to rapid advances in efficiency and distributed training capabilities, and conservative thresholds may be necessary.

One way to frame the goal in this section is that we would like to demonstrate that some AI chips are not being used in a large training run while revealing minimal information about what these chips are doing (in particular, not leaking sensitive model weights or code), and this approach needs to be robust to highly competent adversaries. Partial re-running of chip activities and FlexHEG-based approaches could both serve this purpose. However, they are not yet technologically mature and they could pose security risks due to the amount of access they require—they are zero-knowledge in principle, but obtaining the high degree of security needed in practice could be difficult. Workload classification based on high-level chip measures would be desirable because it does not require very much access, but it may be spoofable with relatively small performance penalties. Interconnect bandwidth requirements appear promising for permitting AI inference while blocking large AI training runs, they seem less spoofable than other high-level chip measures, and they could likely be implemented with existing technology, but this approach has numerous drawbacks.

Other minimal-access methods to verify compute use may exist, but this area is underexplored, and further research is needed to develop effective methods. We emphasize that, given likely advancements in distributed training, it may be difficult to exclude small AI data centers (e.g., hundreds or low-thousands of AI chips) from further monitoring. Excluding such data centers, if possible, is highly desirable for bringing down the burden of compliance and the difficulty of verification. Therefore, developing better methods to verify compute use, ideally with minimal-access and existing technology, is a key area for further research.

Some of the mechanisms in this section would be implemented in a way that is highly specific to the current AI development paradigm of transformer-based large language models and pre-training: inference-only chips, workload classification with high-level chip information (including interconnect bandwidth limits), and some implementations of FlexHEG-based workload classification. On the other hand, some approaches appear more paradigm-invariant, requiring changes to implementation details but very likely remaining relevant under paradigm shifts: partial workload re-running, the general FlexHEG approach, and using inspectors with code access.

While it is not the focus of this section, it is worth discussing how these mechanisms can affect the verification of other policy goals. For instance, it may be desirable to verify particular properties of the models being developed by a country, such as the data or architectures being used. In the appendix, we briefly discuss one such policy goal, verifying that a declared safety case is followed. The mechanisms in this section vary in the granularity of information they provide about AI activities, affecting their usefulness for these other policy goals. High-level chip measures are unlikely to be helpful here. Partial re-running of workloads and code-based analysis, e.g., via FlexHEGs, are likely effective here as they allow access to the code being run in a zero-knowledge manner. These more effective mechanisms rely on strong security and have development times of at least 2 years, so work on them should begin early.

## Building Blocks Preview

11/34 rows displayed. See appendix for complete table and explanation of feasibility estimates.

| Building Block | Mechanism | Details |
|---|---|---|
| Inference-only chips | Inference-only manufacturing | Some AI chips could be manufactured to be especially efficient at inference while not being optimized for training, e.g., Groq, AWS Inferentia, Sohu. Further work may be needed to make such chips sufficiently inefficient for training. |

**Feasibility**

Medium ▾  1–4 years

**Notes**

Some current inference-specialized chips could likely be repurposed for training without significant effort, e.g., as is done in a blog post about AWS Inferentia (Rand, 2023). Therefore, for them to be a credible sign that no large training runs are happening, future chips would have to be designed to be particularly inefficient for training. Some existing inference-specialized chips may be sufficiently incapable of being useful for training, e.g., because they use inflexible ordering of operations. On the other hand, approaches here will need to contend with substantial effort going into overcoming these limitations.

| Building Block | Mechanism | Details |
|---|---|---|
| Emergency chip shutdown | Unplug the data center from its primary and backup power supply | Trivially, a data center operator can claim that its chips are not being used in a large training run if they are not receiving the power needed to operate. This can likely be verified with less invasive methods such as power draw of the data center and knowledge about on-site (e.g., backup) power generation or potentially thermal imaging. |

**Feasibility**

High ▾  <1 year

**Notes**

It would impose negative externalities to unplug an entire data center if there is important non-AI compute in the data center, hence the importance of avoiding co-location.

| Chip use registries | Registry of training transcripts | Data center operators keep a detailed registry of how chips are used. This includes training transcripts that detail model weight checkpoints and everything necessary to replicate a training run (data, hyperparameters, random seed, etc.). |
|---|---|---|

**Feasibility**          **Previous work** (non-comprehensive)

| High ▾ <1 year | Shavit (2023) |
|---|---|

**Notes**

This registry is highly private and likely should only be used in zero-knowledge ways, such as being encrypted except for use on a mutually trusted data center. This registry could be implemented at the data center layer, but it would benefit from individual chips reporting on their activities.

| Partial re-running of chip activities | Partial re-running description | A data center operator can claim its compute was used for a training run, and this claim can be verified by an external party with much less compute by re-running small segments of the full training run to see if they match the declared run. This requires checkpoints and hyperparameters provided in a registry of chip activity. This re-running happens in a zero-knowledge way in a mutually trusted data center where the chance of sensitive information leaking is low. There may be other approaches to verify a given transcript other than partial re-running, such as those explored by Choi et al. (2023). |
|---|---|---|

**Feasibility**          **Previous work** (non-comprehensive)

| Medium ▾ 1–5 years | Shavit (2023); Choi et al. (2023); Baker et al. (Forthcoming) |
|---|---|

**Notes**

Partial re-running enables verification of practically all properties of a training run: model architecture, hyperparameters, data, etc., because the data center operator provides this entire training transcript to a mutually trusted data center for verification.

| Mechanisms for chip limiting, monitoring, or workload classification | FlexHEGs | Discussed [previously](). In addition to the FlexHEG design discussed above, in this section, it is crucial that FlexHEGs have some insight into what a chip is doing. For example, this could look like intercepting the communications this chip makes to other chips and encrypting them (where only specific chips can decrypt them), a way to implement interconnect bandwidth restrictions. This could also include the secure processor viewing the chip's operations and ensuring they do not violate some criteria. Because a FlexHEG design uses a general processor, many verification approaches could be pursued with such a mechanism. |
|---|---|---|

**Feasibility** | **Previous work** (non-comprehensive)
--- | ---
Medium ▾  2–5 years | Petrie et al. (2024)

**Notes**

Requires mass production of new hardware; depending on the implementation, it may require either new AI chips, auxiliary chips with secure enclosures, or something else.

| | Allow list operation stack | An international authority could certify certain operations at the level of "Inference on the Llama 70B model on a DGX H100 server" and issue the specific GPU-level code for such use cases. AI chips would be restricted to only run code approved by this authority. This could be implemented with a [FlexHEG]() mechanism. |
|---|---|---|

**Feasibility**

Medium ▾  2–5 years

| Speculative differences between training and inference with high-level chip measures | Inter-chip interconnect limits (network bandwidth between servers) | Large-scale AI training involves linking thousands of chips with high-bandwidth interconnect. Efficient inference, meanwhile, can be done with dozens to low-hundreds of highly connected chips. Between different "pods" of well-connected chips, the communication requirements are very low for inference (tokens) compared to the activations or gradients that are transferred in other forms of parallelism—this gap is the target of interconnect bandwidth limits. |
|---|---|---|

| Feasibility | Previous work (non-comprehensive) |
|---|---|
| High ▾  <1 year | Discussed below; Kulp et al. (2024); Heim et al. (2024) |

**Notes**

The main distinction here is that inference does not use data parallelism: different batches of data being processed do not need to be closely linked together at all, whereas for training this is needed to share gradients. Advances in distributed training may make this approach ineffective. With well-designed limits and enhanced security, this could be difficult to spoof.

| Chip utilization rates, core utilization rates | Chip utilization, specifically Hardware FLOP Utilization (HFU), is a measure of the actual amount of computations performed by a chip relative to its theoretical maximum. Training and inference often differ along this metric. Current inference for LLMs involves two stages (Verma & Vaidya, 2023): "prefilling" (computing the KV cache on input tokens) and "decoding" (generating additional tokens)—these stages may have distinct effects on power consumption because decoding is more limited by memory bandwidth while prefilling is limited by computational performance (i.e., utilization and power draw are higher during prefilling). Therefore, inference could have distinct fluctuations in utilization. |
|---|---|

| Feasibility | Previous work (non-comprehensive) |
|---|---|
| Medium ▾ <1 year | Heim et al. (2024); Baker et al. (Forthcoming) |

**Notes**

Recently, it has become popular to report Model FLOP Utilization (MFU), a measure of the obtained throughput relative to theoretical max throughput (Chowdhery et al., 2022). MFU calculations are based on the number of tokens processed, so this requires more access than just high-level chip information. On the other hand, HFU is more easily measurable and thus more relevant to the present use case (but for other use cases, such as accurately estimating the model-FLOP size of a workload, this difference is very important). Classification on this basis is likely spoofable with medium cost penalties (e.g., a 20% cost increase via making training less efficient, which is in the budget of the relevant actors).

| Signatures of high-level chip measures | Signatures of high-level chip measures | High-level [information](#) about chip activities (e.g., power draw, numerical precision, network activity, etc.) is likely spoofable in the default setting where one country is trying to classify the workloads of another country's data center with limited access. However, this approach can be made more robust by using high-quality, labeled data from the monitored data center to improve this classification approach. The monitored data center operator would run their most common, internationally approved/permitted workloads (e.g., inference of their popular models) while an inspector has code access to confirm this is happening. The inspector takes readings for high-level chip measures (e.g., power draw, network activity). The inspector can then create a classifier based on this data (e.g., train an ML classifier), which can be applied to high-level chip measures in the future (code access is no longer needed if the high-level chip measures match that of approved activities). If the measurements deviate significantly from their expected patterns or if the common workloads change significantly, the inspector can review the code again. The key insight compared to naively trying to classify workloads is that temporary code access can provide high-quality labeled data from the production system, and the data center operator can be expected to keep its chip activities sufficiently close to these baselines, versus classifiers needing to deal with massive distribution shift in an adversarial environment (if code access is never provided, software stacks and model architectures may differ substantially, and even the AI chips themselves may be different from the classifier's training data). |
|---|---|---|

**Feasibility**                          **Previous work** (non-comprehensive)

High ▾  <2 year                          See [below](#)

**Notes**

It is unclear how consistent these signatures will be over time, but previous results attempting to classify chip activities with high-level information have been effective (Copos & Peisert, 2020; Karimi et al., 2024; Köhler et al., 2021). This approach requires granting temporary code access (including whenever major changes are applied to chip activities), which could pose privacy and security risks. This temporary access could be acceptable if a neutral party or zero-knowledge approach is used (e.g., inspectors who live onsite and have limited communication with their home country). It could also be difficult for inspectors to be confident they are seeing the true code being run.

| Proof-of-Work methods | Proof-of-Work methods for crypto mining | Cryptocurrency mining uses "Proof-of-Work" schemes, which provide confidence that some declared operations have taken place. For example, a standard implementation is for workers to [hash] variants of the same message until one of the resulting hashes matches some criteria (e.g., many leading zeros in the hash). It can be quickly verified that the message variant used produces such a hash. The acceptance criteria (e.g., number of leading zeros) can be varied to control the likelihood of an input message producing an acceptable hash and, thus, the amount of original work confirmed by each presentation of an acceptable hash. Applying Proof-of-Work could verify that compute declared to be doing crypto mining is indeed doing it. |
|---|---|---|

**Feasibility**

| High ▾ | <1 year |
|---|---|

| Compute accounting | Compute accounting | Data center operators can demonstrate that their AI chips are not being used for a large training run by showing that they are being used for other things, as accomplished through [partial re-running of chip activities] and [other] [workload] [classification]. By summing declared and verified compute use and comparing it to total potential compute use, they can show that there is not enough compute left over for a violation. One difficulty here is that the quantity of declared chip uses may not correspond to the actual chip use. For example, chips often do redundant operations during training ("recomputation") in order to save on memory, so a data center operator might claim to have done a training run with 10^25 FLOP, but the chips actually did 1.5*10^25 FLOP because they did substantial recomputing. Forthcoming work from Baker and colleagues makes progress on this. |
|---|---|---|

| **Feasibility** | | **Previous work** (non-comprehensive) |
|---|---|---|
| Medium ▾ | <1 year | Heim et al. (2024); Baker et al. (Forthcoming) |

# Verifying the Authenticity of Model Evaluations

## Why This Policy Goal is Included

Evaluations of AI systems, by their developers or external parties, have numerous use cases. They play a key role in if-then commitments, which AI developers are currently using to make development and deployment decisions (Karnofsky, 2024). They have received substantial attention as a governance node and will likely be part of future AI regulations (EU AI Act, 2024; Executive Order 14110, 2023; "IDAIS-Venice," 2024). In an international AI race context, evaluations of another country's AI systems could reduce race dynamics by building trust that all parties are following agreements to move slowly; they are a more direct measure of AI capabilities progress than compute thresholds (see Story 1 in the appendix for motivation). Due to the broad interest in model evaluations and their numerous use cases, this section focuses on verifying that model evaluations are conducted properly.

## Key Takeaways

- Methods to make external evaluations technically secure, such as having evaluations and deployment both occur in a Trusted Execution Environment and ensuring the model weights and code match, are nearly solved in theory for basic evaluations, but existing hardware may not be sufficiently secure to enable these methods for international verification.

- Substantial progress is still needed in the science of evaluations due to difficulties in knowing what to evaluate, building evaluations, and eliciting a model's full capabilities.

## Verification Approach

There are a few key criteria that must be met to carry out authentic, secure, and effective model evaluation in the international agreement scenario; we mainly discuss authenticity and security. First, there needs to be confidence that the model being evaluated is the model of interest (e.g., the same model that is deployed in some environment or that is trained in a declared training run). Second, the developer and the evaluator need to protect their sensitive information (e.g., the model weights and evaluation prompts). Third, evaluations need to be effective at measuring the desired model property. There are at least a couple of viable technical approaches to achieving the first two of these criteria.

Evaluations could happen in a Trusted Execution Environment (TEE), where mutually verified evaluation code is run, and the inference code is certified to match the code used in deployment or training. The use of a TEE can reduce the risk of sensitive information leaking (though current hardware may not be sufficiently secure to provide guarantees). Additionally, a TEE can facilitate confirming that the model evaluated is the same model that was trained or deployed, specifically by hashing the computation done in the TEE (i.e., the model code and weights), signing it with a chip's private key, and allowing an outside inspector to compare the hashes between two scenarios where they want to ensure the same model was run. Note that this does not require a TEE; it is instead a general technique, but it is unlikely to be secure without TEE-type security. This approach is mostly solved in principle, but there are some remaining implementation challenges (Trask & Bejan, 2024, 48:45). TEEs on NVIDIA H100s (called "Confidential Computing") currently only work on single-GPU setups (NVIDIA Trusted Computing Solutions Release Notes, n.d.). Existing chips may additionally have security vulnerabilities that make this approach insecure against the well-resourced adversaries in our threat model (e.g., leading to evaluation data being stolen). NVIDIA's Confidential Computing is not designed to be secure against advanced physical threats (Nertney, 2023), and more importantly, current AI chips have not undergone the high degree of real-world testing necessary to be confident about their security.

Another approach to confirm that the same model is being evaluated as the one developed or deployed is if AI developers use cryptographically secure audit trails (i.e., version control) to track changes to their models. The actual implementation of this approach may also make use of TEEs. While the basic TEE approach is answering the question "Is model $X$ the same as model $Y$?," more complex audit trails could answer questions such as "Is model $X$ a derivative of model $Y$ with only a small number of changes?" This fine-grained analysis may be useful given the current status quo, where AI developers often modify their models between the version external pre-deployment evaluators have access to and the version released to customers (OpenAI, 2024a). If it is infeasible to do away with such a practice, it may be practical to bound the amount that a model changes before widespread deployment via cryptographically secure audit trails. Audit trails and version controls have precedent in other software development, and there are some attempts to apply them to AI development (Millet, 2024), but there are remaining problems in making them sufficiently secure for the present use case, such as ensuring the integrity of the entire hardware, firmware, and software stack.

There are a couple of other potential approaches to secure evaluations. Evaluations could take place in a mutually trusted data center (where both parties think it is very unlikely sensitive data will be leaked); however, it may be difficult to ensure that the model being evaluated is the correct model. Partial re-running of AI workloads can also be leveraged to confirm that self-reported model evaluations were done correctly, with a verifier repeating some of them and ensuring the same results are achieved. If the model of interest is available to external evaluators during deployment, e.g., via an inference API, evaluators can redo evaluations to check if a model performs sufficiently similar to the model tested earlier, gaining confidence that they are the same model. If there are significant concerns about the security of hardware, inspectors with code access could also carry out evaluations alongside the main AI developers, however, this poses privacy and security risks.

Besides making evaluations secure, it's also crucial that they are effective—actually measuring what they are meant to measure. This problem is not specific to verification, but substantial progress in the science of evaluations is likely needed before evaluations can give strong guarantees in an international verification context. Existing problems include: Evaluators may not be able to elicit the highest level of capability that could occur in deployment, e.g., due to lack of evaluation resources or access (Casper et al., 2024; METR, 2024), undiscovered capability elicitation techniques (Davidson et al., 2023), or either model developers or models themselves purposefully underperforming (Greenblatt, Roger, et al., 2024; Weij et al., 2024). Evaluations may not account for all categories of substantial model risk or may not provide sufficient coverage within each category (Reuel et al., 2024). Some risks arise from a model's interactions with the world and are very difficult to evaluate ahead of time (Mukobi, 2024). There are many more limitations to current model evaluations, and substantial progress is needed (Barnett & Thiergart, 2024; Mukobi, 2024; Reuel et al., 2024).

Beyond these standard evaluation approaches, countries can gain confidence that they have evaluated the proper model by using national intelligence operations to determine if far more capable models are being deployed. In particular, powerful AI systems are likely to have major benefits if deployed in economic or military contexts, and some forms of this deployment may be readily identifiable. However, two key uses for advanced AI systems—AI development and cyber operations—may be difficult to detect.

## Analysis

Trusted external evaluations will almost certainly require participation, but not necessarily honesty, from the model developers. For example, an international agreement could specify that deployed models must be run in TEEs with code to check that this is the same model that was evaluated. If the TEEs and the hardware they are running on are sufficiently secure, such an approach could verify the same model is used and that evaluations are performed properly without leaking sensitive information. However, we are overall uncertain about the difficulty of subverting the verification approaches described here on existing hardware. Currently, external evaluations rely almost entirely on trust that AI developers are not acting adversarially. Given that the present threat model and international verification context involve a substantial departure from this assumption, significant changes are needed. If there are irreparable vulnerabilities in existing AI hardware, it could take multiple years before the seemingly simple guarantees necessary for authentic model evaluations—the correct model is evaluated, and sensitive information is not leaked—can be achieved with high confidence.

In lieu of secure technical mechanisms, other approaches could achieve the goals of secure evaluations, such as inspectors with substantial code access or using cyber espionage to steal and evaluate models. In the long-run, more technical approaches are likely desirable because they can be designed to pose minimal security and privacy risks.

While evaluations are sometimes conceptualized as a static set of questions to which a model generates responses, effective evaluations may require more access, such as fine-tuning or access to model internals (Bucknall & Trager, 2023; Casper et al., 2024). In that case, lower-tech approaches, such as using inspectors in a mutually trusted data center with strong security, may be desirable, as they allow evaluators substantial access. TEE-based approaches could potentially facilitate these more dynamic evaluations, but this would involve substantial complexity in specifying what code is approved and what information evaluators would have access to.

Approaches to make evaluations secure are mostly narrowly scoped technical problems, whereas making evaluations effective is a complex and long-term scientific endeavor. Substantial work is still needed to make evaluations effective, for instance improving threat modeling and deciding what model properties should be measured, figuring out how to measure those properties with evaluations, and achieving sufficient capability elicitation. Due to the risk of under-eliciting capabilities, model evaluations may not be able to assess risks from advanced AI systems, including in a verification context.

## Building Blocks Preview

6/14 rows displayed. See appendix for complete table and explanation of feasibility estimates.

| Building Block | Mechanism | Details |
|---|---|---|
| Self-reported evaluations | Self-reported evaluations | Model developers can report evaluation results for their models without any verification. |
| | | **Feasibility** |
| | | High ▾  <1 year |
| | | **Notes** |
| | | Provides very little confidence on its own. |

| Proof of same model for training, evaluation, and deployment | Verifying same computation in TEE | A TEE can hash key elements of the computation (e.g., the model weights and inference code), and then confirm that the hash at evaluation time matches the hash at training and deployment. The "model from training" could be verified by re-running parts of the training run and comparing the weight hash at the end of the verified run to the inference weights. |
|---|---|---|

**Feasibility** | **Previous work** (non-comprehensive)
--- | ---
Medium ▾ 1–4 years | Trask & Bejan (2024)

**Notes**

This technology is not production-ready yet, but it is very close, and we know how to solve the problem in principle. It is unclear if TEEs with existing AI chips can be made sufficiently secure or if new chips are needed. This general approach could be done without TEEs, but it would be easier to spoof.

| Secure evaluations | Inspectors | In the absence of TEE-based security around external evaluations, but as an improvement on self-report evaluations, international inspectors could observe the evaluations being done alongside the main researchers. |
|---|---|---|

**Feasibility**

High ▾ <1 year

**Notes**

It may be difficult to balance ensuring the inspectors have sufficient access (are sure the right models are being evaluated) with mitigating security risks.

| Effective evaluations | Proof of sufficient capability elicitation | Show that capability evaluations would be reliable even if model developers or models themselves were engaged in sandbagging (purposefully underperforming during evaluation). Proof of training/inference may be needed. |
|---|---|---|

| **Feasibility** | | **Previous work** (non-comprehensive) |
|---|---|---|
| Medium ▾ | 1–5 years | Greenblatt, Roger, et al. (2024); van der Weij et al. (2024) |

**Notes**

This may be impossible for sufficiently advanced AI systems.

| General intelligence gathering to augment evaluations | External deployment monitoring | It may be relatively easy to infer the approximate capability level of AI systems that are deployed in the world, e.g., by looking at their economic impact, the state of automation, or interacting with them to test their capability level. We might expect powerful AI systems to be deployed widely because they could bring massive economic benefits. |
|---|---|---|

**Feasibility**

| Medium ▾ | 1–3 years |
|---|---|

**Notes**

There is likely room for new work, similar to Observational Scaling Laws (Ruan et al., 2024), that allows for the inference of general model capabilities based on particular deployment information. Work aimed at inferring the properties of proprietary models (Carlini et al., 2024) may also be relevant.

| Internal deployment monitoring via HUMINT | Human-based intelligence can help identify the capability level of secret AI projects. This could include spies, unintentional leaks, whistleblowers, and interviews conducted by international authorities. |

**Feasibility**

High ▾  <1 year

**Notes**

Monitoring the capability level of internal AI systems, which are only deployed internally and are kept secret, is more difficult than monitoring externally deployed models.

# Verifying Various Policy Goals

There are many other policy goals that international coordination on AI may aim to achieve. For instance, an international treaty could require every frontier AI development project to write a safety case that is reviewed by an international committee. This section discusses some self-report and verification building blocks that may help with these goals.

## Key Takeaways

- Various human-based data collection methods can help build confidence that agreements are followed, especially whistleblower programs and access to interview relevant personnel.

- In the future, AI systems could assist with verification, e.g., by reviewing code for compliance without leaking sensitive information.

- AI data centers should not be co-located with sensitive military installations or non-AI compute as this would make monitoring and enforcement more difficult.

## Verification Approach

Some general verification mechanisms will be useful for many policy goals. Shared and anonymous reporting channels for AI incidents (e.g., model misuse, cybersecurity incidents) can build trust between countries by increasing transparency into other AI development projects without revealing highly sensitive information. Whistleblower programs are likely a key part of verifying international treaty compliance. Along with whistleblower programs, international authorities could be given the ability to interview the people working on countries' AI projects in order to increase the likelihood of finding covert violations. Whistleblowers and interviews could be used to verify a wide range of claims about AI development (Brundage et al., 2020; Brundage, 2024), such as whether loosely defined safety commitments are being followed, making them more flexible than many technical approaches, which are often limited to verifying narrow claims about AI models.

There are numerous potential mechanisms to help with verification that are enabled by AI systems, such as improved monitoring of the chip supply chain, zero-knowledge code review, and the saving of relevant AI systems' logs. For instance, effective evaluations may require conducting repeated experiments and fine-tuning on a model, with direct weight access to prevent tampering, a task that requires substantial access; while it may be difficult to trust humans working for a foreign government, it may be possible to jointly build an "evaluations bot" that is mutually trusted to carry out such a procedure. Whether AI-enabled approaches are viable will depend on whether such systems can be made competent, trustworthy, and robust to adversaries.

Some policy goals need to verify properties about inference, for instance, whether a model is being used to make weapons or whether certain safety measures are run alongside it. Two main conditions must be met: all copies of the model that are running are known, and all known copies of the model have the intended oversight applied. While it may appear difficult to verify that all copies of the model are known, this could be accomplished with stringent security measures around model weights (e.g., model weights do not leave a particular data center), which developers may implement for standard security reasons, but would need to be made internationally verifiable and sufficiently secure against insider threats (Nevo et al., 2024).

Knowing where copies of a model are could also be accomplished by encrypting model weights so that only certain chips can decrypt them, but this faces a security problem as the chips doing decryption may not be sufficiently secure. Speculatively, a small part of inference computation (e.g., one layer of a model) could happen in a centralized data center which is subject to monitoring, whereas the vast majority of compute does not need to be known or monitored. Because the centralized computation is necessary for inference, all inference would be known; however, this has numerous difficult requirements, such as the centralized computation being very difficult to reproduce. If the vast majority of the AI chips in the world are covered by a governance regime that can restrict their activities (e.g., with FlexHEGs), another option to know where inference is occurring is to broadly restrict chips from running inference on a model of interest, or monitor that inference. Due to the difficulty in ensuring all AI chips are covered by a governance regime, this approach could build confidence but is unlikely to provide strong guarantees.

The actual oversight applied to inference might include input/output classifiers, similar to content filters, an idea we discuss in the appendix. Unfortunately, verifying that inference classifiers are running properly may be difficult, given the adversarial situation. Generally, some mechanisms for limiting or monitoring chips might be applicable to this use case, as the goal is to ensure chips are only running if particular oversight is applied, but they likely require better chip security. Partial re-running of workloads may also be effective, with a mutually trusted data center retroactively confirming declared classifier results on a subset of inference requests (Baker et al., Forthcoming). Goals pertaining to inference may also benefit from monitoring the downstream effects of these AI systems (e.g., economic growth).

Policy goals may also concern the desired behavior of AI systems, e.g., abiding by international agreements or not engaging in military first-strikes. These could be met with substantial progress on Provably Safe AI agendas, using agreed-upon model behavior specifications (an idea we elaborate on in the appendix), or reducing relevant model capabilities; none of these are technologically mature, and all require additional confidence that a specific training process was used and the deployed model is the same model that was trained. Model evaluations are a natural approach to testing if an AI system follows some desired behavior, but current techniques should not be relied on here due to their inability to assess model propensity and the likelihood of corruption by developers. Inference monitoring could potentially be used to ensure deployed AI systems follow desired goals but may require more access.

We note that numerous verification mechanisms require substantial access to data centers, so we encourage AI data centers to be built separately from non-AI data centers and separately from sensitive military installations in order to reduce the privacy cost of closely monitoring AI development and deployment. We discuss numerous ideas from cryptography that will play a role in the various verification mechanisms discussed.

## Analysis

Verifying some property about all inference instances of a model is difficult. This is likely to require substantial access and tight security around AI development (e.g., to ensure all inference of a model is known). Further ensuring that inference oversight is being applied faces difficulties due to the security of AI hardware: new chips or FlexHEG retrofitting of existing chips (e.g., with a secure processor and tamper-proof enclosure) may be necessary. Other approaches that could guarantee that all inference is monitored, such as FlexHEGs applied to all AI chips, are likely much more difficult to achieve. So, while in principle, there are various approaches to monitoring all inference of a particular model, the most promising appears to be securing and centralizing a model's weights to a small number of monitored locations, and this security being robust against insider threats (e.g., subversion by the AI developer). Significant investment in security is needed early to reach this level (Nevo et al., 2024).

One under-explored area is AI-enabled verification mechanisms, which may be especially effective because they could allow zero-knowledge review of sensitive information, however these require favorable developments in AI capabilities and safety properties. Another under-explored approach is training AI systems to abide by shared norms, such as following international agreements. This appears desirable but may be extremely technically difficult.

Whistleblowers and interviews, on the other hand, have been used previously and do not require novel technical advances. Therefore, these low-tech methods are likely to be critical to international verification if verification regimes are needed in the near future.

# Building Blocks Preview

12/29 rows displayed. See appendix for complete table and explanation of feasibility estimates.

| Building Block | Mechanism | Details |
|---|---|---|
| Whistleblower programs | Whistleblower programs | People who whistleblow about violations of international agreements are protected, e.g., with asylum guarantees. Whistleblowing is encouraged by access to reporting channels that are robust to intense state surveillance. |

| | | **Feasibility** | **Previous work** (non-comprehensive) |
|---|---|---|---|
| | | High ▾  <1 year | Baker et al. (Forthcoming) |

| Interviews | Interviews | Countries agree to allow international inspectors to interview personnel working on AI development. These interviews focus on verifying treaty compliance. These interviews should occur in neutral locations with privacy and safety guarantees. |
|---|---|---|

| | | **Feasibility** | **Previous work** (non-comprehensive) |
|---|---|---|---|
| | | High ▾  <1 year | Wasil, Berglund, et al. (2024) |

| AI-enabled verification mechanisms | AI-powered zero-knowledge work | AI systems could be used to do a substantial amount of work with sensitive data and then output only a non-sensitive final result. For instance, an "Evaluations Bot" would be one such system. This could dramatically expand the set of cognitive tasks that could be accomplished in a zero-knowledge way, aiding with many aspects of verification. |
|---|---|---|

**Feasibility**

Medium ▾  1–4 years

**Notes**

The model capabilities will inform what sort of work is possible here, and sufficiently capable models for some task may not be trustworthy. Additionally, models will need improved robustness to jailbreaks in order to be effective.

| Model behavior specification agreements (specifically, the model will not help violate international treaties) | Countries agree to use behavior specifications that are good for other countries. E.g., "The model will engage in $X$ benefit sharing" or "The model will not execute military first-strikes". Verification that a given behavior specification is used can be done with partial re-running of training. Hypothetically, external evaluations could be used to gain confidence that behavior specifications were used, but current evaluations are unable to assess propensity (as is needed) and are unlikely to work in this adversarial setting. Even without being robust, this approach could be cheap to implement and thus worth pursuing. |
| --- | --- |

| Feasibility | Previous work (non-comprehensive) |
| --- | --- |
| Medium ▾<br>1–100? years | Discussed below; OpenAI (2024b) |

**Notes**

The tractability of this approach likely varies significantly with AI capability level. Current approaches for frontier AI systems are insufficient (as evidenced by models' susceptibility to jailbreaks, which could be used by a country when carrying out violations). Robustness to jailbreaks for current models is likely viable within a few years, based on our familiarity with the literature. Proper value alignment for advanced AI systems may be incredibly difficult (Yudkowsky, 2022), and it is difficult to upper bound the time needed.

| | | |
|---|---|---|
| Knowledge of all inference instances of a model | Weights do not leave their training data center | If data centers are [registered](#) and have very good security, it may be possible to quarantine model weights to the data center they were trained on (or a small number of other data centers). If successful, this would allow a monitoring party to know where all inference is happening. While strong security around model weights is generally desired, sufficient effort is needed to make it happen. Additionally, this approach requires that security be sufficient to prevent insider threats from the country operating the data center, e.g., international collaboration on security. |

**Feasibility**  Previous work (non-comprehensive)

| Medium ▾ | 1–5 years | Nevo et al. (2024); Greenblatt (2024); Harack et al. (Forthcoming) |
|---|---|---|

**Notes**
Requires very good security and likely continuous monitoring of security but is otherwise straightforward.

| | | |
|---|---|---|
| | Widely applied chip restrictions | If nearly all AI chips in the world are under a governance authority with tools like FlexHEGs, various mechanisms for chip limiting or monitoring could be applied, as discussed previously. For instance, chips could be restricted to only running inference or training on certain models which have been approved by an international authority. If a substantial fraction of all AI chips were covered by such controls, this could increase confidence that inference was not being done on an unapproved model. |

**Feasibility**

Medium ▾   2–6 years

**Notes**

A key uncertainty is defining which chips these restrictions would need to apply to, and a key difficulty would be implementing sufficiently secure restrictions on those chips. Inference typically has lower compute requirements than training; if these are still high (e.g., >64 cutting-edge AI chips), this may be feasible but would be very difficult. If the model of interest could be run on consumer hardware, it is likely infeasible to bring all such hardware under a governance authority. It may also matter both whether the model can be run, at all, on some hardware, and whether it can be run efficiently (e.g., if consumer hardware can run a model, but does so much slower than data center AI chips, restrictions on data center AI chips could be effective).

| Inference-time oversight | Inference classifiers | Lightweight classifiers run locally on inference workloads, enabling verifiers to check that deployment follows the rules while limiting access. |
|---|---|---|

**Feasibility**          **Previous work** (non-comprehensive)

Medium ▾   1–2 years       Discussed below

MIRI TECHNICAL GOVERNANCE TEAM

| Provably Safe AI | Provably Safe AI | This is a family of research agendas. Formal verification of AI agents and AIs that oversee them such that deployment of the agent will provably fall within some safety specifications designed with human input. |
|---|---|---|

**Feasibility** | **Previous work** (non-comprehensive)
---|---
Medium ▾  3–15 years | Dalrymple et al. (2024); Dalrymple (2024)

| Reduce model capabilities in relevant domains | Knowledge unlearning | After training, apply knowledge unlearning techniques to reduce the model's capabilities in the dangerous domain. Verification that countries are following this commitment could include partial re-running of chip activities. |
|---|---|---|

**Feasibility** | **Previous work** (non-comprehensive)
---|---
Medium ▾  1–5 years | Casper (2023); Tamirisa et al. (2024)

**Notes**

Current techniques are insufficient. May be intractable for advanced AIs due to knowledge collisions. This unlearning may need to be robust to fine-tuning, or not, depending on the risk there.

| Non-AI monitoring | General intelligence gathering (economic) | Publicly available discussions about AI integration, scientific studies, general economic measures, and private financial data are all likely to indicate when a country is getting substantial economic returns from its AI systems and automation. |
|---|---|---|

**Feasibility**

High ▾  1–2 years

| No data center colocation | No colocation with non-AI compute | Data centers built only for AI workloads allow numerous crucial security, verification, and enforcement mechanisms to be applied without collaterally affecting non-AI compute. For example, verification approaches based on data-center-wide power draw would be ineffective if there were power-hungry non-AI workloads happening in the same data center. |
|---|---|---|

**Feasibility**

High ▾  1–2 years

- - - - - - - - - - - - - - - - - - - - - - - - - - - - - - - - - - - - - - - - - - - - - - - - - - - - - - - - - -

| | No colocation with sensitive military facilities | Verification mechanisms that make use of physical inspections, or compute tracking in general, may be much more difficult if AI data centers are part of military installations, as there is an increased risk of sensitive information leaking. |
|---|---|---|

**Feasibility**

High ▾  1–2 years

# Conclusion

This report gives an overview of mechanisms that can help verify that countries are not breaking international agreements about AI development. Such agreements and confidence-building mechanisms may be crucial to avoiding societal-scale risk from advanced AI systems. There are numerous barriers to the verification approaches in this report working:

- Substantial international political will (or domestic political will, if applied to national regulation) is needed.
- Most of the specific mechanisms are not yet technologically mature.
- Substantial time is needed for testing and implementation of mechanisms.
- Future advancements in algorithmic progress or distributed training could make verification much harder.
- The mechanisms in this report are unlikely to be robust to superintelligent AI capabilities, given the large jump from the current technological landscape to that one.
- The implementation details of many mechanisms are somewhat specific to the current language model, transformer, pre-training paradigm and may require adjustment for other paradigms.

Many of the mechanisms described in this report are in their early stages—frequently the ideation stage—but appear feasible. If they were a priority for a national government, minimal versions could be achieved in months and more robust versions in a few years. However, if development projects only happen when risks are acute (e.g., following a catastrophic AI event), we may not have that time. Therefore, we must begin this development early, especially for mechanisms that require serial time. We provide a list of such mechanisms in the appendix, including FlexHEG mechanisms, tracking existing AI chips, and centralizing the chip supply chain.

The main threat model in this report is motivated and well-resourced state actors putting substantial effort into bypassing verification mechanisms. While this is a critical threat model for international verification to deal with, one should not let perfect be the enemy of good: many of the mechanisms discussed here, while not comprehensive on their own, could be used together as an effective verification regime.

Verification of AI treaties is a large area of research and practice for which this report provides only a high-level overview. As such, there are numerous directions for future work that are not tackled here, for example:

- What verification mechanisms are most relevant and useful for domestic, rather than international, regulation and verification?

- How can verification mechanisms be put together to accomplish comprehensive and ambitious goals, such as all frontier AI development following a safety case (discussed in the appendix), or enforcing a global pause on frontier AI development and deployment?

- Prototyping and stress-testing the mechanisms discussed in this report (e.g., developing novel methods for verifying compute use with minimal-access, improving methods for partial re-running of chip activities, making TEEs more secure).

- Carving the space of verification mechanisms by their robustness to different types of adversaries, as Aarne et al. (2024) do for some on-chip mechanisms.

- Building political will at the international, national, and corporate levels to develop and implement these mechanisms.

We conclude by noting that effective verification mechanisms could be a catalyst for international agreements, as has been the case previously (Toivanen, 2017). Strong verification is crucial, regardless of whether the international situation is one of "trust but verify" or "distrust and verify".

# Appendix

Table of contents:

# Acknowledgments

We are thankful to the following people for useful discussion or feedback during this project:

Ben Harack, Mauricio Baker, James Petrie, Ben Bucknall, Romeo Dean, Davis Brown, Eli Tyre, Jacob Lagerros, Lennart Heim, Gretta Duleba, Alex Vermeer, Joseph Rogero, Thomas Larsen, Peter Barnett, David Abecassis, Misha Gerovitch, Eli Lifland, Caleb Withers, Jason Hausenloy, Malo Bourgon, Nick Gabrieli, David Schneider-Joseph, Gabriel Kulp, Tom Reed, Tom Shlomi, Joe Collman, Matteo Pistillo, Yashvardhan Sharma, Olivia Jimenez, Manos Seferis, Timothy Chauvin, Siméon Campos, Jonathan Happel, Michael Byun, Janet Eagan, Josh Clymer

These individuals do not necessarily agree with the views in the paper, and all mistakes are our own.

# Author Contributions Statement

Lisa was the Research Lead for the project, came up with the initial ideas and directions, contributed some original ideas for mechanisms, and did some editing. Aaron did the majority of the writing and did most of the analysis and mechanisms research.

# Feasibility Estimates

All the feasibility estimates assume significant political will to curb risks from frontier AI development (i.e., similar to the U.S. response to 9/11). The time estimates reflect our estimation of how long it would take to **design and implement** the given measure. They are dominated by technical R&D time for most mechanisms due to the assumption of strong political will. Approaches that require new AI chips tend to have long implementation timelines due to the time it takes for new chips to become a significant share of computing power (Heim, 2024a).

A feasibility rating of "High" indicates that there are no major technological hurdles we are aware of ("the world basically knows how to do this, even if there are some finicky details"), a rating of "Medium" indicates some technological innovation is needed, but it could likely be accomplished with <200 experts working for a year; a rating of "Low" corresponds to a mechanism being particularly politically intractable or a very difficult technological project. The actual implementation of many mechanisms will require substantial effort from thousands of people, but we assume this is easy given the political will and technological maturity. These are **very optimistic assumptions**.

We have low confidence in most of the feasibility estimates. They are preliminary, quick, estimates, rather than a thorough analysis.

# Building Blocks

## Locating Compute Building Blocks

| Building Block | Mechanism | Details |
|---|---|---|
| Chip ownership and location registry | Chip ownership and location registry | AI chip owners / data center operators must register the chips they own, and where those chips are located, with a central regulator. Chip sales require updating the registry with the new owner and location. |

**Feasibility** — **Previous work** (non-comprehensive)

High ▾ <1 year — Jones (2024); Shavit (2023); Baker (2023); Fist & Grunewald (2023)

---

**Data center registry**

Data centers that do not have AI chips but could later be repurposed for AI use may need to be registered with a central regulator. "Data center" can be defined quite broadly, and it is unclear what data centers would need to be registered.

**Feasibility**

High ▾ <1 year

**Notes**

Perhaps an initial requirement for registration would be if a data center currently consumes >1 MW—this would correspond to ~1,000 H100 GPUs. There are likely thousands to tens of thousands of such data centers currently (Pilz & Heim, 2023).

| Flexible Hardware-Enabled Guarantee (FlexHEG) mechanisms | Main FlexHEG design | A FlexHEG is a mechanism added to an AI accelerator that allows privacy-preserving verification of chip compliance with varied and flexible policy goals. This could be retrofitted to existing chips by inspectors or added to chips as they are shipped to data centers. The key technical requirements of such a mechanism are a secure enclosure to detect physical tampering, a self-disablement mechanism, a secure processor, an accelerator interlock (control over GPU sufficient for enforcement), a mechanism update process, and confidential communication between devices. If these component pieces are built, they could be added to existing AI chips or servers. |
|---|---|---|

**Feasibility** | **Previous work** (non-comprehensive)
---|---
Medium ▾  2–5 years | Petrie et al. (2024)

**Notes**
Strong and secure FlexHEG mechanisms would allow the implementation of many different policy goals that do not need to be decided in advance, making them highly desirable. Key difficulties are likely to include designing and verifying a secure processor and designing tamper-proof enclosures.

|  | On-chip cryptographic capabilities | Chips can execute basic cryptographic operations securely, such as key signing. This should probably be done with a dedicated hardware security module. |
|---|---|---|

**Feasibility** | **Previous work** (non-comprehensive)
---|---
Medium ▾  1–3 years | Aarne et al. (2024)

**Notes**
Many AI chips can already do this, but their adversarial robustness is unclear, so new AI chips could be needed.

| Tamper-evident chips | Includes: [video surveillance](#), tamper-evident seals or packaging on chips, and potentially remote attestation. Helps detect violators but does not on its own prevent tampering—needs to be paired with enforcement. |
|---|---|

| Feasibility | | Previous work (non-comprehensive) |
|---|---|---|
| High ▾ | 1–5 years | Aarne et al. (2024); TamperSec (n.d.) |

**Notes**
Some approaches here require replacing AI chips, but some may be applied to existing chips (with temporary physical access) or at the data center level.

| Tamper-proof chips | Tampering results in destroying chip capabilities, e.g., via a secure enclosure of a chip that destroys the chip if broken. In effect, this mechanism is tamper-evident chips with enforcement built in. Helps directly prevent violations. |
|---|---|

| Feasibility | | Previous work (non-comprehensive) |
|---|---|---|
| Medium ▾ | 1–5 years | Aarne et al. (2024); TamperSec (n.d.) |

**Notes**
Some approaches here require replacing AI chips, but some may be applied to existing chips (with temporary physical access).

| Chip identity verification | Each chip has a unique identifier, perhaps secured as a private key or a physical unclonable function ("Physical unclonable function", 2024). This enables other chip mechanisms. Basic serial numbers are not adversarially robust. |
|---|---|

| Feasibility | | Previous work (non-comprehensive) |
|---|---|---|
| High ▾ | 1–3 years | Aarne et al. (2024); Reuel et al. (2024) |

**Notes**
Many AI chips can already do some versions of this, but their adversarial robustness is unclear, so new AI chips or modifying existing chips could be needed.

| Trusted Execution Environment (TEE) | Code run in TEEs is secure against most external surveillance, and only approved programs can interact with the TEE. To ensure integrity, TEEs should use trusted code, likely starting with [secure boot](). "Confidential Computing" is the name of NVIDIA's TEE implementation. See also "Secure Enclave" from Apple. |
|---|---|

| **Feasibility** | | **Previous work** (non-comprehensive) |
|---|---|---|
| Medium ▾ | 1–4 years | Aarne et al. (2024); Kulp et al. (2024) |

**Notes**

H100s already have confidential computing for single-chip setups, and it may not be too difficult to get this for multi-chip setups (Nertney, 2023). The adversarial robustness of current implementations is unclear, so new AI chips could be needed.

| Firmware roll-back protection | Once a firmware update has been applied, previous firmware versions cannot be used. This is a critical component to firmware-based regulations changing over time. |
|---|---|

| **Feasibility** | |
|---|---|
| Medium ▾ | 2–5 years |

**Notes**

Some AI chips, such as H100s, appear to have firmware roll-back protection (*NVIDIA H100 NVL GPU Product Brief*, 2024), but the current implementation is likely not sufficiently secure, so new AI chips could be needed.

| | FlexHEG applied to location verification | The secure processor that is part of a FlexHEG could implement various chip location tracking protocols and be trusted to do so given that it is part of a FlexHEG. On the other hand, strong FlexHEGs would obviate the need for location verification as governance could be conducted remotely with trust that chips will not be tampered with. |

**Feasibility**

Medium ⌄  2–5 years

**Notes**

If a FlexHEG is in place, additionally implementing location verification is likely easy and could be done in less than a year.

---

| Chip location tracking | Ping-based location tracking | Set up a series of servers that ping chips and triangulate chip location based on response time to different servers. This is relatively low precision. |

**Feasibility** | **Previous work** (non-comprehensive)

High ⌄  2–4 years | Aarne et al. (2024); Brass & Aarne (2024)

**Notes**

May not require replacing hardware. The main security requirement is chips having a private key, however, it is unclear if current chips are secure against physical attacks to extract such a key, as would be needed here. May require chips to be connected to the internet.

---

| | Topology-based location tracking | Chips locate themselves based on signals from emitters with known locations, such as cell towers. |

**Feasibility** | **Previous work** (non-comprehensive)

Medium ⌄  2–4 years | Brass & Aarne (2024)

**Notes**

Requires an antenna on chips. May be difficult to make sufficiently robust.

| GPS-based location tracking | Chips locate themselves using GPS or similar technology and report this to a monitoring server. |
|---|---|

**Feasibility** | **Previous work** (non-comprehensive)

| Medium ▾ 2–4 years | Brass & Aarne (2024) |

**Notes**

GPS can be easily spoofed. Requires an antenna on chips. May be difficult to make sufficiently robust.

| Firmware-based chip operating licenses for current chips | It may be possible to implement operating licenses on current AI chips via firmware updates and secure boot. This is very unlikely to be secure against state actors, but augmenting with tamper-evident mechanisms such as security cameras may be effective (this requires physical access). |
|---|---|

This license could be set to a fixed amount of FLOP or clock time. The license could include the chip's identity, the license expiration, and it could be signed by the private key of a regulator or counterparty. Chip does not need direct internet connection to renew license (i.e., could be air-gapped and the license delivered via a physical drive). License renewal can be connected to numerous governance goals, such as chips reporting their activity, their location, or their compliance with on-chip limitations.

**Feasibility** | **Previous work** (non-comprehensive)

| Medium ▾ 1–3 years | Aarne et al. (2024); Kulp et al. (2024); Petrie (2024) |

**Notes**

The protocols for location tracking in this section are relatively straightforward, but they could be spoofable either by tampering with the mechanisms on the chips (current chips likely aren't sufficiently secure) or elsewhere, as in GPS spoofing.

| National intelligence operations | Satellites (visual) | Satellite imagery may be used to detect the construction of data centers, fabs, and power infrastructure. |
|---|---|---|

| **Feasibility** | **Previous work** (non-comprehensive) |
|---|---|
| High ▾ <1 year | Wasil, Reed, et al. (2024); Pilz & Heim (2024) |

**Notes**
Satellite information is unlikely to differentiate AI vs. non-AI data centers. It also may struggle to differentiate data centers from other industrial buildings (cooling may be a key differentiator).

| | Satellites (infrared) | Thermal imaging may be used to detect data centers in use. In particular, data centers may remain in a narrow temperature range throughout the day due to chips constantly running and being cooled. |
|---|---|---|

| **Feasibility** | **Previous work** (non-comprehensive) |
|---|---|
| High ▾ <1 year | Wasil, Reed, et al. (2024) |

| | Geophysical MASINT to detect underground construction | Various approaches are already in use to detect underground construction (National MASINT Office, 2022). This could be applied to reduce the likelihood of underground data centers being built secretly. |
|---|---|---|

| **Feasibility** |
|---|
| High ▾ <1 year |

**Notes**
These approaches may not, on their own, differentiate between data centers and other underground construction.

| HUMINT | Human-based intelligence such as spies, unintentional leaks, whistleblowers, and interviews conducted by international authorities can add confidence that secret AI development, data center, or chip production are not taking place. |

| Feasibility | Previous work (non-comprehensive) |
| --- | --- |
| High ▾ <1 year | Baker et al. (Forthcoming) |

| Power grid analysis | Large AI data centers may have substantial power requirements. Surveillance of existing power grids and off-grid generation could, therefore, detect such data centers. |

| Feasibility |
| --- |
| Medium ▾ <1 year |

**Notes**

The power draw of 100,000 H100s is ~130 MW ("NVIDIA DGX H100 Datasheet"). Given the scale of current military nuclear reactors (up to low hundreds of MW electrical output on some aircraft carriers) ("United States naval reactors"), covert generation of substantial power may be possible. Chip performance per watt is increasing at a rate of ~1.6x per year (Epoch AI, n.d.). It's estimated that there are currently ~500 data centers globally with >10 MW capacity (Pilz & Heim, 2023).

| Cyber operations and communication surveillance | Signals-based intelligence is already used. For detecting AI compute, such approaches might focus on AI developers, data centers, and chip producers. |

| Feasibility |
| --- |
| High ▾ <1 year |

- - - - - - - - - - - - - - - - - - - - - - - - - - - - - - - - - - - - - -

| FININT | AI development projects are very expensive, and covert AI projects might leave a substantial money trail. Intelligence gathering here might focus on purchases of AI-relevant raw materials, chips, and other data center components. |
|--------|---|

| **Feasibility** | **Previous work** (non-comprehensive) |
|---|---|
| High ▾ <1 year | Wasil, Reed, et al. (2024) |

- - - - - - - - - - - - - - - - - - - - - - - - - - - - - - - - - - - - - -

| Tracking of relevant personnel | There are hundreds to thousands of experts who have the knowledge and skills to contribute to frontier AI development, chip design, or chip fabrication. Building state-of-the-art domestic chip production would be very difficult without these people. Various measures could be used to confirm that these people are not contributing to covert national projects in violation of an international agreement. |
|---|---|

| **Feasibility** |
|---|
| High ▾ <1 year |

| Security | AI data centers are secure against chip theft | Data center operators and their countries collaborate to boost data center security. Ideally, this would prevent the theft of model weights and algorithmic secrets, but preventing the theft of physical chips is the focus of this section and is probably much easier (current security may be sufficient if done with [international collaboration](#)). |
|---|---|---|

| **Feasibility** |
|---|
| High ▾ <1 year |

| Physical inspection of data centers | Verify chip identity and count | Use unique chip identifiers to ensure the right chips are present; count to make sure all are accounted for. |
|---|---|---|

**Feasibility**

High ▾ <1 year

---

Verify non-tampering of chips (only relevant if there are mechanisms that could be tampered with)

Make sure chips have not been modified in violation of agreements.

| **Feasibility** | **Previous work** (non-comprehensive) |
|---|---|
| High ▾ <1 year | Aarne et al. (2024) |

**Notes**
Easy if we have good tamper-evident chips to begin with.

---

Audit data center security

Check that the data center is at the security level claimed.

| **Feasibility** | **Previous work** (non-comprehensive) |
|---|---|
| Medium ▾ 1–4 years | Nevo et al. (2024) |

**Notes**
It may be easy to verify that the security in a data center is strong enough to prevent the theft of chips, but it may be more difficult to verify the much higher level of security needed to secure model weights.

---

Power accounting

Compare the expected power draw of a data center, based on the type and number of chips, to the actual power draw to build confidence that there are not undeclared chips.

**Feasibility**

Medium ▾ 1–2 years

**Notes**
If a data center has a combination of older (less energy efficient) and newer chips, this is difficult. If a data center has a combination of AI and non-AI chips, this is difficult.

| Continuous monitoring of AI data centers | Perimeter and portal continuous monitoring | Materials coming in and out of data centers (especially AI chips) flow through portals, which are jointly monitored to avoid chip theft. Existing physical security measures such as fencing and cameras are likely to be sufficient if they can be jointly monitored. |
|---|---|---|

**Feasibility**      **Previous work** (non-comprehensive)

| High ▾ | <2 year | *START: Annex to Protocol on Inspection and Continuous Monitoring Activities* (n.d.) |
|---|---|---|

**Notes**

It may also be desirable for network traffic in and out of data centers to be jointly monitored (this can reduce the risk of model weights being improperly moved), but that is out of scope for this section.

| | Security cameras inside data centers | This is standard (Google Cloud Tech, 2020) but would involve giving international inspectors access or (more likely) having international inspectors install their own cameras. The primary goal of security cameras in data centers for verification is to ensure chips are not being removed, added, or modified to bypass governance mechanisms, so cameras should focus on chips. |
|---|---|---|

**Feasibility**

| High ▾ | <1 year |
|---|---|

**Notes**

Cameras might confirm that unauthorized chips aren't being substituted into a data center, that interconnect limits aren't being changed, or that chips are not being tampered with.

| | | |
|---|---|---|
| | Sensors for chip activities (not relevant to locating chips) | Various [workload classification methods](#) involve gathering information about chips, such as their power draw or network traffic. Specific equipment would be installed to gather this information. |

| **Feasibility** | **Previous work** (non-comprehensive) |
|---|---|
| High ▾ 1–3 years | Baker et al. (Forthcoming) |

**Notes**
There are likely low-millions of high-performance AI chips in the world (Epoch AI, n.d.), so mass-production of these sensors would be necessary, which could take many months or a couple years.

| | | |
|---|---|---|
| Register chip production | Supply chain registry | Key parts of the chip supply chain must register their production facilities with a central verifier. This should likely be based on a thorough assessment of the current bottlenecks and crucial actors for chip production. |

**Feasibility**

High ▾ <1 year

| | | |
|---|---|---|
| Tracking chip production | Chain-of-custody | Chain-of-custody (CISA, n.d.) is implemented for the AI chip supply chain: firms document the transfer and storage of components along the chip supply chain. |

**Feasibility**

Medium ▾ 1–3 years

| | | |
|---|---|---|
| | Tracking raw materials and component parts | Major parts of the chip supply chain, from raw materials to EUV machines, to clean room components, are tracked by intelligence agencies and an international body. |

**Feasibility**

Medium ▾  1–3 years

**Notes**

It's unclear if the *raw materials* are sufficiently AI-specific that it makes sense to track them. There are many details and approaches to tracking chip production, and we leave them to future work.

| | | |
|---|---|---|
| | General intelligence gathering | National intelligence agencies use satellites, HUMINT, and other sources of information to identify the construction and operation of unregistered chip production facilities. |

**Feasibility**

High ▾  <1 year

**Notes**

Covertly building a chip supply chain appears very difficult, and international regulations with whistleblower programs may be sufficient for detection.

| | | |
|---|---|---|
| Physical inspection of fabs | Physical inspection and continuous monitoring of fabs | Check that chip production facilities are implementing the agreed-upon on-chip measures, have the manufacturing capacity claimed, and are not doing unauthorized production or distribution. Includes human inspectors, cameras, interviews with employees, etc. |

**Feasibility**

High ▾  <1 year

**Notes**

Includes inventory audits.

MIRI TECHNICAL GOVERNANCE TEAM

| Evaluation of chip designs and testing of chips in development | If on-chip mechanisms are used, there needs to be assurance that the manufactured chips actually have these mechanisms correctly implemented. To ensure chip designs have not been backdoored to circumvent governance measures, it may be necessary for international inspectors to evaluate these designs and confirm that a small number of randomly selected fabricated chips are implementing the intended design. Numerous tests like this may be necessary throughout the chip design process to ensure the on-chip mechanisms are implemented correctly. |
|---|---|

**Feasibility**

Medium ▾  1–4 years

**Notes**

This might involve a small contingent of experts from each country working on this verification with strong information security to avoid the leaking of sensitive data about chip designs.

| Multilateral export controls | Multilateral export controls | Members of the international agreement have a shared list of export controlled countries for AI chips and major components, in effect requiring a license from an international authority. |
|---|---|---|

**Feasibility**

High ▾  <2 year

**Notes**

These rules could be "disallow" or "allow" based, depending on risk. Countries might agree to reduce the proliferation of AI chips via shared export controls. It appears difficult to make such agreements robust to a country later deciding to break such an agreement.

## Compute Use Building Blocks

| Building Block | Mechanism | Details |
|---|---|---|
| Chip ownership and location registry | Chip ownership and location registry | AI chip owners / data center operators must register the chips they own, and where those chips are located, with a central regulator. Chip sales require updating the registry with the new owner and location. |

**Feasibility**

High ▾   <1 year

**Previous work** (non-comprehensive)

Jones (2024); Shavit (2023); Baker (2023); Fist & Grunewald (2023)

| Building Block | Mechanism | Details |
|---|---|---|
| Inference-only chips | Inference-only manufacturing | Some AI chips could be manufactured to be especially efficient at inference while not being optimized for training, e.g., Groq, AWS Inferentia, Sohu. Further work may be needed to make such chips sufficiently inefficient for training. |

**Feasibility**

Medium ▾   1–4 years

**Notes**

Some current inference-specialized chips could likely be repurposed for training without significant effort, e.g., as is done in a blog post about AWS Inferentia (Rand, 2023). Therefore, for them to be a credible sign that no large training runs are happening, future chips would have to be designed to be particularly inefficient for training. Some existing inference-specialized chips may be sufficiently incapable of being useful for training, e.g., because they use inflexible ordering of operations. On the other hand, approaches here will need to contend with substantial effort going into overcoming these limitations.

| | Modify chips to be inference-only | Using a combination of firmware updates, improved chip security, and likely operating licenses, modify chips to only be useful for AI inference, even after manufacturing. This could potentially be implemented with [FlexHEG](#) mechanisms. |
|---|---|---|

**Feasibility**

Medium ▾  2–5 years

**Notes**

This likely requires that chips be hardened during manufacturing and equipped with flexible hardware mechanisms like operating licenses. Mechanisms that resist adversaries may be difficult to design.

| Emergency chip shutdown | Unplug the data center from its primary and backup power supply | Trivially, a data center operator can claim that its chips are not being used in a large training run if they are not receiving the power needed to operate. This can likely be verified with less invasive methods such as power draw of the data center and knowledge about on-site (e.g., backup) power generation or potentially thermal imaging. |
|---|---|---|

**Feasibility**

High ▾  <1 year

**Notes**

It would impose negative externalities to unplug an entire data center if there is important non-AI compute in the data center, hence the importance of [avoiding co-location](#).

| | Unplug particular chips or servers | A data center could unplug certain AI chips while letting the rest of the data center receive power, and this could be verified with [cameras](#) or physical inspections. If there are substantial differences in the external resource requirements for AI and non-AI chips, this could be verified without access to data center internals, e.g., using power draw or cooling information. |
|---|---|---|

**Feasibility**

High ▾   <1 year

**Notes**

AI chips require specific infrastructure (e.g., server racks, cooling, networking equipment) to operate, so there are many ways to show such chips are not operational, e.g., all the AI chips are stacked on a table instead of in server racks.

| Chip use registries | Low-detail registry | Data center operators keep track of high-level information about different AI workloads run, e.g., training/inference, which model, which project, and the size of workloads. This type of registry would ideally account for all chip use in a data center. For some of these uses, [high-level chip information](#) could add confidence to this registry. Successfully spoofing a low-detail registry and the associated high-level chip measures is likely possible given the lack of workload re-running, so this registry does not provide strong guarantees. |
|---|---|---|

**Feasibility**

High ▾   <1 year

**Notes**

A key issue is that an adversary could split a single workload so that it appears to be multiple workloads. Know Your Customer (KYC) rules are a step toward such a registry, but they are unlikely to be applicable in the context of international verification.

--------------------------------------------------------

| Registry of training transcripts | Data center operators keep a detailed registry of how chips are used. This includes training transcripts that detail model weight checkpoints and everything necessary to replicate a training run (data, hyperparameters, random seed, etc.). |
|---|---|

**Feasibility**                    **Previous work** (non-comprehensive)

High ▾  <1 year                    Shavit (2023)

**Notes**

This registry is highly private and likely should only be used in zero-knowledge ways, such as being encrypted except for use on a mutually trusted data center. This registry could be implemented at the data center layer, but it would benefit from individual chips reporting on their activities.

--------------------------------------------------------

**Partial re-running of chip activities**

| Partial re-running description | A data center operator can claim its compute was used for a training run, and this claim can be verified by an external party with much less compute by re-running small segments of the full training run to see if they match the declared run. This requires checkpoints and hyperparameters provided in a registry of chip activity. This re-running happens in a zero-knowledge way in a mutually trusted data center where the chance of sensitive information leaking is low. There may be other approaches to verify a given transcript other than partial re-running, such as those explored by Choi et al. (2023). |
|---|---|

**Feasibility**                    **Previous work** (non-comprehensive)

Medium ▾  1–5 years                Shavit (2023); Choi et al. (2023); Baker et al. (Forthcoming)

**Notes**

Partial re-running enables verification of practically all properties of a training run: model architecture, hyperparameters, data, etc., because the data center operator provides this entire training transcript to a mutually trusted data center for verification.

--------------------------------------------------------

| On-chip weight logging | Individual chips occasionally save a snapshot of their memory (this is a "shard," or subset, of all model weights). This could be a (short) hash of the weights stored locally or the (larger) full-weights stored externally, ideally cryptographically signed by the chip to ensure authenticity. |

| **Feasibility** | | **Previous work** (non-comprehensive) |
|---|---|---|
| Medium ▾ | 1–5 years | Shavit (2023) |

**Notes**

Logging on individual chips may add security relative to just using the data center level registry. This is especially true if chips have strong security, but it is likely still useful to attempt this with existing chips.

| Proof-of-Training-Transcript | Given a training transcript and weight shards from individual chips, prove that this training transcript is the only feasible one that leads to the observed weight shards. Ideally this is done in a zero-knowledge way on a mutually trusted data center. Methods include re-running segments of training to ensure they reach the claimed end-point and testing memorization of recent data points at various checkpoints. |

| **Feasibility** | | **Previous work** (non-comprehensive) |
|---|---|---|
| Medium ▾ | 1–3 years | Shavit (2023); Jia et al. (2021); Choi et al. (2023) |

**Notes**

Robustness to adversaries is unclear. It may be impossible to know that the given training transcript is the only feasible one that leads to the weight shards.

| Mutually trusted data center | To show that a particular training transcript leads to the expected weight shards, run a small section of the transcript in a trusted data center, verify that the weight shards are as expected, and communicate this to the verifying party without leaking model details. The trust requirement here is that sensitive information used in the data center (e.g., model weights, model architectures) is highly unlikely to leak. This confidence might come from a combination of improved chip security (it is likely difficult to gain mutual confidence in existing chips) and other security measures, such as reducing the amount of information that can leave the data center (Greenblatt, 2024) and using onsite staff who have limited external contact. While stronger chip security could take substantial R&D effort, lower-tech solutions which require more political will could be viable in the short-term. |
|---|---|

| **Feasibility** | | **Previous work** (non-comprehensive) |
|---|---|---|
| Medium ▾ | 1–6 years | Shavit (2023); Baker et al. (Forthcoming); Nevo et al. (2024); Heim (2024b) |

**Notes**

Ensuring this data center is sufficiently secure may be difficult. New AI chips could be needed.

| Key detail: Provably random initialization | One major risk is that an AI developer will split up a large training run into numerous small training runs to avoid detection. This can perhaps be avoided by having model training start by initializing model weights randomly, e.g., according to a set seed, which a verifier can then replicate to ensure the beginning weights are random. It may also be necessary to require that the order of training data be decided through a certified-random process. |
|---|---|

| **Feasibility** | | **Previous work** (non-comprehensive) |
|---|---|---|
| High ▾ | <1 year | Shavit (2023); Choi et al. (2023) |

| Mechanisms for chip limiting, monitoring, or workload classification | FlexHEGs | Discussed previously. In addition to the FlexHEG design discussed above, in this section, it is crucial that FlexHEGs have some insight into what a chip is doing. For example, this could look like intercepting the communications this chip makes to other chips and encrypting them (where only specific chips can decrypt them), a way to implement interconnect bandwidth restrictions. This could also include the secure processor viewing the chip's operations and ensuring they do not violate some criteria. Because a FlexHEG design uses a general processor, many verification approaches could be pursued with such a mechanism. |
|---|---|---|

**Feasibility**          **Previous work** (non-comprehensive)

| Medium ▾ | 2–5 years | Petrie et al. (2024) |
|---|---|---|

**Notes**
Requires mass production of new hardware; depending on the implementation, it may require either new AI chips, auxiliary chips with secure enclosures, or something else.

- - - - - - - - - - - - - - - - - - - - - - - - - - - - - - - - - - - - - - - - - - - -

| | Workload approval | Individual workloads are pre-approved by a verifier as being in accordance with the rules, and chips only work if a valid signature is given. This is a more fine-grained version of operating licenses and might be implemented via a FlexHEG mechanism. |
|---|---|---|

**Feasibility**          **Previous work** (non-comprehensive)

| Medium ▾ | 2–5 years | Baker et al. (Forthcoming) |
|---|---|---|

**Notes**
It may be difficult for the verifier to do this without substantial access to algorithms. Vulnerability to tampering would be a property of the FlexHEG design used, and it could be very difficult to make this limitation robust on existing hardware.

- - - - - - - - - - - - - - - - - - - - - - - - - - - - - - - - - - - - - - - - - - - -

| | |
|---|---|
| Software and firmware restrictions | Chips are limited (via firmware and software [attestation](#)) to only use approved functionality. For example, no training if parallelized across >16 GPUs, or only run training on models approved by an international authority. Whether this is more of an [allow list](#) or a [disallow list](#) likely affects feasibility. This could be implemented via a firmware update that specifies approved functionality (combined with [rollback protection](#)). This could be implemented with a [FlexHEG](#) mechanism. |

**Feasibility**

Medium ▾   2–5 years

**Notes**

A key difficulty is making these restrictions adversarially robust, especially if they are not tied to expiring licenses (and therefore being updated frequently). Vulnerability to tampering would be a property of the FlexHEG design used, and it could be very difficult to make these restrictions robust on existing hardware.

| | |
|---|---|
| Inspectors reviewing code | Inspectors are given substantial access to the AI development projects, including reviewing code, allowing them to verify that large training runs are not taking place. |

**Feasibility**

Medium ▾   1–2 years

**Notes**

Requires significant access and might pose privacy or security risks, but it is not particularly difficult, technically. In the future, it may be possible to use AIs to [assist with tasks like](#) this in a zero-knowledge manner.

MIRI TECHNICAL GOVERNANCE TEAM

| Allow list operation stack | An international authority could certify certain operations at the level of "Inference on the Llama 70B model on a DGX H100 server" and issue the specific GPU-level code for such use cases. AI chips would be restricted to only run code approved by this authority. This could be implemented with a [FlexHEG](#) mechanism. |
|---|---|

**Feasibility**

| Medium ▾ | 2–5 years |
|---|---|

---

| Disallow list operation stack | This could include blocking certain training operations such as backpropagation. While it is mathematically similar to other operations, it may nevertheless be possible to isolate and prohibit backpropagation given substantial control of the AI stack. This could be implemented with a [FlexHEG](#) mechanism. |
|---|---|

**Feasibility**

| Low ▾ | 2–15 years |
|---|---|

**Notes**

This may be impossible. If successful, it would likely involve replacing the AI software stack with approved libraries and restricting developers from writing code outside of these frameworks. This would be a major engineering endeavor.

---

| On-chip interconnect limits ("Fixed Set") | Chips are manufactured to only have high interconnect with a small fixed set (e.g., 128) of other chips, making them difficult to use for training large models. This could be implemented with a [FlexHEG](#) mechanism. |
|---|---|

| **Feasibility** | | **Previous work** (non-comprehensive) |
|---|---|---|
| Medium ▾ | 3–5 years | Aarne et al. (2024); Kulp et al. (2024); Discussed [below](#) |

**Notes**

Requires replacing hardware. "However, this level of interoperability could be at least 2 years away, based on an interview with an industry expert" (Aarne et al., 2024).

---

------------------------------------------------------------

| Networking equipment interconnect limits | Regardless of how they are manufactured, chips could be arranged in small pods with high intra-pod interconnect and very small inter-pod interconnect. Physical inspections combined with continuous monitoring (e.g., cameras) of data centers can confirm inter-pod interconnect. |

**Feasibility** | **Previous work** (non-comprehensive)
--- | ---
High ▾  1–2 years | Discussed [below](below)

------------------------------------------------------------

| Sensors for other properties of chips | Various [workload classification methods](workload-classification-methods) involve gathering information about chips, such as their power draw or network traffic. Many of these properties could be measured with specialized devices. |

**Feasibility** | **Previous work** (non-comprehensive)
--- | ---
High ▾  1–2 years | Baker et al. (Forthcoming)

**Notes**

Development time would depend on the sensors needed.

| Speculative differences between training and inference with high-level chip measures | Inter-chip interconnect limits (network bandwidth between servers) | Large-scale AI training involves linking thousands of chips with high-bandwidth interconnect. Efficient inference, meanwhile, can be done with dozens to low-hundreds of highly connected chips. Between different "pods" of well-connected chips, the communication requirements are very low for inference (tokens) compared to the activations or gradients that are transferred in other forms of parallelism—this gap is the target of interconnect bandwidth limits. |

| Feasibility | Previous work (non-comprehensive) |
| --- | --- |
| High ▾  <1 year | Discussed below; Kulp et al. (2024); Heim et al. (2024) |

**Notes**

The main distinction here is that inference does not use data parallelism: different batches of data being processed do not need to be closely linked together at all, whereas for training this is needed to share gradients. Advances in distributed training may make this approach ineffective. With well-designed limits and enhanced security, this could be difficult to spoof.

| | Synchronous changes in many GPUs, e.g., beginning of a training run, logging of intermediate checkpoints | The beginning of a training run may involve many chips doing nothing, which then suddenly begin processing inputs, showing a spike in energy use, network traffic, and memory use across many chips simultaneously. Similar patterns may occur at other points in training. |

**Feasibility**

| Low ▾  <1 year |
| --- |

**Notes**

This was observed in the Llama 3.1 405B training, resulting in "instant fluctuations of power consumption across the data center on the order of tens of megawatts" (Dubey et al., 2024). This is likely spoofable with small performance penalties, e.g., by slowly adding more servers to a workload or by using various distributed training techniques.

| Chip utilization rates, core utilization rates | Chip utilization, specifically Hardware FLOP Utilization (HFU), is a measure of the actual amount of computations performed by a chip relative to its theoretical maximum. Training and inference often differ along this metric. Current inference for LLMs involves two stages (Verma & Vaidya, 2023): "prefilling" (computing the KV cache on input tokens) and "decoding" (generating additional tokens)—these stages may have distinct effects on power consumption because decoding is more limited by memory bandwidth while prefilling is limited by computational performance (i.e., utilization and power draw are higher during prefilling). Therefore, inference could have distinct fluctuations in utilization. |
|---|---|

| Feasibility | Previous work (non-comprehensive) |
|---|---|
| Medium ▾  <1 year | Heim et al. (2024); Baker et al. (Forthcoming) |

**Notes**

Recently, it has become popular to report Model FLOP Utilization (MFU), a measure of the obtained throughput relative to theoretical max throughput (Chowdhery et al., 2022). MFU calculations are based on the number of tokens processed, so this requires more access than just high-level chip information. On the other hand, HFU is more easily measurable and thus more relevant to the present use case (but for other use cases, such as accurately estimating the model-FLOP size of a workload, this difference is very important). Classification on this basis is likely spoofable with medium cost penalties (e.g., a 20% cost increase via making training less efficient, which is in the budget of the relevant actors).

| Power draw associated with training and inference differences | Power draw may act as a proxy for chip utilization. |
|---|---|

| Feasibility | Previous work (non-comprehensive) |
|---|---|
| Medium ▾  <1 year | Heim et al. (2024); Patel et al. (2023) |

**Notes**

This has similar problems as chip utilization. It is unclear how closely power draw correlates with HFU (especially in an adversarial environment).

---

| Between-chip data flows could indicate training | Inference does not require propagating gradients backward in a model, so a cluster could be configured for pipeline parallelism and only permit data to flow one direction through chips. |
|---|---|

**Feasibility**

| Medium ▾ | <2 year |
|---|---|

**Notes**

For example, a data center operator declares that chips A are storing early model layers while chips B are storing later layers, and data flow is only permitted to flow from A to B but not backward. This likely requires designing custom networking equipment. The spoofability is unclear.

---

| Network bandwidth within servers | Generally, the communication pattern between chips in a server could classify the server's activity and forms of parallelism being used. |
|---|---|

**Feasibility**      **Previous work** (non-comprehensive)

| Medium ▾ | <1 year | Heim et al. (2024) |
|---|---|---|

**Notes**

Given that both training and inference benefit from within-pod/server parallelism, this may not be a useful differentiator.

---

| Quantity of accelerators used in a workload | Large AI training is distinct in its use of thousands of AI chips for many days, so it may be possible to differentiate workloads based on their size. However, this is unlikely to work in the international verification context because there is effectively one customer who can break a large workload into smaller, innocuous-seeming workloads. |
|---|---|

**Feasibility**      **Previous work** (non-comprehensive)

| Low ▾ | <1 year | Heim et al. (2024) |
|---|---|---|

**Notes**

In the context of cloud compute providers, there are numerous customers submitting workloads which are likely different. Given the threat model in this report, we should assume these may all be coordinated by a single actor. Workloads could be split up sequentially or in parallel.

---

| Numerical precision | Lower precision data formats may be more common in AI than non-AI workloads. However, precision is unlikely to differentiate AI training and inference reliably. |
|---|---|

| **Feasibility** | **Previous work** (non-comprehensive) |
|---|---|
| Low ▾ <1 year | Heim et al. (2024) |

**Notes**

It is sometimes thought that training requires higher precision than inference, however, this difference is unlikely to be reliable enough for the present use case. There are known cases of using low precision (e.g., fp8) for large LLM training (Inflection AI, 2023).

| Modification of weights in memory | Training requires repeatedly updating model weights, so it may be possible to monitor changes to values in memory to differentiate training and non-training workloads. |
|---|---|

| **Feasibility** | **Previous work** (non-comprehensive) |
|---|---|
| Medium ▾ <3 year | Heim et al. (2024) |

**Notes**

It is unclear how this would be implemented and how reliable it would be.

| Signatures of high-level chip measures | Signatures of high-level chip measures | High-level information about chip activities (e.g., power draw, numerical precision, network activity, etc.) is likely spoofable in the default setting where one country is trying to classify the workloads of another country's data center with limited access. However, this approach can be made more robust by using high-quality, labeled data from the monitored data center to improve this classification approach. The monitored data center operator would run their most common, internationally approved/permitted workloads (e.g., inference of their popular models) while an inspector has code access to confirm this is happening. The inspector takes readings for high-level chip measures (e.g., power draw, network activity). The inspector can then create a classifier based on this data (e.g., train an ML classifier), which can be applied to high-level chip measures in the future (code access is no longer needed if the high-level chip measures match that of approved activities). If the measurements deviate significantly from their expected patterns or if the common workloads change significantly, the inspector can review the code again. The key insight compared to naively trying to classify workloads is that temporary code access can provide high-quality labeled data from the production system, and the data center operator can be expected to keep its chip activities sufficiently close to these baselines, versus classifiers needing to deal with massive distribution shift in an adversarial environment (if code access is never provided, software stacks and model architectures may differ substantially, and even the AI chips themselves may be different from the classifier's training data). |
|---|---|---|

**Feasibility**                      **Previous work** (non-comprehensive)

High ▾   <2 year                See below

**Notes**

It is unclear how consistent these signatures will be over time, but previous results attempting to classify chip activities with high-level information have been effective (Copos & Peisert, 2020; Karimi et al., 2024; Köhler et al., 2021). This approach requires granting temporary code access (including whenever major changes are applied to chip activities), which could pose privacy and security risks. This temporary access could be acceptable if a neutral party or zero-knowledge approach is used (e.g., inspectors who live onsite and have limited communication with their home country). It could also be difficult for inspectors to be confident they are seeing the true code being run.

| | | |
|---|---|---|
| Proof-of-Work methods | Proof-of-Work methods for crypto mining | Cryptocurrency mining uses "Proof-of-Work" schemes, which provide confidence that some declared operations have taken place. For example, a standard implementation is for workers to hash variants of the same message until one of the resulting hashes matches some criteria (e.g., many leading zeros in the hash). It can be quickly verified that the message variant used produces such a hash. The acceptance criteria (e.g., number of leading zeros) can be varied to control the likelihood of an input message producing an acceptable hash and, thus, the amount of original work confirmed by each presentation of an acceptable hash. Applying Proof-of-Work could verify that compute declared to be doing crypto mining is indeed doing it. |

**Feasibility**

High ▾  <1 year

| | | |
|---|---|---|
| | Proof-of-Work methods for other common uses of AI compute | Ideally, Proof-of-Work-style approaches could also be used to validate many declared uses of AI compute. The methods discussed for partial re-running of chip activities include some versions of this for AI. |

**Feasibility**

Medium ▾  2–6 years

| Compute accounting | Compute accounting | Data center operators can demonstrate that their AI chips are not being used for a large training run by showing that they are being used for other things, as accomplished through [partial re-running of chip activities](#) and [other](#) [workload](#) [classification](#). By summing declared and verified compute use and comparing it to total potential compute use, they can show that there is not enough compute left over for a violation. One difficulty here is that the quantity of declared chip uses may not correspond to the actual chip use. For example, chips often do redundant operations during training ("recomputation") in order to save on memory, so a data center operator might claim to have done a training run with $10^{25}$ FLOP, but the chips actually did $1.5*10^{25}$ FLOP because they did substantial recomputing. Forthcoming work from Baker and colleagues makes progress on this. |

| **Feasibility** | | **Previous work** (non-comprehensive) |
|---|---|---|
| Medium ▾ | <1 year | Heim et al. (2024); Baker et al. (Forthcoming) |

# Evaluations Building Blocks

| Building Block | Mechanism | Details |
|---|---|---|
| **Self-reported evaluations** | Self-reported evaluations | Model developers can report evaluation results for their models without any verification. |

**Feasibility**

| High ▾ | <1 year |
|---|---|

**Notes**
Provides very little confidence on its own.

| Building Block | Mechanism | Details |
|---|---|---|
| Proof of same model for training, evaluation, and deployment | Verifying same computation in TEE | A TEE can hash key elements of the computation (e.g., the model weights and inference code), and then confirm that the hash at evaluation time matches the hash at training and deployment. The "model from training" could be verified by re-running parts of the training run and comparing the weight hash at the end of the verified run to the inference weights. |

**Feasibility**        **Previous work** (non-comprehensive)

| Medium ▾ | 1–4 years | Trask & Bejan (2024) |
|---|---|---|

**Notes**
This technology is not production-ready yet, but it is very close, and we know how to solve the problem in principle. It is unclear if TEEs with existing AI chips can be made sufficiently secure or if new chips are needed. This general approach could be done without TEEs, but it would be easier to spoof.

| Cryptographically secure audit trail | An audit trail demonstrating that deployments of a model are using the same model that was evaluated by tracking the development of the model in a cryptographically secure way. This is basically version control which is cryptographically signed to be legitimate. |
|---|---|

| Feasibility | | Previous work (non-comprehensive) |
|---|---|---|
| Medium ▾ | 1–3 years | Reuel et al. (2024, Section 5.4.1); Brundage et al. (2020); Millet (2024) |

**Notes**

Software audit trails are a standard tool. However, there may be some difficulties in applying them to AI deployment, such as including various efficiency optimizations in the audit trail. It is also unclear how effectively these could be made secure for the present use case.

- - - - - - - - - - - - - - - - - - - - - - - - - - - - - - - - - - - - - - - - - - - - - - - - - - - - -

| Redo evaluations in deployment | One way for an external auditor to gain confidence that the model they evaluated is the same model that is deployed via a public API is simply to re-run the evaluations during deployment and check if the results match those from testing (or are similar enough, given that inference often uses sampling). This mechanism is less relevant to the international verification context because it relies on public API access, which is somewhat unlikely for frontier models in the international context (these models seem much more likely to only be deployed internally, with publicly available capabilities lagging behind). It is also unclear how effective this would be against adversarial efforts from developers. |
|---|---|

| Feasibility | | Previous work (non-comprehensive) |
|---|---|---|
| Medium ▾ | <1 year | Gao et al. (2024) |

**Notes**

Likely not relevant to the international verification context. One potential application to the international verification context would be to allow an evaluator to query a model that is not generally available externally and treat these queries like other requests. For instance, if an external auditor had access to the accounts of an AI developer employee, they could potentially determine if the inference model being served changes over time by analyzing the model responses.

| Secure evaluations | TEEs for evaluation | On-chip and data center-level security mechanisms allow for secure external evaluations. The model remains with the developer, an inspector can be confident that the evaluations were properly run and not spoofed, and the developer cannot see what evaluations were run. |
|---|---|---|

**Feasibility**                          **Previous work** (non-comprehensive)

| Medium ▾  1–4 years | Aarne et al. (2024); Kulp et al. (2024); Trask et al. (2024) |
|---|---|

**Notes**

H100s already have confidential computing for single-chip setups, and it may not be too difficult to get this for multi-chip setups (Nertney, 2023). The adversarial robustness of current implementations is unclear, so new AI chips could be needed.

| | Mutually trusted data center | It may be necessary to run evaluations in a trusted data center where model details cannot leak, evaluations cannot leak, nor could a model developer spoof evaluation results. This level of security might come from a combination of improved chip security (it is likely difficult to gain mutual confidence in existing chips) and other security measures, such as reducing the amount of information that can leave the data center (Greenblatt, 2024) and using onsite staff who have limited external contact. While stronger chip security could take substantial R&D effort, lower-tech solutions that require more political will could be viable in the short-term. |
|---|---|---|

**Feasibility**                          **Previous work** (non-comprehensive)

| Medium ▾  1–6 years | Shavit (2023); Baker et al. (Forthcoming); Nevo et al. (2024); Heim (2024b) |
|---|---|

**Notes**

Ensuring this data center is sufficiently secure may be difficult.

| | Inspectors | In the absence of TEE-based security around external evaluations, but as an improvement on self-report evaluations, international inspectors could observe the evaluations being done alongside the main researchers. |
|---|---|---|

**Feasibility**

High ▾ <1 year

**Notes**
It may be difficult to balance ensuring the inspectors have sufficient access (are sure the right models are being evaluated) with mitigating security risks.

| | Evaluations Bot | Instead of human inspectors who may pose security risks, it may be possible for countries to jointly design an AI system to run zero-knowledge evaluations on each other's models. Such a bot could e.g., carry out dynamic red-teaming in a mutually trusted data center. |
|---|---|---|

**Feasibility**

Medium ▾ 1–6 years

**Notes**
Given the design process for the Evaluations Bot, both countries could have faith that the model is not backdoored to leak sensitive information, and that the model is sufficiently competent to complete the task.

| Effective evaluations | Better science of evaluations | Model evaluations are widely believed to be somewhat insufficient for assessing model capabilities and very insufficient for assessing model propensities for advanced AI systems. Substantial progress in the science of evaluations is needed to remedy these shortcomings and make evaluations effective for assessing risk. |
|---|---|---|

**Feasibility**          **Previous work** (non-comprehensive)

Medium ▾ 1–5 years      Bengio et al. (2024); Mukobi (2024); Barnett & Thiergart (2024)

| Proof of sufficient capability elicitation | Show that capability evaluations would be reliable even if model developers or models themselves were engaged in sandbagging (purposefully underperforming during evaluation). Proof of training/inference may be needed. |
| --- | --- |

| **Feasibility** | | **Previous work** (non-comprehensive) |
| --- | --- | --- |
| Medium ▾ | 1–5 years | Greenblatt, Roger, et al. (2024); van der Weij et al. (2024) |

**Notes**

This may be impossible for sufficiently advanced AI systems.

| Dynamic evaluations | As argued in previous work (Bucknall & Trager, 2023; Casper et al., 2024), effective evaluations may require that evaluators can iterate on model prompting strategies, fine-tune models, access model families, and more. Evaluating certain capabilities may also involve a model repeatedly interacting with the real world, e.g., running code on computers, as in METR (2024). These forms of access may be more difficult to facilitate in a secure way than merely prompting a model with a standard set of queries. If more access is needed, the difficulty of designing secure environments (e.g., TEEs) for this evaluation could rise substantially. |
| --- | --- |

**Feasibility**

| Medium ▾ | 1–4 years |
| --- | --- |

| General intelligence gathering to augment evaluations | External deployment monitoring | It may be relatively easy to infer the approximate capability level of AI systems that are deployed in the world, e.g., by looking at their economic impact, the state of automation, or interacting with them to test their capability level. We might expect powerful AI systems to be deployed widely because they could bring massive economic benefits. |

### Feasibility

Medium ▾  1–3 years

### Notes

There is likely room for new work, similar to Observational Scaling Laws (Ruan et al., 2024), that allows for the inference of general model capabilities based on particular deployment information. Work aimed at inferring the properties of proprietary models (Carlini et al., 2024) may also be relevant.

| | Internal deployment monitoring via cyber espionage | Cyber offensive tools may be applied to better understand the state of AI capabilities in a secret AI project. |

### Feasibility

Medium ▾  1–3 years

### Notes

Monitoring the capability level of internal AI systems, which are only deployed internally and are kept secret, is more difficult than monitoring externally deployed models.

| Internal deployment monitoring via HUMINT | Human-based intelligence can help identify the capability level of secret AI projects. This could include spies, unintentional leaks, whistleblowers, and interviews conducted by international authorities. |

**Feasibility**

High ▾ <1 year

**Notes**

Monitoring the capability level of internal AI systems, which are only deployed internally and are kept secret, is more difficult than monitoring externally deployed models.

## Other Building Blocks

| Building Block | Mechanism | Details |
|---|---|---|
| Incident reporting | Incident reporting | Worrying AI incidents can be anonymously reported to a central body that helps gauge risk. |

**Feasibility**

High ▾  <1 year

| Whistleblower programs | Whistleblower programs | People who whistleblow about violations of international agreements are protected, e.g., with asylum guarantees. Whistleblowing is encouraged by access to reporting channels that are robust to intense state surveillance. |

**Feasibility**  **Previous work** (non-comprehensive)

High ▾  <1 year  Baker et al. (Forthcoming)

| Interviews | Interviews | Countries agree to allow international inspectors to interview personnel working on AI development. These interviews focus on verifying treaty compliance. These interviews should occur in neutral locations with privacy and safety guarantees. |

**Feasibility**  **Previous work** (non-comprehensive)

High ▾  <1 year  Wasil, Berglund, et al. (2024)

| | | |
|---|---|---|
| AI-enabled verification mechanisms | Logs of relevant AI activities | Any AI systems used in chip development (e.g., robots in factories) or AI algorithm development have logs of their activities and observations stored and accessible by a trusted verifier. These logs could be used to confirm that these AI systems are not being used to violate international agreements. |

**Feasibility**

Medium ▾  1–5 years

**Notes**

Minimal versions of this could be developed quickly, but ensuring they are robust and apply to all inference would require more time and potentially new hardware.

| | | |
|---|---|---|
| | AI-powered zero-knowledge work | AI systems could be used to do a substantial amount of work with sensitive data and then output only a non-sensitive final result. For instance, an "Evaluations Bot" would be one such system. This could dramatically expand the set of cognitive tasks that could be accomplished in a zero-knowledge way, aiding with many aspects of verification. |

**Feasibility**

Medium ▾  1–4 years

**Notes**

The model capabilities will inform what sort of work is possible here, and sufficiently capable models for some task may not be trustworthy. Additionally, models will need improved robustness to jailbreaks in order to be effective.

| AI-powered data processing | Trusted AI models can be used to classify large amounts of data (e.g., video surveillance of data centers) to determine whether it contains violations. |
|---|---|

**Feasibility**

High ▾ <1 year

**Notes**

This is a subset of "AI-powered zero-knowledge work", and it is additionally worth noting because it increases the amount of labor available (i.e., this is useful for privacy-preservation reasons and for total labor supply reasons).

| AI-powered supply chain tracking | Crucial parts of the chip supply chain can be monitored with drones in addition to standard security measures. |
|---|---|

**Feasibility**

High ▾ 1–3 years

**Notes**

For example, EUV machines and fabricated chips can be tracked in transit by drones from multiple countries to ensure they reach their intended destination.

| Model behavior specification agreements (specifically, the model will not help violate international treaties) | Countries agree to use behavior specifications that are good for other countries. E.g., "The model will engage in $X$ benefit sharing" or "The model will not execute military first-strikes". Verification that a given behavior specification is used can be done with partial re-running of training. Hypothetically, external evaluations could be used to gain confidence that behavior specifications were used, but current evaluations are unable to assess propensity (as is needed) and are unlikely to work in this adversarial setting. Even without being robust, this approach could be cheap to implement and thus worth pursuing. |
|---|---|

| Feasibility | Previous work (non-comprehensive) |
|---|---|
| Medium ▾<br>1–100? years | Discussed below; OpenAI (2024b) |

**Notes**

The tractability of this approach likely varies significantly with AI capability level. Current approaches for frontier AI systems are insufficient (as evidenced by models' susceptibility to jailbreaks, which could be used by a country when carrying out violations). Robustness to jailbreaks for current models is likely viable within a few years, based on our familiarity with the literature. Proper value alignment for advanced AI systems may be incredibly difficult (Yudkowsky, 2022), and it is difficult to upper bound the time needed.

| Knowledge of all inference instances of a model | Weights do not leave their training data center | If data centers are [registered](#) and have very good security, it may be possible to quarantine model weights to the data center they were trained on (or a small number of other data centers). If successful, this would allow a monitoring party to know where all inference is happening. While strong security around model weights is generally desired, sufficient effort is needed to make it happen. Additionally, this approach requires that security be sufficient to prevent insider threats from the country operating the data center, e.g., international collaboration on security. |
|---|---|---|

| Feasibility | | Previous work (non-comprehensive) |
|---|---|---|
| Medium ▾ | 1–5 years | Nevo et al. (2024); Greenblatt (2024); Harack et al. (Forthcoming) |

**Notes**
Requires very good security and likely continuous monitoring of security but is otherwise straightforward.

| | Model weights only decryptable by certain chips | If chips are partially hardened, at least to the point of having private keys and TEEs, model weights can be encrypted with the public key of the chips they will later run on, allowing only those chips to decrypt the model weights and run the model. This could potentially be implemented with [FlexHEG](#) designs, with the secure processor providing encryption and decryption while the secure enclosure makes it difficult to extract model weights or private keys. |
|---|---|---|

**Feasibility**

| Medium ▾ | 2–4 years |
|---|---|

**Notes**
This is unlikely to be robust to physical attacks on chips (i.e., extraction of chip private key) unless substantial hardening measures are taken. If only pre-specified chips can run inference for a model, it is much easier to locate those chips and thus know where all inference is occurring.

--------------------------------------------------------------------------------

| Bottlenecked, distributed inference | One party or data center has some of the model weights and another has the other model weights. Both parties are required for inference to take place. A country could have one centralized and internationally monitored data center that performs a small but crucial part of inference computation (such as one layer of a model), ensuring international regulations are followed, but a majority of compute would not be directly monitored. This approach is useful because it reduces the amount of compute that needs to be monitored (i.e., cheaper and easier), and unmonitored compute could be located in secret data centers. |

| Feasibility | Previous work (non-comprehensive) |
| --- | --- |
| Medium ▾   1–5 years | Borzunov et al. (2023) |

**Notes**

If carrying out a forward pass on a model requires multiple parties, each of those parties is a node for knowing inference is taking place. There are a few requirements: The centralized data center must be secure enough (according to not just the host country, e.g., including against insider threats) to protect the centralized computation (Nevo et al., 2024). The centralized computation must also be necessary for inference and very difficult to reverse engineer (Carlini et al., 2024) or replace (e.g., by training a module to match its behavior). Additionally, monitoring may be more difficult because the centralized data center primarily receives intermediate activations from a model rather than inputs and outputs; this could be addressed by having decentralized clusters send all inputs and outputs (tokens) to the centralized data center and the centralized data center re-running a small fraction of the full model inference to ensure the submitted inputs, activations, and outputs match.

--------------------------------------------------------------------------------

| | Widely applied chip restrictions | If nearly all AI chips in the world are under a governance authority with tools like FlexHEGs, various mechanisms for chip limiting or monitoring could be applied, as discussed previously. For instance, chips could be restricted to only running inference or training on certain models which have been approved by an international authority. If a substantial fraction of all AI chips were covered by such controls, this could increase confidence that inference was not being done on an unapproved model. |
|---|---|---|

**Feasibility**

| Medium ▾ | 2–6 years |
|---|---|

**Notes**

A key uncertainty is defining which chips these restrictions would need to apply to, and a key difficulty would be implementing sufficiently secure restrictions on those chips. Inference typically has lower compute requirements than training; if these are still high (e.g., >64 cutting-edge AI chips), this may be feasible but would be very difficult. If the model of interest could be run on consumer hardware, it is likely infeasible to bring all such hardware under a governance authority. It may also matter both whether the model can be run, at all, on some hardware, and whether it can be run efficiently (e.g., if consumer hardware can run a model, but does so much slower than data center AI chips, restrictions on data center AI chips could be effective).

| Inference-time oversight | Inference classifiers | Lightweight classifiers run locally on inference workloads, enabling verifiers to check that deployment follows the rules while limiting access. |
|---|---|---|

**Feasibility**        **Previous work** (non-comprehensive)

| Medium ▾ | 1–2 years | Discussed below |
|---|---|---|

MIRI TECHNICAL GOVERNANCE TEAM

| | | |
|---|---|---|
| | Verification that inference classifiers are running | Inference software stacks could be designed to ensure classifiers are running correctly. This could use one of the approaches discussed for [limiting or monitoring chip activities](#) (and chip restrictions being tied to the presence of specific oversight). Chips likely need to be made more secure. [Partial re-running](#) of chip activities could also be applied here, with a verifier checking that the classifier results from declared inference are correct. |

**Feasibility**

Medium ▾  1–4 years

**Notes**

E.g., a classifier which is approved by an international body runs on inputs and outputs, a CPU-based TEE attests to this fact (with tamper-evidence to avoid private-key exfiltration), and AI chips are restricted to only perform computations when such a signature is present.

| | | |
|---|---|---|
| Provably Safe AI | Provably Safe AI | This is a family of research agendas. Formal verification of AI agents and AIs that oversee them such that deployment of the agent will provably fall within some safety specifications designed with human input. |

**Feasibility**                    **Previous work** (non-comprehensive)

Medium ▾  3–15 years    Dalrymple et al. (2024); Dalrymple (2024)

| Reduce model capabilities in relevant domains | Train without dangerous information | Countries agree on particularly dangerous knowledge they do not want models to know; information about these topics is filtered out of training data. This would be moderately easy to do but harder to verify: verification that countries are following this commitment could include partial re-running of chip activities. |
|---|---|---|

**Feasibility**

| Medium ▾ | <1 year |
|---|---|

**Notes**

May be intractable for advanced AIs due to them learning knowledge without explicitly training on it. Requires defining dangerous knowledge before training and thus runs some risk of proliferating that knowledge.

| | Knowledge unlearning | After training, apply knowledge unlearning techniques to reduce the model's capabilities in the dangerous domain. Verification that countries are following this commitment could include partial re-running of chip activities. |
|---|---|---|

| **Feasibility** | **Previous work** (non-comprehensive) |
|---|---|
| Medium ▾  1–5 years | Casper (2023); Tamirisa et al. (2024) |

**Notes**

Current techniques are insufficient. May be intractable for advanced AIs due to knowledge collisions. This unlearning may need to be robust to fine-tuning, or not, depending on the risk there.

| Non-AI monitoring | General intelligence gathering (military) | Countries already attempt to gain insight about each other's military capabilities via numerous intelligence gathering methods. |
|---|---|---|

**Feasibility**

| High ▾ | <1 year |
|---|---|

| Physical inspections (military) | Inspectors monitor military production facilities to ensure particularly powerful weapons (e.g., novel WMDs) are not being developed. Ideally, this happens in a privacy-preserving manner. |

**Feasibility**

High ▾ · 2–4 years

**Notes**

Political will and privacy costs are likely a large implementation barrier. Given the uncertainty about what novel weapons could be developed, it also may be difficult to reliably monitor for them.

| General intelligence gathering (economic) | Publicly available discussions about AI integration, scientific studies, general economic measures, and private financial data are all likely to indicate when a country is getting substantial economic returns from its AI systems and automation. |

**Feasibility**

High ▾ · 1–2 years

| Monitoring of automation and economic capacities | Countries could agree to monitor the production and distribution of robots (and their components), as a way of providing early warning for many types of automation. This could involve access to otherwise-private high-level financial data accompanied by occasional physical inspections. |

**Feasibility**

High ▾ · 1–2 years

**Notes**

Because the focus is on economic applications and capabilities, deployment is likely to be widespread, making monitoring easier (as compared to monitoring secret robotics programs).

| | | |
|---|---|---|
| No data center colocation | No colocation with non-AI compute | Data centers built only for AI workloads allow numerous crucial security, verification, and enforcement mechanisms to be applied without collaterally affecting non-AI compute. For example, verification approaches based on data-center-wide power draw would be ineffective if there were power-hungry non-AI workloads happening in the same data center. |
| | | **Feasibility** |
| | | High ▾  1–2 years |
| | No colocation with sensitive military facilities | Verification mechanisms that make use of physical inspections, or compute tracking in general, may be much more difficult if AI data centers are part of military installations, as there is an increased risk of sensitive information leaking. |
| | | **Feasibility** |
| | | High ▾  1–2 years |
| Cryptography concepts | Public-key cryptography | A pair of keys, known as a private key and public key. A message can be encrypted by a reader's public key, allowing only the reader (with their own private key) to decrypt and read the message. |
| | Cryptographic signature | Using public-key cryptography, a party can sign a message with their private key, allowing readers (anybody with the corresponding public key) to know the signer was the true author of the message. |
| | | **Notes** |
| | | This is commonly used to verify that the author of a software update was the true company: the company releases their public key to the public and then signs software updates with their private key. |
| | Hashing | A message can be encoded into a short representation (a "hash") with a known algorithm. Only the same original message would lead to that hash, and the original message cannot be reverse-engineered from the hash. |

MIRI TECHNICAL GOVERNANCE TEAM

| | |
|---|---|
| Secure Boot | When powering on, a computer only runs software that is cryptographically signed by trusted authorities (e.g., the device manufacturer). Allows a computer to boot into a trusted environment. |
| Zero-knowledge proof | A zero-knowledge proof is a protocol where a prover demonstrates to a verifier that some claim is true without revealing sensitive information about the claim to the verifier besides its truth value. This report uses the term "zero-knowledge" to refer to protocols with this property, even though they may not be strictly zero-knowledge proofs. |
| Cryptographic Escrow | A party to declare and encrypt a set of information and then selectively show/decrypt subsets of that information to a verifying party later. In doing so, the proving party must commit ("Commitment Scheme," 2024) to their full story at the beginning but only needs to reveal the story incrementally, e.g., as diplomatic talks progress. |

**Previous work** (non-comprehensive)

Philippe et al. (2019)

**Notes**

This approach is conceptually similar to a prover privately implementing a blockchain ("Blockchain," 2024) that sequentially includes all the data they may later want to reveal and then being able to selectively reveal the true value of any particular block.

# Elaboration on Particular Mechanisms

This section elaborates on a few of the mechanisms described above that benefit from more clarity. These are largely derivative of existing ideas (we do not claim originality), but placing them in the context of international treaties may provide value to some readers.

## Inference Classifiers

### Summary

Inputs and outputs during inference can be seen by a trusted classification model which asserts that they do not contain any prohibited content.

### Longer Description

This is a similar approach to ChatGPT and Claude having secondary filters for prohibited content other than the LLM itself refusing; it is also similar to content moderation generally. It can likely be done in a privacy-preserving way where country $X$ develops a classifier that country $Y$ must run, and country $Y$ can verify that the classifier can only transmit back a small amount of information (e.g., "permitted" or "not permitted") to country $X$, to avoid leaking sensitive data. The classifier itself, and its judgments, should be obfuscated from country $Y$ to make it more difficult for country $Y$ to nefariously get around the classifier's judgments. Inputs/outputs classified as "not permitted" can undergo additional inspection, e.g., by an air-gapped multi-national committee.

### Feasibility

Medium; 1-2 years. We have very similar content moderation systems already in deployment for numerous applications and hundreds to thousands of experts. Making the classifier opaque to country $Y$ seems like a standard obfuscation problem like that of obfuscating proprietary code or proprietary ML models—it would be aided by having Confidential Computing for AI chips.

One major hurdle is ensuring that all relevant inputs and outputs are passed through the classifier. One way to achieve this is to both: prevent weights from leaving the data center, and have close monitoring of all chips in the data center. The second of these might involve FlexHEG mechanisms such that a chip is only allowed to operate if it has received cluster-level authorization for a particular batch of inputs that has been classified.

Another hurdle is having clear definitions of permitted content and turning these into classifiers: a rule like "no inputs/outputs about cybersecurity" may be easier to enforce than "no inputs/outputs used for offensive cyber operations", as the second requires more sophisticated classifiers.

A classifier like this may be able to tolerate low sensitivity (a high false negative rate). This is because effective violations will likely require many queries to models, but finding even one such query would be a substantial violation of the treaty with significant consequences.

It may be difficult (for political and security reasons) to develop a method for verifying the queries classified as violations in a way that does not leak sensitive information. Checking these queries in some way (e.g., human inspectors) is a critical part of this approach being effective because the classifier will likely not be perfectly reliable.

## Implementation Cost

Such a classifier need not be very compute intensive, especially relative to the models being run. This could likely be done well for <10% of the model inference cost. Classifiers like these are useful for model developers as well, e.g., as part of an AI Control scheme (Clymer et al., 2024; Greenblatt, Shlegeris, et al., 2024), or to monitor for misuse, so there may be efforts to make these classifiers cheap and effective, independent of their use for international verification.

# Shared Model Behavior Specifications: AIs Do Not Violate International Agreements

## Summary

Specifications for safe AI behavior should include AIs abiding by international agreements, and this should be required for development, similar to other safety-relevant behavior specifications.

## Longer Description

An international verification regime could take place after there are AI systems capable of contributing to novel research and engineering. In more extreme cases, AI systems may be doing a substantial fraction of intellectual labor. If this is the case, there is considerable worry that AI systems would be used to help violate international agreements, e.g., doing distributed training research to get around interconnect limits. One potential approach to this problem is to decrease the relevant AI systems' propensity to violate international agreements. This motivates the present approach of using shared model behavior specifications to prevent AI systems from violating international agreements.

Current AIs are trained with the developer's intention of AIs following some behavior specification; see, for example, OpenAI's "Model Spec" (OpenAI, 2024b). There is ongoing research aimed at ensuring models actually follow these specifications. If that research is successful, it may be possible to incorporate specifications such as "abide by international agreements about AI development" as part of AI model behavior goals, along with other likely goals such as "abide by local laws" and "uphold the UN Declaration of Human Rights".

Ensuring that models always follow these specifications and are resistant to jailbreaking is an unsolved problem for current models. Getting future AI systems to follow behavior specifications may be far harder, akin to solving the AI intent alignment problem (Christiano, 2018). Based on our familiarity with the field, we think that jailbreak resistance could be achieved for *current* AI systems in a few years if there is a substantial research effort (an unrealistic expectation, given that increasing model capabilities will change where research is directed).

For sufficiently advanced models (e.g., AGI), getting any behavior specifications correctly into their goal systems may be incredibly hard and infeasible on the relevant timescales.

Furthermore, our threat model involves models being trained and deployed by adversaries who may wish to bypass a particular behavior specification. Thus, there is a worry that developers might attempt to subvert this mechanism. They could do this via purposefully backdooring a model in training to, e.g., ignore its behavior rules when a certain password (Greenblatt, Roger, et al., 2024) is used; by fine-tuning (Volkov, 2024) a model to remove behavior guardrails; or by using any number of other techniques to bypass behavior specifications. This could be far more difficult than preventing jailbreaking in the black-box setting. While there is some work on reducing open-weight models' risk of being fine-tuned for malicious use, e.g., Tamirisa et al. (2024), Deng et al. (2024), Henderson et al. (2023), this problem is unsolved. We further note that our threat model includes adversaries being willing to spend substantial resources to bypass a verification regime and, thus, potentially to bypass behavior specifications. Work on open-weight fine-tuning prevention is mostly aimed at attackers with <1% of the compute resources as original training, whereas the international verification context requires we consider attackers who may be willing to spend more than 100% of original training costs if they can do so covertly (the covert constraint means that 1% may still be a relevant target, depending on the verification regime).

It is unclear how such a behavior specification would be translated into AI training processes, as this would be context-dependent. For instance, it could be instituted via specific training data being included or in the "constitution" used by AIs providing feedback to a model being trained (Bai et al., 2022). These details are critical to how another party would verify that the behavior specification is correctly being applied. For many of those implementation details, the verification approach is likely to re-run portions of training to confirm a training transcript and then inspect that transcript to ensure it uses the behavior specification correctly (and does not include obvious backdooring). This would be accompanied by attestations that the deployed model (in monitored deployments) matches that from the declared training run, in order to avoid fine-tuning attacks. Inference classifiers may also be used to reduce the likelihood of jailbreaking. While, hypothetically, evaluations (e.g., adversarial prompting to encourage a model to break the agreement) could potentially help verify that trained models are following the desired specification, current evaluations are unable to assess model propensity and would thus be insufficient. This is especially true given how porous model behavior specifications can be, e.g., getting models to follow the specification precisely and interpret ambiguous cases correctly.

We note that the goal "follow international agreements about AI" may be more difficult to train into models than many other behavior specifications because these international agreements are subject to change over time. A goal such as "do not make biological weapons" has no need to change over time, but the specific rules that are imposed by international agreements might change, so there is a need for this goal to be represented in a flexible but secure (jailbreak-resistant) manner in AI systems.

### Feasibility

Medium (but maybe very difficult); 1-100? years. As discussed, there are numerous potential issues which could render this infeasible. However, it may also parallel the feasibility of getting other behavior goal specifications (OpenAI, 2024b) into AI systems, such as getting an AI system to follow applicable laws.

One issue is that countries may not agree to this specification as there could be negative side-effects (or the fear of them) from a model not prioritizing national interests. Furthermore, confirming that a behavior specification was used (and not backdoored) likely requires re-running portions of training or similar methods which involve substantial access.

### Implementation Cost

This mechanism does not add inference overhead. There is likely minimal training implementation cost, assuming a behavior specification is used already. This could add substantial complexity to default training behavior objectives due to the goal change problem discussed, but it is unclear.

## Networking Equipment Interconnect Limits

### Summary

AI chips can be physically arranged to have high bandwidth interconnect with only a small number of other chips (a "pod") and very low bandwidth interconnect to chips outside this pod, thus preventing their use for large training runs or (for sufficiently small pods) large model inference. This can be physically verified without access to details about what the chips are doing (e.g., with security cameras and special networking equipment). While this approach does not make large training impossible, it can make the time cost sufficiently burdensome to act as a strong deterrent.

This approach draws inspiration from the approach of chips being manufactured in a "fixed set", discussed in Kulp et al. (2024). It is distinct because it can be retrofitted to chips already in use via networking equipment, but it requires physical access to data centers for verification.

## Longer Description

AI chips could be networked together so that they are ineffective for large training. In particular, it is currently much easier to train large models if a developer is parallelizing across thousands of chips with high interconnect bandwidth between them. Therefore, a data center could assert that it is not doing a large training run because the physical connection of chips would make this highly inefficient. There are numerous ways to implement interconnect bandwidth limits; the main one would likely be using special networking equipment with a physical switch that only allows either very low bandwidth (e.g., 1 Mb/s) or standard high bandwidth (e.g., 400 Gb/s) communication with a pod, in a way that is clearly physically verifiable. The low bandwidth mode would correspond to a pod very likely conducting inference, where the key observation is that input and output tokens are a smaller communication payload than those used in various forms of parallelism (e.g., gradients for all weights in the case of data parallelism). This equipment's presence and switch state can be verified through occasional physical inspections and security cameras. This is a convenient verification mechanism because the regulator does not need to access the most sensitive information: the algorithms, weights, and data being used. We now discuss some of the details:

This mechanism requires physical verification, e.g., regular random physical inspections and security cameras in data centers to detect any modifications to interconnect, and monitoring of the supply chain for interconnect equipment (to ensure the approved low-interconnect cables have not been tampered with to make them too performant).

There is an important distinction between training and inference for interconnect limits: training is typically parallelized across thousands of chips, so restrictions on the order of a hundred chips may be effective. On the other hand, efficient inference may be possible with a smaller number of chips. Inference can be carried out with roughly as many chips as it takes to fit the model into DRAM, which is currently around ten chips for some of the best models (Dubey et al., 2024). However, there are significant efficiency gains from using tensor parallelism and larger pods for inference, perhaps around 128 chips.

Interconnect limits should be at least high enough to allow tokens to quickly pass in and out of a pod for inference, but they should be lower than would allow this pod to engage in parallel training with other pods. Additionally, it would be a major hassle if it took a very long time to load models (or other data) into GPU memory. Ideally, these limits should be tailored individually to input and output communication from a pod, based on the claimed activities of the pod (e.g., very low output bandwidth for chips that are doing long-generation inference, as this is a slow process that can work at low output bandwidth). The hassle of interconnect limits significantly affecting the time it takes to load a model into memory could be avoided by having a physical switch such that a pod is permitted to use high bandwidth interconnect every so often (e.g., once every 48 hours).

We note that interconnect limits could also be applied to limit inference on especially large models. It may be very slow to run inference for large models without a high bandwidth connection between many chips; for instance, the Llama 3 405B (Dubey et al., 2024) model cannot perform inference at BF16 precision on a server with eight H100 GPUs. Therefore, a data center could claim that it is not doing large model inference on the basis that it does not have the physical interconnect to do this efficiently—e.g., it has pods of just four chips with limited interconnect. Presently, we focus on interconnect limits to restrict training.

The key bounds on the specific inter-pod limits should be based on the requirements for efficient inference and efficient training. For instance, pod sizes should be large enough to enable efficient inference (e.g., 128 chips). Meanwhile, inter-pod interconnect limits should be set so that input and output tokens can be quickly shared for inference, but inter-pod communication is too bandwidth-limited for training. The primary goal of interconnect limits is to make large training prohibitively slow, e.g., take 100x longer than if the same GPUs were networked with high bandwidth interconnect. If the between-pod communication is of full-model gradients (in the case of data parallelism), then these interconnect limits are taking advantage of the difference between the size of the model weights (which is approximately equal to the size of gradients) needed for training and the size of input and output tokens needed for inference. If the between-pod communication is of activations (in the case of pipeline parallelism), then these interconnect limits are taking advantage of each token being represented by a high dimensional vector between layers, i.e., the activations for each token at intermediate layers are larger than the representation of that token's index in the vocabulary. We discuss pipeline parallelism below.

Inter-GPU interconnect can often exceed 400 Gb/s within a data center. In the present case, we would like to enable small pods of chips to communicate quickly (e.g., >400 Gb/s), but have extremely low communication to other chips (e.g., 1 Mb/s) in order to be prohibitively slow for training, even given advances in distributed training efficiency and subversion efforts. These limits would likely be stricter than merely requiring the use of standard ethernet cables (which are typically at least 1 Gb/s).

Because relatively low bandwidth amounts, e.g., 500 Mb/s, a similar speed to 5G phones ("5G," 2024), could enable a pod to be used in a distributed training run (see below), additional measures are needed to restrict covert pod communication, other than blocking the main communication channels. For instance, security measures may include enclosing each pod in a Faraday cage in order to block radio signals from a device covertly placed in the pod.

This interconnect bandwidth approach (at least as oriented around preventing data parallelism between pods) is somewhat analogous to Greenblatt's (2024) idea to use upload limits to prevent model exfiltration. The approach in that work is to leverage this difference between the size of model parameters and inference tokens for improved security by limiting the amount of information that leaves a data center (such that stealing a model would take prohibitively long). The thresholds of worry are different: in the present case, if the inter-pod bandwidth limits mean it takes 10 hours to transfer gradients, that may be sufficient to dissuade distributed training, but if the model weights could be stolen in 10 hours, that would clearly not be a sufficient security measure. The upload limits approach is also more likely to apply at a higher level (e.g., to an entire data center), whereas the present approach will apply to pods of a small number of chips within a data center. Both approaches may face difficulty if it is easy to compress model weights or gradients significantly. Preliminary work from Brown & Mazieka (2024) finds that, in the upload security limit context, compressing to ~1 bit/parameter is likely feasible (and potentially lower), but in their threat model, attackers can take advantage of retraining the model after compressed-weight exfiltration. Retraining is less of a concern in the present case because such retraining would add considerable time cost to training (the goal of using distributed workers was to accomplish more training in parallel, and individual workers needing to retrain adds serial time).

One potential benefit of this interconnect-restriction approach is that it's flexible: physically verifiable interconnect limits can be changed relatively easily (rather than manufactured-in limits), so it may not be a major issue that the pod-size limits for training and inference will be very different. Flexibility is also convenient if there are changes to the safety landscape and it becomes possible to safely conduct large training runs; it is easy to do this by switching networking equipment. The flexibility also gives treaty signatories an easy out if they choose to leave the treaty and its requirements.

However, the flexible nature of this mechanism is also a limitation: it would be very easy to break such a treaty agreement if, for instance, there is a domestic change in priorities (e.g., the U.S. pulling out of the JCPOA). If we need mechanisms that are likely to last decades, interconnect bandwidth restrictions would need to be more permanent than just being based on networking equipment. But this issue is not specific to interconnect limits: many of the mechanisms for verification would face major hurdles if they needed to be robust to years of overt subversion, e.g., on-chip mechanisms would also struggle with the threat of adversaries building their own unregulated AI chips on such timescales.

### Example Limits Calculation

Here we suggest tentative inter-pod bandwidth limits, but these suggestions should be seen as illustrative and hypothetical, rather than final—properly setting limits would require extensive engagement with AI developers to understand their constraints and state of the art methods, as well as potential future developments. We base these calculations on preventing *data parallel training*, which is only one method of making such an estimate; a thorough limit needs to involve other forms of parallelism.

As discussed, there are two key implementation details, the pod size and the interconnect limits; we provide example calculations for how to set these limits, based largely on the Llama 3.1 405B model (Dubey et al., 2024).

The pod size should be large enough for efficient inference but not too much larger (as that would run the risk of large training runs). The majority of the pre-training for the Llama 3.1 405B model uses pods (total GPUs / data parallel count) of 128 H100s. We choose this as our pod size.

Interconnect limits take advantage of lower inference than training bandwidth requirements, so in this example we want to make assumptions which lower-bound training bandwidth and upper-bound inference bandwidth, to see if this gap continues to exist under realistic, but pessimistic, conditions.

*Training Communication Requirements*

In data parallel training, the total communication requirement (send and receive) per gradient step per worker is at least 2*size_of_gradients. Conceptually, this reflects that each worker must send the set of gradients from their data and receive the average set of gradients from the other data parallel workers' data. This can reach 2*size_of_gradients in the case of a Parameter Server whose job is to receive gradients from all workers, reduce them, and broadcast the average gradients; the Parameter Server requires substantial bandwidth, but each worker only needs to both send and receive size_of_gradients. In practice, other methods for doing this all_reduce operation are used, such as ring all_reduce, which require more total bandwidth per worker but are overall more efficient. Here, we use the Parameter Server baseline where for each gradient step, each data parallel worker must send size_of_gradients data and receive size_of_gradients data.

The Llama 3.1 405B model has 405 billion parameters (corresponding to the same number of gradients, for this model). Ignoring issues with this specific model's training, which will likely be solved in the future, we assume gradients are stored in BF16, 16 bits. Per training batch, the amount of information transferred for each pod would then be (405B params * 16 bits/param * 0.125 bytes/bit) = 812 GB per batch (i.e., approximately the storage size of the model) in each direction.

Time to process a training batch: (131072 tokens * 6 FLOP/token-parameter * 405000000000 parameters / (400e12 FLOP/s/GPU achieved * 128 GPUs)) = 6.2 seconds per batch. We can further check this answer against the ~60 days that training was reported to take (different configurations were used, making these estimates uncertain). Given that there is a global batch size of 16M, training on a total of 15.6T tokens would take ((15.6T / 16M) * (6.2 seconds per batch ) / (60 * 60 * 24 seconds per day)) = 70 days, close enough for our purposes.

So our total required communication for (synchronous data parallel training) is approximately: (812 GB / 6.2 seconds) = 131 GB/s each direction.

*Token Communication Requirements Prelude*

The vocabulary size of the model is 128,000, and each token can thus (conservatively) be transferred with 17 bits/token ($2^{17} > 128,000$); in practice, tokens could be compressed much more than this, likely to less than one bit per token. We provide estimates for the communication requirements of the model doing inference using data from both inference throughput (most relevant) and training throughput (additional sanity check). Note that the inference data is based on a smaller pod size, whereas throughput per GPU may be much higher for a larger pod size (one reason to also consider training throughput data that uses the desired pod size).

### Throughput Communication Requirements Based on Inference Data

Inference throughput reported in the Llama 3.1 paper (Dubey et al., 2024, Figure 24a) is about 5,000 tokens/second for prefill tokens in a 16 GPU setup. Based on this, we can calculate the approximate (input) data requirements as: (5000 tokens/second / 16 GPUs * 128 GPUs/pod * 17 bits/token * 0.125 bytes/bit) = 85,000 bytes per second = 85 KB/s per pod.

Throughput is typically much lower for decoding (generating tokens), e.g., about 1,000 tokens/second for Llama (Dubey et al., 2024, Figure 24b). This would correspond to 17 KB/s (upload) per pod. As mentioned, it may be ideal for interconnect limits to separately target upload and download. Presently, we'll assume the conservative case that inference is almost entirely prefilling (e.g., one token being generated) as this has the highest communication requirements.

### Throughput Communication Requirements Based on Training Data

Much of the Llama 3.1 405B model training (Dubey et al., 2024, Table 4, Row 2) uses a sequence length of 8,192 and a batch size of 16 per data parallel worker (pod). Therefore, the size of tokens transferred into a pod per batch is (8,192 tokens/sequence * 16 sequences * 17 bits/token * 0.125 bytes/bit) = 278 KB per batch per pod of 128 H100s.

So our estimated communication for tokens needed per pod during training is: (278 KB / 6.2 seconds—calculated earlier as the time to process a training batch) = 45 KB/s.

### Naive Results

To repeat the number calculated above:

- Training gradient communication bandwidth: 131 GB/s (each direction)
- Inference token communication bandwidth: 85 KB/s (download). We will use this figure as our inference communication requirement.
- Inference token communication bandwidth based on training throughput: 45 KB/s (download)

So our initial assessment is that the per-batch communication requirements for tokens would be (131 GB/s / 85 KB/s) = 1.5 million times less than for gradients, assuming no compression of either. This large difference is the key to interconnect bandwidth being a useful governance tool in our approach.

*Caveats*

There are numerous ways in which these calculations are imprecise. For example,

- Generating inference tokens, "decoding", is typically much slower than "pre-filling" (which makes up all of training and some of inference requests)—the above inference throughput is thus an over-estimate.
- Non-text inputs and outputs may have different token requirements.
- Importantly, throughput per GPU likely changes with pod size such that the "tokens per GPU per second" number extrapolated from a 16 GPU setup is insufficient. This is a major source of error, and, unfortunately, there is little public data available about large inference clusters.
- In practice, many techniques are used to make inference more efficient (including increasing throughput), such as using quantization, speculative decoding, pruning, etc. The way to deal with such techniques is likely to adjust pod size downward when a compute operator wants to do higher-throughput inference.
- The inference throughput numbers from the Llama paper appear to be higher than those in various blog posts (Comly et al., 2024; Snowflake AI Research, 2024), but this is acceptable given our goal of upper-bounding inference communication requirements.
- During data parallel training, a pod must also receive tokens to process, so the actual communication requirements during training are (tokens + gradients) in, (gradients) out, per batch.
- Etc.

Fortunately, there is a large space between these two bounds.

Another key limitation is that this calculation does not deal with distributed training, a topic we discuss in more depth in a later section. Based on that discussion, we make a couple of adjustments.

*Adjustments Based on Distributed Training*

In a traditional data parallel approach such as that used in the Llama training, gradients need to be aggregated every training step (i.e., every batch), e.g., every 6.2 seconds for the above estimate. In a high-bandwidth setting, this aggregation is relatively fast, e.g., less than 6 seconds. Data parallel training can, in principle, be bottlenecked by the speed at which these gradients can be aggregated, as this time is idle GPU time (i.e., every 6.2 seconds, the GPUs have to wait for the global gradient update before processing the next batch, assuming these aren't done simultaneously as they sometimes are). In this case, the goal of interconnect limits would be to extend this aggregation delay by sufficiently long, such that training is highly inefficient. Naively, if transferring a copy of gradients took 60 seconds, this would increase the required training time by ~10x compared to near-instantaneous synchronization (from ~6 to ~60 seconds per batch).

Given the discussion of distributed training [below](#), we consider a case where training Llama requires 100x less data to move per gradient synchronization and 100x less frequent gradient synchronizations: 8.12GB every 620 seconds; we also assume gradient aggregation occurs in parallel with computation. So, we consider the case of distributed training that can parallelize across workers with a bandwidth of 8.12 GB / 620 seconds, or 13.1 MB/s, per direction, as the training run we would like to prevent. This adjustment for distributed training is rough and a large source of error.

*Limits Suggested by This Example Calculation*

The distributed training bandwidth requirements are 13.1 MB/s. Recall that our tokens would like to move at approximately 85 KB/s for inference. Based on these two bounds, and wanting to significantly slow training (e.g., 100x), we recommend an inter-pod interconnect limit of 125 KB/s = 1,000 Kb/s = 1 Mb/s for each direction.

This limit is intended for demonstration purposes. It does not account for future distributed training methods (it only uses approximations of current methods), it assumes no token-compression (whereas token-compression will be very helpful for widening the gap), it is based on rough guesses at key values, and it does not involve knowledge of proprietary infrastructure.

Compute would be unusable if this limit was imposed naively—it would take ~75 days to load the full Llama model into memory! Therefore, these interconnect limits should be implemented with modular networking equipment, such that high bandwidth is available occasionally for large data transfers—e.g., once every 48 hours. We leave those details to future work.

## Interconnect Limits and Distributed Training

We now discuss the vulnerability of this approach to advances in distributed training. Distributed training is a research sub-field with the goal of efficiently training large AI systems across many "islands of compute" (also referred to as: "workers", "pods", "nodes") which have limited connection between them. For instance, the organization Prime Intellect (Prime Intellect Team, 2024) aims to enable training between compute nodes on different continents. Here, we discuss what distributed training means for interconnect bandwidth limits aimed at preventing training. While predicting future advances is difficult, we can examine previous research to obtain rough estimates for how much this might change the interconnect limits necessary. Given our suggested implementation of interconnect limits, the key factor to consider is how much communication each pod requires in a given setup; we assume a central server without strong interconnect limits (e.g., any unmonitored server outside a data center) that stores a central copy of a model, and most processing is occurring in pods which are under a governance regime and have limited interconnect bandwidth.

Federated learning schemes such as DiLoCo (Douillard et al., 2024) take a single model, copy it to multiple pods, have each pod train on some data, and then aggregate across pods. Compared to standard data parallel training, this reduces each pod's communication requirements by a multiple of the number of inner steps taken, e.g., if each pod does 100 steps of training before aggregating across pods, the communication of model weight differences ("pseudo gradients") would happen 100x less frequently.

Douillard et al. (2024, Figure 2) demonstrate similar performance to standard data parallel training for a 150M parameter scale LLM, using 500 local steps (i.e., a 500x decrease in communication frequency).

In follow-up work, Jaghouar et al. (2024, Figure 7) scale to 1.1B parameters and find somewhat worse but still strong performance: shows (Open)DiLoCo slightly underperforming the total-FLOP-adjusted baseline with 125 local steps, while being less performance at 500 local steps.

Similar work from Sani et al. (2024) shows similar federated learning approaches working at the 7B scale, however they do not appear to do the requisite FLOP-adjustment for these results to be directly relevant.

Together, these results indicate that at least a 100x reduction in communication frequency is likely possible via workers training for 100 inner steps before sharing gradients.

Another source of communication reduction is gradient compression. The CocktailSGD method introduced by Wang et al. (2023) combines random sparsification, top-k sparsification, and quantization to compress gradients by 117x without hurting fine-tuning performance.

Some distributed training techniques will stack with one another, likely including DiLoCo-like approaches and gradient-compression approaches. While it's imprecise to simply multiply the estimates from above (e.g., because the gradient-compression results are for fine-tuning), our goal is to obtain rough estimates, so we do this. Stacking two 100x efficiency gains leads to an improvement of 10,000x, which we expect is likely possible, compared to the naive, 16-bit-precision, data parallel, case.

Note that directly decreasing inter-pod communication requirements by 10,000x would not be particularly significant in the standard data parallel setting because this communication is not a major bottleneck. Per Amdahl's Law ("Amdahl's Law," 2024), speeding up one part of a system that only imposes 10% of the time cost can only result in a maximum total time reduction of ~10%. However, the governance mechanism being discussed involves restricting communication between pods, making communication responsible for >99% of the time cost. Therefore, these huge communication reductions directly affect the viability of training in a between-pod bandwidth-restricted setting.

Again, this discussion of specific limits based on existing distributed training research is intended as an example. We leave a more thorough assessment of the relationship between distributed training and interconnect limits to future work.

While the above discussion focuses primarily on data parallel training, we also need to discuss forms of parallelism with smaller communication requirements, specifically pipeline parallelism.

*Pipeline Parallelism*

Pipeline parallelism shards a model vertically across different workers, e.g., the first two layers are on one worker, the next two on another worker, etc. So inter-pod communication corresponds to the activations or gradients for each token. As discussed by Erdil & Schneider-Joseph (2024), data parallelism is more effective when network latency is a limitation (due to only needing to communicate twice per batch), whereas pipeline parallelism or tensor parallelism are advantaged when bandwidth is the limitation.

Relative to token inputs and outputs, pipeline parallelism has only a modest increase in communication requirements: inference inputs and outputs are of the size (batch_size * sequence_length * bits_per_token), while pipeline parallelism involves communicating the activations for tokens between workers: (batch_size * sequence_length * activation_size * bits_per_activation), and technically also (*2 for both forward pass activations and backward pass gradients). While pipeline parallelism often uses mini-batches, this reduces to simply batch_size for our use case because mini-batches serve the purpose of more efficiently processing the same original batch size across pipeline parallel workers.

Assuming batch_size and sequence_length are consistent, the increase by a factor of (activation_size * bits_per_activation * 2 / bits_per_token) is likely large, naively. For instance, in the case of Llama 3.1 405B, the activation_size is 16,384, and activations and gradients are in 16-bit precision. As noted previously, 17 bits per token is a conservative estimate. While there appears to be a difference in communication requirements of at least 1,000x, this range is much narrower than that of data parallelism (recall about 1,500,000x for naive data parallelism).

We consider an example: an AI developer might claim to be doing inference with 128 pods of 128 H100s each, with inter-pod bandwidth of 125 KB/s in each direction. In reality, they could be training a large model with 128-way pipeline parallelism (e.g., a model 128x deeper than Llama). To determine if pipeline parallel training is viable in this setup, we will calculate the throughput of this training process. The training process is restricted to the throughput between workers, which is itself constrained by 125 KB/s interconnect limits. If we assume the model being trained is somewhat optimized for pipeline parallelism, perhaps its d_model is 5,000 (i.e., a very narrow residual stream compared to other components of this huge model). The allowable throughput is (125 KB/s * 8 bit/byte * 1/16 act/bit * 1/5000 tokens/act) = 12.5 tokens/second. This is far too slow for frontier model training. Even with a 100x adjustment to account for factors like gradient compression, 1,250 tokens/second is quite slow: approximately 2000x slower than Llama training throughput (16M tokens / 6.2 seconds = 2.6 million tokens/second).

Efforts to increase the throughput of this system would likely bump into the interconnect bandwidth restrictions which are set only slightly higher than is needed for token-inference, e.g., much lower than that required to transfer activations between pipeline stages.

We note a few potential issues with interconnect limits that target pipeline parallelism:

- Improvements in sparse architectures may allow for much more compressed activation transfers.
- For frontier LLMs, the size of the residual stream scales sub-linearly with total parameter count (Kaplan et al. 2020; Duvet et al., 2024)—much larger models will have only modestly larger residual streams, and thus modestly larger communication requirements for pipeline parallel training. Purposeful efforts to circumvent interconnect bandwidth restrictions could of course expand the possibilities here.
- Additionally, because pipeline parallelism communication depends on context length and batch size, it may be possible to decrease the value of these hyperparameters in order to obscure training (e.g., claim to be doing inference with a batch size of 16,000 but actually be doing training with a batch size of 1); it still seems possible for limits here to substantially slow training, as we saw in the example above.

Previous work has shown promising results in applying pipeline parallelism to distributed training. For instance, Ryabinin et al. (2023) are able to obtain good GPU efficiency when applying a version of pipeline parallelism that they call SWARM parallelism to train 1B models with 500 Mb/s inter-worker bandwidth. This data point indicates that bandwidth requirements will likely need to be below hundreds of Mb/s to be effective (this is a small model and not the focus of interconnect bandwidth restrictions, but many more optimizations are likely possible, so this provides some evidence about larger models).

Feasibility

Feasibility: High; <2 years. This would require downgrading existing interconnect between pods and likely creating new hardware that can modulate interconnect in a verifiable way (merely plugging/unplugging cables would work but may not be a desirable long-term solution). Verification with physical inspections and security cameras appears straightforward. It may be less straightforward to deal with the various security issues with such an approach—namely that pods must only be able to communicate via the throttled channel, and there are no secret communication devices present. We expect these security issues are resolvable.

As discussed, improvements in distributed training could pose an issue for interconnect limits. We think it is unlikely that improvements here will be sufficient to close the gradient—token gap, so interconnect limits could still be useful. However, this means that the particular interconnect limits used may need to change with time. Other than this interconnect limit mechanism, distributed training may pose major problems for governing AI compute, in general.

Restricting large AI training runs by using interconnect limits would also be vulnerable to substantial [algorithmic progress](#) as the compute requirements for some dangerous level of training may drop over time. Interconnect limits may also lose efficacy if there are certain substantial shifts in training methods or the AI development paradigm. For example, it is possible that, in the future, training costs could become dominated by the *inference costs of creating high-quality synthetic data* that models then train on. That is, the model weight updates could be a small fraction of the total training compute. Interconnect limits would not prevent this inference and, therefore, may not slow down training substantially in such a regime (depending on the "training" costs). A change like this might be deemed "algorithmic progress", but it is worth explicitly mentioning because it could involve a similar total compute-efficiency of training, but with more of that compute happening in an un-throttled (inference) manner. The specific implementation details of interconnect limits would have to change in response to changes in AI development practices, and some such changes might render interconnect limits ineffective (e.g., if the between-pod interconnect requirements for training drop below those for inference).

One detail is how to set the interconnect limits based on the context length and batch size desired. Memory footprint is based on numerous factors, including the size of the model, the model architecture, data type/precision, batch size, context length, and whether gradients and optimizer states need to be stored (training). For the case of using interconnect limits to restrict large model inference, these hyperparameters could potentially be rearranged to do a violation, e.g., running inference on a small model with a large context size may be similarly memory intensive to inference on a large model with a smaller context size (thus enabling a potential violation through use of the large model). Therefore, setting the right limits will likely be an ongoing process.

Another issue is larger storage inputs and outputs such as images and video. Interconnect limits may prohibit the use of models that generate images, audio, or videos, if these require large outputs (above the thresholds). We are unsure how to mitigate this issue; one approach would be for many pods to simply not do these activities (i.e., be text-only), and certain pods can have additional oversight applied while dealing with larger data; another approach would be to compress and decompress this information on each side of the bottleneck, but this could be quite difficult. If these limits are set in a way that is aware of both inputs and outputs, these restrictions could apply only to model generations but not inputs (e.g., allowing models to take many modalities as inputs but only produce text).

## Implementation Cost

There are some costs in hiring physical inspectors. Poorly set limits could make inference less efficient. Additional physical security could be expensive, depending on the necessary level. The cost of developing the necessary interconnect and monitoring hardware is likely small, as existing technology could likely be repurposed with minimal modification. Interconnect limits aimed at preventing large training runs likely do not pose a substantial cost on text-inference or smaller training if the limits are set in an informed way and adapted to respond to advances in training efficiency. However, it may be necessary to block the generation of images and other high-bandwidth-requiring content, which would be costly; this may be an acceptable cost given the high-stakes nature of the situation.

## Signatures of High-Level Chip Measures

### Summary

Grant inspectors temporary code access to gather labeled data, enabling classifiers (based on high-level chip activity) to detect deviations without ongoing code access.

### Longer Description

One way to improve on classifying workloads based on high-level chip measures (e.g., power draw, memory bandwidth utilization, etc.) is to use temporary code access in order to get high-quality, labeled data about chip activities, train a classifier with this data, and then use this classifier for workload classification based on high-level chip measures (e.g., including applying it to other data centers). Temporary code access means inspectors can have high confidence that a particular workload is the one that occurred when they took a measurement of chip activity, and then the classifier can compare this chip activity to chip activity at other points in time and check whether other chip activity is sufficiently close to the approved activity.

This mechanism follows from two problems: first, one country would like to show that some chips are not doing prohibited activities but would not like to allow access to sensitive information about those chips, such as the model weights and data; second, classifiers based on high-level chip activity, such as power draw or data flow between chips, which are an initial solution to the first problem, will likely not be robust in the adversarial environment in which they are needed—they will likely be spoofable with minor performance penalties. A solution to both of these problems is to use classification based on high-level chip activity but ensure the classifiers are high-quality by giving them access to in-distribution, labeled data, requiring temporary code access for inspectors. Conceptually, this is similar to gathering a fingerprint or signature of chips doing some particular activity (e.g., "inference on the Llama 3 70B model quantized to 8-bit and run on a 128 H100 cluster", "training GPT-2 on different data distributions"), and then ensuring chip activity stays close to one of the permitted signatures later. This is not fundamentally different from a normal classification approach, but it is likely more robust in the context of international verification.

The default approach to classifying workloads based on high-level chip information is for one country to train a classifier on labeled data they generate themselves, across a wide variety of workloads, on AI chips they have access to. This classifier takes various high-level chip measures (e.g., power draw, network traffic, bandwidth utilization, etc.) as input and outputs the general workload type or perhaps just whether the workload is permitted or not. This classifier would then be deployed to classify workloads being carried out by another country, potentially on different AI chip designs and using different model architectures. By contrast, the present mechanism of using signatures of high-level chip information involves the classifier incorporating data from the country being monitored and their AI activities, where this data is labeled via inspectors with code access.

In the default case, there are two reasons it is difficult to do classification based on high-level chip information. First, there is likely a significant distribution shift from the training data of the classifier to the deployment distribution; for example, the country being monitored could be using different AI model architectures, different software stacks, or potentially even different (proprietary) AI chips than the classifier is based on. Second, this is an adversarial situation where competent state actors may be able to spoof chip activities effectively with limited costs: these actors can run arbitrary code on chips, and many of the chip activity differences are relatively small. For instance, inference often has lower chip utilization than training; production numbers are generally not public, but, for example, training may involve an average hardware utilization of 40%, and inference may involve an average hardware utilization of 20%. In such an example, artificially dropping training utilization to match inference would make training take twice as long (or twice as many chips), which may not be a prohibitive slowdown given the threat model of nation-states being willing to spend substantial resources to violate agreements. Whether it is possible to modulate chip behavior in such a precise manner to trick a classifier is unclear, but given the relevant actors are highly competent and can run arbitrary code, this appears somewhat likely (for spoofing high-level chip activity).

Classifying AI workloads based on chip activities has been successful in previous literature, but this work has not involved an adversarial environment (Copos & Peisert, 2020; Karimi et al., 2024; Köhler et al., 2021). Therefore, it is unclear whether such classifiers would work in an international verification context. Giving the classifier access to on-distribution, labeled data could change the effectiveness substantially in the face of adversaries, who now have a much more narrow target of chip activity they must hit to avoid detection.

This mechanism involves granting inspectors temporary code access, which could pose substantial privacy and security risks. These inspectors would need sufficient access to know that chips are performing permitted activities while their measurements are taken, e.g., that the code they believe is being run is, in fact, the code being run. It is currently unclear how much access would be required to gain confidence here. Because inspectors would have access to highly sensitive information, extreme measures may be necessary to reduce the risk of them leaking this information. For example, they could be required to live onsite and have their communication to their home country restricted for a period of time. If AI development progress is moving quickly, the secrets that inspectors have access to may be obsolete in months or a couple of years.

Once a classifier has been trained, this mechanism would function similarly to general classification on the basis of high-level chip information, except the range of expected values could be quite narrow. Data center operators in the monitored country could claim they are doing some permitted activity by showing that their high-level chip activity matches that of one of the approved signatures, e.g., "inference on the Llama 3 70B model quantized to 8-bit and run on a 128 H100 cluster" or "training GPT-2 on different data distributions". In particular, a large fraction of AI chips will likely be used for inference on a relatively small number of types of workloads (e.g., the best models at the time and the models that provide the best performance by cost at the time), and these large-scale inference workloads could each be matched to a particular signature (if this inference is permitted under the agreement). If the classifier identifies chip activities that do not match an approved signature, inspectors could review the relevant code to see if a treaty violation has occurred or whether the classifier needs to be updated. Generally, the classifier needs to be updated (requiring code access) every time a new type of workload with a unique signature is introduced.

One notable point of uncertainty is that it is unclear how much particular workloads will vary in their chip activity signatures, a crucial aspect of this mechanism being effective. Given the success of previous workload classification attempts (Copos & Peisert, 2020; Karimi et al., 2024; Köhler et al., 2021), we are optimistic that variation will be manageable. This question will depend on the definition of a workload type, i.e., the granularity of a workload. For example, "inference on the Llama 3 70B model quantized to 8-bit and run on a 128 H100 cluster" could have a similar signature to "inference on the Llama 3 70B model quantized to 8-bit and run on a 16 H100 cluster", but this is unclear and would depend on various other factors (such as the other possible workloads). How to implement such a classifier and define workload types is largely an empirical question based on the performance of different approaches, and we leave it to future work.

Given that this approach requires temporary code access, it is also worth discussing why this approach is better than merely providing substantial code access all the time. This approach preserves more sensitive data than general code access. Again, the depth to which a workload needs to be defined is an open question, but high-level chip activities will likely be data agnostic; for example, inference with an AI model answering questions about sports will look very similar to inference of the same model answering questions about cybersecurity. If these are sufficiently similar, inspectors could be given access to the code (including input data) for a model running inference on benign data, but code access would not be needed when this model is conducting inference on sensitive data, such as that relating to national security. If the classifier detects potential violations, it may be necessary to reveal the more sensitive data, but this would not be the case by default.

Feasibility

High; <2 years. Key uncertainties include which high-level chip activity to measure, how to define the granularity of workloads, whether the distribution of chip activity for each workload type will be sufficiently specific, and how to enable inspectors to have enough access without substantial risk of leaking sensitive data. An additional uncertainty is how to respond to the classifier indicating a violation, e.g., how much access inspectors are allowed in such a case or how frequently inspectors can investigate such results. Previous results overall provide optimism about workload classification based on information like a chip's power draw and communication with other chips, and supplementing this with on-distribution data from the production systems being monitored would likely make such an approach effective, plausibly even against competent adversaries.

## Implementation Cost

It could be somewhat expensive and take months to years to mass produce the necessary monitoring equipment (e.g., which measures a chip's power draw or communication to other chips), but this likely does not require inventing substantially new technologies. Generally, permitting code access for international verification is extremely costly; in this case such access is temporary, which is a slight improvement due to data privacy, and we are optimistic that inspectors could pose little risk of leaking sensitive information when various precautions are taken.

# Mechanisms Needed Early

One way to think about the space of verification mechanisms is to ask: which mechanisms could be quickly implemented if there was substantial political will, and which need substantial work in advance or significant serial time? This section lays out which mechanisms are likely to need substantial work in advance of when they are needed:

- FlexHEGs
- Mutually trusted data center and computing infrastructure
- Centralize and track existing chips
- Centralize and track the chip supply chain
- Inference-only chips
- Strong security of model weights (this may be the only chance to monitor inference)

We now discuss each in slightly more depth:

## FlexHEGs

Numerous verification approaches would benefit or require that AI chips be able to implement various functionalities in a secure way. This can often be done on the AI chip itself or via another processor located with the AI chip and enclosed in a tamper-proof enclosure. In both cases, work is needed in advance to design, implement, and test these mechanisms. In the case of adding a secure processor and tamper-proof enclosure to existing chips or servers, the lead time is likely 1-3 years. In the case of fabricating new chips with stronger security, the lead time is likely 2-5 years, including the time it takes to saturate the compute stock. Designing a highly performant processor that is mutually trusted may be a difficult task, and there is limited precedent for it (OpenTitan, n.d.).

One advantage of using an auxiliary processor for governance, other than the main AI chip, is that the performance requirements are much lower.

## Mutually Trusted Data Center and Computing Infrastructure

The general mechanism of partially re-running AI workloads in a trusted data center requires that such a data center exists. Beyond the obvious time needed to build such a facility (e.g., potentially new construction in a neutral country), there are likely to be security difficulties. Numerous aspects of the hardware infrastructure in a facility are likely untrusted by default. While hardware backdoors are not necessarily a concern for most security situations, they are highly relevant here due to the capability of relevant actors, e.g., OC5 in Nevo et al. (2024). There needs to be confidence that none of the many actors along the chip supply chain (e.g., NVIDIA, TSMC, Micron, Intel, etc.) have introduced vulnerabilities to the AI chips in this data center. It is unclear what would be needed to obtain mutual confidence in the security of hardware and software in such a data center—there is some precedent in open-source processors, but they are far less performant than state-of-the-art AI hardware (OpenTitan, n.d.). This might be a major technical lift, including years of collaboration. It is possible that multiple layers of standard security could be sufficient, even with insecure AI hardware, as discussed by Baker and colleagues (Forthcoming).

## Centralize and Track Existing Chips

The general approach to locating AI chips requires locating existing chips that lack sophisticated on-chip mechanisms. Governments should, therefore, begin closely tracking AI chips, both domestically and internationally, where possible. U.S. export controls legally prevent the sale of high-end AI chips to many countries; however, these controls could be supplemented with stronger monitoring and enforcement, e.g., to closely track the resale of H100s throughout 2025.

While it may not be politically feasible, it could assist a future verification regime to also centralize compute, both domestically and internationally, as soon as possible due to fewer countries and data center operators needing to participate. This could include measures such as:

- Expand existing export controls on semiconductors and apply a presumption of denial to countries which are not close U.S. allies (regardless of whether a U.S.-based entity is involved). We note that this could have significant negative effects, e.g., decreasing U.S. influence over AI development in restricted countries, accelerating China's chip supply—it is unclear to what extent these effects would occur.
- Offer tax incentives or infrastructure support to the owners of small quantities of AI chips (e.g., academic institutions, small startups) if they co-locate their chips in large data centers. This could be partially achieved by building strong compute infrastructure that is cheaply accessible to such institutions, obviating the need for them to own their own chips, e.g., through a National AI Research Resource (Executive Order 14110, 2023).

- Require large compute providers to report the location of their AI data centers to domestic regulators. This may be required by the Biden Executive Order on AI (Executive Order 14110, 2023). For the purpose of effectively tracking compute, the threshold for "large data center" likely needs to be lowered, compared to that suggested by the EO, and additional measures should be taken to ensure these reports are accurate (e.g., inspections).

As discussed, putting AI data centers in sensitive military facilities could make verification much more difficult, so we recommend against it.

## Centralize and Track the Chip Supply Chain

Many verification approaches require that secret AI chip production be infeasible. Therefore, efforts to centralize and track the chip supply chain should start early. We leave a more thorough discussion of achieving this to other work. Steps might include: strengthen existing export controls on the semiconductor supply chain and their enforcement; work with countries and companies throughout the chain to ensure high quality information about this supply chain reaches the correct decision-makers.

## Inference-Only Chips

One hope for future verification regimes is that *some* AI activities will be permitted even while others are restricted. For instance, this report discusses restrictions on large training runs. Ideally, it would be very easy to verify that some compute is abiding by an agreement, but this is difficult in part because many AI chips are general processors that can perform a range of AI activities. There are efforts to make AI chips that *excel* at either training or inference, but they are typically not purposefully built to be inefficient for other activities. The present use case has the additional constraint that such chips *cannot practically* be used for prohibited activities. Work is needed to make inference-optimized chips sufficiently inefficient for training that they would serve the use case discussed in this report. This may prove extremely difficult or effectively impossible, especially given the need to be robust to future improvements in training methods (which may target these chips).

If feasible, inference-only chips would present an excellent opportunity to get value out of AI inference without imposing risks (if risks come from training but not this inference). The lead time on developing and testing new chips like this is at least 2 years, likely more, and further time would be needed to mass-produce these chips (i.e., for them to play a key role in a verification regime, they have to comprise a substantial portion of compute).

Notably, the feasibility of making inference-only chips increases substantially if we assume a less capable and covert adversary. Specifically, some current inference-optimized chips are likely difficult to use for training because they conduct operations in a specific order that is used for inference, but they lack the corresponding operations needed for backpropagation in training. A highly competent adversary could likely find ways to bypass this limitation, such as offloading certain operations to other chips or approximating the necessary training operations. However, such chips could still be very useful for domestic regulation, as the AI developers are generally less capable and more law-abiding (e.g., likely to have whistleblowers) than in the international verification context.

Investment in AI chips, in general, may have negative downstream consequences, such as enabling rapid and dangerous AI progress or increasing the risk of AI proliferation to dangerous actors. Investment in specialized AI chips, which can only do some operations (or FlexHEG-type mechanisms, which could flexibly do only certain operations), is likely better as these chips could potentially pose minimal risk while enabling society to benefit from advancements in AI capabilities.

## Strong Security of Model Weights (This May Be the Only Chance to Monitor Inference)

Monitoring AI inference is difficult due to the requirement that *all* inference instances be monitored. One promising approach is to contain the weights of an AI system during training and deployment to a small number of physical locations (e.g., the data center it was trained in and maybe a few other data centers). This, in turn, requires extremely strong security around model weights, and this security being verifiable by another party to be secure against insider threats. As discussed by Nevo et al. (2024), model weight security robust against the most capable nation-state actors could take many years, so it is crucial that progress begins early.

# What Compute Needs to Be Monitored?

A key question for verification regimes is what compute they need to monitor, e.g., only data centers with more than 1,000 H100 equivalents? 10,000? 100,000? While it is infeasible to answer this question conclusively with current information, we aim to describe the key factors in this section.

## Which AI Activities

A key variable is whether policy goals concern pre-training runs, post-training, or inference, as these have different compute requirements. Which of these activities is targeted by policy goals likely depends on their potential for enabling dangerous AI activities. For example, if a potentially dangerous AI model is widely available (e.g., its weights are available for download, or it can be accessed via a public API), monitoring inference may be necessary. On the other hand, if there are models that are generally capable but not specifically able to do the relevant dangerous tasks, risks might arise from fine-tuning (post-training). Hopefully, dangerous models have not been developed, and the risk from them must first route through pre-training. This would be desirable because pre-training tends to have the largest compute requirements, making monitoring and verification easier.

In the future, there are likely to be times when substantial risk arises from inference on existing models. In such cases, effective AI governance is much more difficult than if risks must first route through expensive pre-training. For example, it is currently feasible to do inference on small, non-frontier AI systems using a personal laptop, so if inference on small models is a concern (e.g., if there are highly capable small models that pose catastrophic risks), such compute may need to be monitored.

This report's focus on restricting large pre-training runs reflects that pre-training compute is sometimes a proxy for danger from AI systems, it is a major focus in AI governance research and policy, and it allows for a clear decomposition of the space of verification mechanisms. Pre-training compute is a highly imperfect proxy for risk because risk could arise from other AI activities; this section and report are nevertheless focused primarily on pre-training.

## Algorithmic Progress

As discussed by Pilz et al. (2024) and Tucker et al. (2020), compute becomes more efficient over time due to improvements in *hardware price performance* (i.e., FLOP purchasable per dollar) and *algorithmic efficiency* (i.e., performance per FLOP, also called "algorithmic progress"). These effects mean that, for a given capability level of an AI system, less hardware is needed to train a model to this capability level over time. The rate of hardware price performance improvement is estimated at ~1.3x per year (Epoch AI, n.d.).

The rate of algorithmic efficiency improvements is estimated at ~3x per year (Ho et al., 2024), an extremely rapid rate of progress. Such a rate implies that if training some model requires 10,000 H100s in 2022, it would take about only 1,000 H100s just two years later in 2024. Algorithmic efficiency is especially of note because the development loops are relatively fast: the creation of better training algorithms can be directly translated into the next pre-training runs, and increased efficiency for inference can immediately bring down inference costs. Furthermore, algorithmic progress is largely the result of standard AI research and engineering: intellectual work that may be automated relatively early. Automation of AI R&D could cause algorithmic progress to be even faster than the current rapid pace (Davidson, 2023). This could quickly lead to much lower compute requirements to train AI systems to some capability level. Hardware and algorithmic efficiency improvements indicate that the amount of compute being monitored needs to expand (or begin large) in order to detect training to a given capability level.

## Performance Effect

There is additionally a "*performance effect*" (Pilz et al., 2024; Tucker et al., 2020) where using more compute translates to better AI capabilities, so actors with a relatively large amount of compute may continue to have the most capable AI systems (assuming those efficiency gains are broadly available). Therefore, if policy goals are concerned with the highest capabilities at a given point in time, it may be sufficient to focus on actors who direct large amounts of compute.

## Distributed Training

While algorithmic progress, hardware progress, and performance effects all affect the total amount of compute that may be worrisome, another dimension is how this compute is arranged. Naively, one might expect that compute can be discussed at the data center level, e.g., "Does this data center have enough AI chips to do a large training run?" However, this approach is imprecise because compute which is not geographically close can be pooled to do various AI activities. The field of *distributed training* (Douillard et al., 2024; Jaghouar et al., 2024)—see also "decentralized training" and "federated learning"—is specifically aimed at making it easier to train AI systems across data centers that are not geographically close or which have limited communication between them. Currently, large AI training runs usually occur in one or a small number of data centers (Fist & Datta, 2024): Gemini 1 and 1.5 are trained across multiple data centers (Gemini Team, 2024), the Llama 2 and 3 series of models are trained in one or two data centers (Dubey et al., 2024). Google does not report how many superpods, each with 4,096 chips, are used in that Gemini training, but public estimates place this at 12-19 superpods across an unknown number of data centers (Besiroglu, n.d.).

However, using only a few data centers is not a fundamental constraint, especially given advances in distributed training. This field has demonstrated substantial progress in small-scale training (e.g., academic research), and the evidence indicates that distributed training will also be possible at large scale. Per the Gemini paper (Gemini Team, 2024), emphasis ours: "TPU accelerators primarily communicate over the high speed inter-chip-interconnect, but at Gemini Ultra scale, we combine SuperPods in multiple datacenters using Google's intra-cluster and inter-cluster network (Poutievski et al., 2022; Wetherall et al., 2023; yao Hong et al., 2018). Google's network latencies and bandwidths are sufficient to support the commonly used **synchronous** training paradigm, exploiting model parallelism within superpods and **data parallelism across superpods**."

Notably, further reductions in the bandwidth requirements are [known](#) via federated learning (data parallel workers synchronize gradients less often) (Douillard et al., 2024) or tensor parallelism (Ryabinin et al., 2023). We, therefore, expect distributed training to be viable in the future, at least for some AI developers, especially given the current investment in inter-data center bandwidth from multiple compute providers (Fist & Datta, 2024).

Distributed training advances would make it such that, even if a large amount of total compute (e.g., 100,000 H100s) is needed for a large training run, this compute could be made up of many smaller amounts of compute (e.g., 20 different data centers each with 5,000 H100s). Without distributed training, a monitoring and verification regime would only need to be concerned with data centers that house 100,000 H100s—a large and likely detectable quantity. However, if many smaller compute clusters could be pooled, these smaller clusters also need to be monitored to ensure they are not taking part in a distributed training run. Previous work has termed this method of bypassing compute limits "structured training", following similar approaches in the finance industry (Heim et al., 2024). We leave a thorough literature review of distributed training to further work, but based on a cursory understanding of the literature, we expect distributed training will be feasible with performance penalties that are acceptable for the relevant threat actors in this report (e.g., if the 20 data center split described above cost 30% more than a centralized training run, this would be well within the budget of nation-states hoping to violate an international agreement). Counteracting this threat would require reducing the size of data centers being monitored substantially. It is unclear how much success the field of distributed training will have in the future, but even small progress here could pose a major problem for verification regimes, requiring very invasive monitoring to ensure agreements are not violated.

## Difficulty in Defining AI Chips

This report focuses on AI compute, typically thought of as including high-end data center GPUs and TPUs. For the sake of simplicity, we defer to the criteria for defining advanced (AI) chips as used by the current U.S. export controls (Dohmen & Feldgoise, 2023), based on total performance, performance density, or marketing for data center use. Note that these criteria were updated from their initial specification, following the difficulty in defining AI chips in a way that is robust to a changing landscape of chip production and AI development.

Future advances could enable more hardware to contribute to AI activities. If, for example, consumer gaming GPUs (Grunewald, 2024) could be pooled to effectively carry out a distributed training run, this would pose major problems for a verification regime, and reducing such a risk could pose major issues for privacy. There is precedent for the pooling of compute resources in crypto mining. Fortunately, attempts to aggregate large amounts of consumer compute would likely be detectable via whistleblowers or state intelligence.

While it is currently the case that specialized hardware is used for most frontier AI activities, we caution that this may not remain the case in the future. We leave a comprehensive analysis of which chips should be considered "AI chips" to future work, especially as it depends on technological advancements which are difficult to foresee.

## Costs of Monitoring

Monitoring more compute will be more costly than monitoring less compute. Depending on the mechanisms being used, this monitoring could infringe on privacy rights (e.g., if it is necessary to monitor consumer hardware) or pose security risks (e.g., if it is necessary to monitor military data centers). Specifically, mechanisms like partial re-running of workloads could be quite invasive, whereas security cameras could be much less invasive. The decision about what compute needs to be monitored should be based both on the risk of dangerous AI activities and the costs of monitoring; however, risks from dangerous AI activities are likely an overwhelming consideration in most cases.

We are unsure what compute will need to be monitored in the future, and this is likely to change over time. The key factors include which AI activities pose relevant risks, hardware price performance, algorithmic efficiency, performance effects (more compute leads to better capabilities, all else equal), distributed training, which types of chips can be used for AI workloads, and the costs of monitoring. If these factors develop favorably, monitoring and verification could be relatively easy, e.g., only needing to focus on the 50 largest data centers in the world. However, it seems more likely that, especially over time, monitoring much more compute will be needed to verify treaty compliance.

# Track AI Chips, Not Data Centers

A common theme in compute governance is to discuss "data centers" as the prime target; after all, data centers house AI compute. However, in the context of international verification, we believe it is more efficacious to track AI chips themselves (e.g., from production and monitoring resale) rather than data centers. We lay out a few arguments:

## Argument 1: Worrying Data Centers Will Not Necessarily Be Detectable With High Reliability

Various factors discussed, especially distributed training, mean that small amounts of compute will likely be of concern (e.g., the combination of many geographically separated clusters of 5,000 AI chips). It is likely feasible for a nation-state to hide each cluster of 5,000 AI chips, if they have the chips to begin with. AI chips are not particularly physically distinct.

Some of the newest AI chips require liquid cooling, which might be distinct. However, it seems plausible that there will be high-performance chips which can be air cooled (e.g., some H100s are air cooled). Regardless of whether cooling is liquid or air based, cooling infrastructure might still be distinguishable, relative to data centers that do not house AI chips. This is possible, but we expect it is relatively easy to hide the differential in cooling infrastructure, given the threat model in this report.

Given the current large capital expenditures on data center power infrastructure, some have hypothesized that power may be an effective metric for detecting secret data centers (Wasil, Reed, et al., 2024). This is likely not the case for the distributed training threat model—a data center with 5,000 H100 GPUs has a power requirement of ~6 MW ("NVIDIA DGX H100 Datasheet"; Pilz &

Heim, 2023), similar to ~5,500 U.S. homes (U.S. Energy Information Administration, 2024). This is likely too small of a footprint to serve as a useful tool for detecting secret data centers.

There are many existing data centers with enough power capacity to host these distributed clusters. It's estimated (Pilz & Heim, 2023) that there are currently ~500 data centers globally with >10 MW capacity. Furthermore, U.S. naval reactors have tens to low hundreds of megawatts of electrical output, and there are at least 100 of these reactors ("United States naval reactors"). While it would be non-trivial for a state actor to covertly set up hundreds of megawatts of power for a distributed training run, it seems feasible.

Power-based detection also must deal with improvements in energy efficiency over time. Chip performance per watt is increasing at a rate of ~1.6x per year (Epoch AI, n.d.). Over time, a constant amount of FLOP requires substantially less power. By default, the largest data centers will continue to require large amounts of power, but the distributed training issue compounds with improving hardware efficiency to make small pockets of AI compute even harder to detect.

## Argument 2: AI Chips Are More Precisely the Thing Being Regulated

One option for regulation would be to require that a country tell you about the location of all of its medium or large data centers (e.g., >5 MW). Unfortunately, this might include data centers that are highly sensitive (e.g., running the government's classified networks), so it may be politically difficult to have physical inspections. As mentioned, there are hundreds or thousands of such data centers, so inspecting them all is difficult. Even if physical inspections are permitted, they will end up focusing on whether there are AI chips present and what those AI chips are doing (as this is what's relevant to the AI-related international agreement). Perhaps it would have been easier to simply monitor the chips, to begin with.

Crucially, the data center industry is much larger than just AI. The vast majority (>80%) of data center power consumption is non-AI (Goldman Sachs, 2024). Most data centers that would be captured by a focus on the data center level will simply not have AI chips. It's better to not over-monitor if one can help it.

## Argument 3: The Chip Supply Chain Is Amenable to Regulation

The supply chain for AI chips is fairly narrow (Khan, 2021; Sastry et al., 2024). There are only a few companies that can fabricate high-end chips, and there are numerous other bottlenecks in the supply chain. Therefore, it is comparatively easy to monitor the supply of chips by monitoring, e.g., the <20 fabrication facilities necessary for AI chip production (Sastry et al., 2024; *TSMC Fabs - Taiwan Semiconductor Manufacturing Company Limited*, n.d.).

## Outlook

Overall, international agreements would likely benefit **both** from monitoring data center construction and from continuous monitoring of AI chips from production. If the situation is not very adversarial (e.g., domestic regulation), focusing on data centers may be sufficient, but distributed training and the ability to make secret data centers would likely pose an issue in the international verification context. Therefore, we think focusing specifically on tracking AI chips is likely more useful for the context of international verification.

# Would Highly Capable AI Render Verification Mechanisms Ineffective?

Discussion of the near-term future would be remiss to not discuss the impact of increasingly powerful AI systems. While many of the verification mechanisms discussed in this report could work in a world with technology similar to today's, that might not be the environment in a few years. It is plausible (Davidson, 2023) that AI systems will begin contributing substantially to scientific progress in AI and other fields, the result of which may be rapid improvement in AI capabilities. The rate of technological progress in such a world may be much faster than society is used to, which could pose major challenges for a verification regime.

While this AI-enabled feedback loop may appear far away, so does ambitious international coordination to regulate AI development. If international coordination only happens in such a future, when there are already highly capable AIs—e.g., as effective as expert humans across many tasks—verification could be quite difficult. Many of the verification mechanisms described in this document are being planned, designed, and implemented in a fundamentally different threat landscape: the current one, without crazy-powerful AIs. It seems likely that they will fail to provide protection in this new regime.

For instance, Nevo et al. (2024) focus on data center security and define the highest-level threat actors as "Top-priority operations by the top cyber-capable institutions… Operations roughly less capable than or comparable to 1,000 individuals…spending years with a total budget of up to $1 billion on the specific operation…". While there are very few of these actors in the world today, this kind of operation could be far more common in a world with human-expert-level AI systems. In such a world, top-priority state operations might (Finnveden, 2023) look more like 100,000 AIs that are near or exceed human experts.

Fortunately, advanced AI capabilities may be dual-use: some verification mechanisms can be updated with time, and AIs can also be used to strengthen them. For instance, techniques for capability elicitation (assisting with evaluation efficacy) may benefit from work by automated researchers. However, many mechanisms rely on slow-to-replace hardware, such as on-chip mechanisms. Even mechanisms that have high trust currently—such as cryptography-based

methods—may be vulnerable in this new threat landscape (e.g., due to the development of quantum computers and whatever comes after them, breaking post-quantum cryptography). As another example, superintelligent AI systems may be able to develop new methods of producing advanced semiconductors, bypassing monitoring of the existing AI chip supply chain. It is difficult to elucidate the possibilities here: people alive today are used to GDP growth on the order of 2% per year and the technology R&D speeds that come with that, but a regime where decades worth of progress happens in months would bring major challenges which are difficult to foresee. Additionally, advanced AI capabilities may fail to be dual-use because of misalignment problems: a misaligned AI system deployed to secure a verification regime may act to sabotage (Benton et al., 2024) such a regime if this is better for its goals.

If superintelligent AI capabilities exist in the world where international verification is being conducted, new mechanisms will be needed. It is likely futile to design approaches now which will be robust to such a threat model.

## Verification Approach for All Frontier AI Development Following a Safety Case

Various mechanisms discussed in this report can contribute to verifying that an AI development process follows a safety case. A safety case is a structured argument that AI systems are unlikely to cause a catastrophe. It is made by developers before the development and deployment of AI systems (Buhl et al., 2024; Clymer et al., 2024). Per previous work, there are four types of arguments that are likely to be made in a safety case: inability (the AI system is not capable enough to cause a catastrophe), control (there are control measures around the AI system that render it incapable of causing catastrophe), trustworthiness (the AI system will behave desirably despite having the capacity to cause catastrophe), and deference (trustworthy AI advisors assert that the AI system is safe) (Clymer et al., 2024). Safety cases are likely to be somewhat specific to their deployment environment; for example, a developer might present one safety case for the use case "use this AI system to help with AI R&D internally" and another safety case for the use case "provide this AI to the general public via API".

This section discusses what it might look like to have safety cases be part of an international agreement and, particularly, what mechanisms could be applied to verify that a safety case is followed. It may not be immediately obvious why an AI developer would deviate from a stated safety case. For clarity, some reasons include: it may be technically difficult to follow a safety case, it may be costly to follow a safety case (e.g., requiring many experiments to develop the case and potentially expensive monitoring during deployment), and following a safety case prohibits deployment outside of domains covered by the safety case (e.g., if a safety case applies only to using AIs for AI development, this AI system could not be used to run a pharmaceutical company, even though that may be lucrative).

First, developers would make such a case, and an international body would approve or reject this case. In making such a case, developers will likely present a plan for training, evaluations, and control of their AI system. This will involve aspects such as what model architecture is used, which evaluation datasets will be run during training, what training datasets will be used, what classifiers will be applied to model outputs during deployment, and what the deployment use cases will be. Such a safety case will be the result of many experiments conducted by the model developer.

The overall training process matching the stated training process can occur via partial re-running of AI workloads. Evaluations can also take advantage of partial re-running or be authenticated via TEEs. As discussed previously, partial re-running and using TEEs for evaluations might both have security problems on current hardware: existing AI chips may have vulnerabilities that make them insufficient for this. If that were the case, more access would be needed to ensure the declared training process is followed, e.g., having international inspectors review the code. One difficulty is ensuring various experiments claimed in a safety case were actually run correctly; this could potentially be done by re-running or via audit-trail style approaches that keep an experiment log secure against tampering.

Control arguments in a safety case will be based on inference-time classifiers running. As discussed, one promising approach to verify the implementation of such classifiers is to prevent model weights from leaving a small number of secure locations, and ensure control protocols are properly applied in this small number of data centers.

The other key issue to deal with is ensuring no other AI development happens that does not follow a safety case. If AI compute has been sufficiently secured and large pre-training runs are still a central point of governance, this could be achieved via the previously discussed methods for demonstrating some compute is not being used for large training runs. On the other hand, if frontier AI development does not have substantial compute requirements or compute is not sufficiently monitored, this would be considerably more difficult. In such a scenario, whistleblowers, interviews, and national intelligence operations could be used, but it is unclear how effective they would be at ensuring no prohibited AI development is happening. Tracking relevant personnel, in particular, could be a promising approach, but it poses privacy and sovereignty concerns.

The above mechanisms are generally not technologically mature, so if verification of safety cases is required quickly, it would likely require a large amount of access, such as inspectors directly reviewing code. Nevertheless, the above approach would likely succeed at verifying that AI developers follow a safety case—if the right technologies were developed. On the other hand, as in the case of model evaluations, there is still a need to make the safety cases good, such that following them would entail minimal risk from a particular AI system.

# Hypothetical Futures

Forward-looking AI governance is difficult due to the deep uncertainty about the future of AI development and society. In order to motivate some of the framing choices in this report, e.g., the focus on international agreements, we provide the reader with a few short descriptions of possible futures where this work could be relevant. These are hypothetical stories which seem plausible, not necessarily our mainline predictions.

## Story 1

In the late 2020s, AI systems are able to replace humans in significant portions of AI research and development. It is clear that AI systems will soon have massive impacts on economic productivity and military technology development, motivating the U.S. Government (USG) and the People's Republic of China (PRC) to each consolidate their AI companies and launch national AI projects. One frontier AI system breaks out of the security measures placed around it and begins operating autonomously and copying itself to many rentable servers. This rogue AI system successfully executes a large cyber attack, taking down critical infrastructure and causing hundreds of deaths. The international community correctly identifies that this is a major problem and is able to successfully shut down and delete all of the copies of this particular AI system. The international community further recognizes that this "warning shot" is an early indicator for much larger problems to come and that the current path of AI development is likely to lead to catastrophic harm from advanced AI systems. These risks are concerning, and the ideal situation would be to stop entirely; however, the USG and the PRC are both worried that the other will continue racing ahead. These countries agree to measures to reduce the race pressures they are facing, for instance, restricting the size of new AI training runs (this could be a mild restriction—slowing down—or an intense restriction—fully stopping). Slowing down or stopping enables both sides to spend more time solving technical safety problems and promoting societal adaptation and resilience, but verification measures are needed to ensure both sides uphold this agreement.

## Story 2

An AI system capable of automating almost all human remote work is developed in early 2026 by a U.S. company in collaboration with USG. The U.S. has ~80% of the world's AI chips, and this AI system is not stolen by other nation-states. Due to the significant capabilities of this AI system, the PRC is worried about actors in the U.S. (USG or companies) using this system to undergo an intelligence explosion—rapidly advance AI capabilities to significantly above human level. The PRC threatens a hot war if U.S. AI development continues, due to the strategic risks of an adversary having superintelligent AI systems. In order to avoid a potentially nuclear war, the USG and PRC enter a mostly one-directional agreement where the PRC aims to verify that U.S. AI development is moving slowly. Due to China's limited access to compute, monitoring is mostly applied to the U.S. It's possible such an agreement could also include provisions for benefit sharing, e.g., China gets access to the previous generation of AI models or receives direct economic benefits from U.S. growth.

## Story 3

A U.S. company develops an AI system capable of automating AI development in the 2030s and successfully uses this AI system to solve all key technical AI alignment challenges while improving AI capabilities to a superhuman level. This superhuman AI system is tasked with implementing a global monitoring regime to prevent the development of misaligned AI systems while allowing democratic access to advanced AI capabilities. While the ideas in this report are superseded by much better ideas developed by this superintelligence, actually implementing this monitoring and verification regime requires substantial trust and political buy-in, along with serial R&D time. Reports like this one moved the conversation in the correct direction early and prompted useful technical work that enables better monitoring regimes in such a future. The key point of this story is that some version of an international governance regime to prevent dangerous AI activities (albeit potentially very different from the ideas discussed here) seems inevitable, conditional on humanity developing powerful AI systems while avoiding catastrophe.